\documentclass[paper]{JHEP3}
\usepackage{epsfig,subfigure}
\usepackage{graphicx}
\usepackage{amssymb}
\usepackage{amsmath}
\usepackage{booktabs,multirow,tabularx}
\usepackage{slashed}
\usepackage{cite}
\usepackage{caption}
\usepackage{float}
\usepackage{placeins}
\usepackage{rotating}
\usepackage{lscape}
\usepackage[Q=yes,pverb-linebreak=no]{examplep}

\DeclareGraphicsExtensions{.eps}
\graphicspath{{./}}

\newcommand{\perc}{\%}
\newcommand\sss{\scriptscriptstyle}

\newcommand{\gev}{\,\textrm{GeV}}

\newcommand{\pb}{\,\textrm{pb}}
\newcommand{\tev}{\,\textrm{TeV}}

\newcommand{\ttz}{t\bar{t}Z}
\newcommand{\ttwp}{t\bar{t}W^+}
\newcommand{\ttwm}{t\bar{t}W^-}
\newcommand{\ttw}{t\bar{t}W^\pm}
\newcommand{\tth}{t\bar{t}H}
\newcommand{\ttH}{\tth}
\newcommand{\ttv}{t\bar{t}V}
\newcommand{\ttV}{\ttv}
\newcommand{\ttww}{t\bar{t}W^+W^-}
\newcommand{\tthwp}{t\bar{t}HW^+}
\newcommand{\tthwm}{t\bar{t}HW^-}
\newcommand{\ttzwp}{t\bar{t}ZW^+}
\newcommand{\ttzwm}{t\bar{t}ZW^-}
\newcommand{\tthz}{t\bar{t}HZ}
\newcommand{\tthh}{t\bar{t}HH}
\newcommand{\ttzz}{t\bar{t}ZZ}

\def\beq{\begin{equation}}
\def\beqn{\begin{eqnarray}}
\def\eeq{\end{equation}}
\def\eeqn{\end{eqnarray}}
\def\beal{\begin{align}}
\def\endal{\end{align}}
\newcommand\bb{\bar{b}}

\newcommand\as{\alpha_{\sss S}}

\newcommand\aem{\alpha}

\newcommand\muF{\mu_{\sss F}}
\newcommand\muR{\mu_{\sss R}}

\newcommand\bt{\bar{t}}
\newcommand\bq{\bar{q}}

\newcommand\aNLO{{\sc\small MadGraph5\_aMC@NLO}}
\newcommand\UFO{{\sc\small UFO}}

\newcommand\ML{{\sc\small MadLoop}}
\newcommand\CutTools{{\sc\small CutTools}}
\newcommand\OL{{\sc\small OpenLoops}}
\newcommand\MadFKS{{\sc\small MadFKS}}
\newcommand\IREGI{{\sc\small IREGI}}
\newcommand{\pt}{p_{\sss T}}
\newcommand{\Ht}{H_{\sss T}}
\newcommand{\LOo}{{\rm LO},1}
\newcommand{\LOt}{{\rm LO},2}
\newcommand{\LOth}{{\rm LO},3}
\newcommand{\NLOo}{{\rm NLO},1}
\newcommand{\NLOt}{{\rm NLO},2}
\newcommand{\NLOth}{{\rm NLO},3}
\newcommand{\NLOf}{{\rm NLO},4}
\newcommand{\HBR}{{\rm HBR}}
\newcommand{\xLOo}{{\rm LO~QCD}}
\newcommand{\xLOt}{{\rm LO~EW}}
\newcommand{\xNLOo}{{\rm NLO~QCD}}
\newcommand{\xNLOt}{{\rm NLO~EW}}
\author{S.~Frixione$^a$, V.~Hirschi$^b$, D.~Pagani$^c$, H.-S.~Shao$^a$, M.~Zaro$^{de}$\\
$^a$ PH Department, TH Unit, CERN, CH-1211 Geneva 23, Switzerland\\
$^b$ SLAC, National Accelerator Laboratory\\
$\phantom{^b}$ 2575 Sand Hill Road, Menlo Park, CA 94025-7090, USA\\
$^c$ Centre for Cosmology,  Particle Physics and Phenomenology (CP3),\\
$\phantom{^c}$ Universit\'e Catholique de Louvain, B-1348 Louvain-la-Neuve, Belgium\\
$^d$ Sorbonne Universit\'es, UPMC Univ. Paris 06, UMR 7589, LPTHE,\\
$\phantom{^d}$ F-75005, Paris, France\\
$^e$ CNRS, UMR 7589, LPTHE, F-75005, Paris, France
}
\title{Electroweak and QCD corrections to top-pair hadroproduction 
in association with heavy bosons}
\abstract{We compute the contribution of order $\as^2\aem^2$ to 
the cross section of a top-antitop pair in association with
at least one heavy Standard Model boson -- $Z$, $W^\pm$, and Higgs --
by including all effects of QCD, QED, and weak origin and by working
in the automated {\sc\small MadGraph5\_aMC@NLO} framework.
This next-to-leading order contribution is then combined with that of
order $\as^3\aem$, and with the two dominant lowest-order ones, $\as^2\aem$
and $\as\aem^2$, to obtain phenomenological results relevant to
a 8, 13, and 100~TeV $pp$ collider.
}
\preprint{
 CERN-PH-TH-2015-083\\
 CP3-15-10\\
 LPN15-022\\
 SLAC-PUB-16253\\
 }
\begin{document}

\section{Introduction\label{sec:intro}}
The hadroproduction cross sections for a top pair in association with 
a Standard Model (SM) heavy boson $V$ are interesting in many respects, 
and are thus actively studied by theorists and experimenters alike.
In the context of the SM, they constitute excellent tests for the
EW theory and for its perturbative predictions, and a unique and direct 
probe of the $\ttV$ couplings $\lambda_{\ttV}$ in the cases $V=Z$ and 
$V=H$. On the other hand, by assuming the correctness of the SM description
of $\ttw$ and $\ttz$ production, these processes feature prominently as 
backgrounds in several searches for beyond-the-SM (BSM) signals,
typically characterised by multi-lepton signatures (e.g.~same-sign 
di-leptons or tri-leptons final states), including SUSY~\cite{Barnett:1993ea,
Guchait:1994zk,Baer:1995va}, extra dimensions~\cite{Cheng:2002ab},
and models with heavy top-quark partners~\cite{Contino:2008hi}.
Needless to say, they are also backgrounds to $\ttH$ production itself,
again in the case of multi-lepton signatures.

Experimentally, evidence of $\ttV$ production is hard to obtain
even at the very high LHC c.m.~energies, owing to the smallness of 
the rates and/or to background contaminations. Measurements of
the $\ttw$ and $\ttz$ cross sections extracted from Run-I data have 
been reported by both ATLAS and CMS~\cite{ATLAS-CONF-2012-126,1205.5764,
1303.3239,1406.7830,ATLAS-CONF-2014-038}; they are affected by very
large uncertainties, which are statistically dominated (although not
overwhelmingly so). Conversely, the $\ttH$ cross section has not yet
been measured at the LHC; however, exclusion limits on the SM value 
have been established with several searches that employ a variety of 
Higgs decay channels~\cite{ATLAS-CONF-2012-135,ATLAS-COM-CONF-2013-089,
CMS-PAS-HIG-12-025,CMS-PAS-HIG-13-015,CMS-PAS-HIG-13-020,CMS-PAS-HIG-13-019,
1409.3122,1502.02485}.

From the theoretical point of view, the calculation of $\ttV$ cross 
sections used to represent a very challenging problem beyond the 
leading order (LO); such higher-order calculations have a strong
phenomenological motivation, since these processes are characterised
by large $K$ factors. These computational problems have been 
fully solved at the next-to-leading order (NLO) by the advent of 
modern automation techniques; nowadays, results accurate to the NLO
in QCD are straightforwardly available (see e.g.~refs.~\cite{Lazopoulos:2008de,
Hirschi:2011pa,Garzelli:2011is,Kardos:2011na,Garzelli:2012bn} for $\ttz$
production, refs.~\cite{Hirschi:2011pa,Campbell:2012dh,Garzelli:2012bn}
for $\ttw$ production, and refs.~\cite{Beenakker:2001rj,Beenakker:2002nc,
Dawson:2002tg,Dawson:2003zu,Frederix:2011zi,Garzelli:2011vp} for
$\tth$ production); the most recent among them include matching
to parton showers.

While at present the precision of the theoretical predictions is thus
sufficient for all kind of phenomenological applications, since one is
dominated by experimental uncertainties, the Run II of the LHC, with both
energy and luminosity larger than in Run I, might soon change the situation,
whence the need to increase the precision of the perturbative predictions.
At fixed order, this can be done in two ways: by computing either the 
NNLO QCD corrections, or the NLO electroweak (EW) ones; these, as a rule of 
thumb, are believed to have comparable numerical impacts. Calculations at 
the NNLO in QCD for processes of the complexity of $\ttV$ production
are beyond the scope of the currently available technology. This is
not the case for NLO EW corrections, and the aim of this paper is 
to compute them. We point out that, for what concerns $\ttH$ production,
purely-weak and EW corrections have been presented in 
refs.~\cite{Frixione:2014qaa,Yu:2014cka} respectively, while they are 
currently unknown for $\ttz$ and $\ttw$ production, and are considered 
in this paper for the first time.

We remind the reader that, in the context of a perturbative expansion
where both $\as$ and $\aem$ are treated as small parameters (called
a mixed-coupling scenario in ref.~\cite{Alwall:2014hca}), what is 
meant by ``NLO EW corrections'' is conventional, and might lead to
ambiguities. A more precise terminology, advocated in 
ref.~\cite{Frixione:2014qaa}, uses ``leading NLO'' and ``second-leading NLO''
contributions to denote what are traditionally called NLO QCD and 
NLO EW corrections\footnote{The standard definition of the latter 
neglects the contributions due to the radiation of an extra heavy
boson. See sect.~\ref{sec:setup} and ref.~\cite{Frixione:2014qaa}
for more details.}, respectively. In the case of $\ttV$ production,
these correspond to the ${\cal O}(\as^3\aem)$ and ${\cal O}(\as^2\aem^2)$ 
contributions to the cross sections. In the phenomenological results
to be presented below, we shall include the leading and second-leading
LO terms as well, of ${\cal O}(\as^2\aem)$ and ${\cal O}(\as\aem^2)$ 
respectively.

Our work is performed within the \aNLO\ framework~\cite{Alwall:2014hca}.
Apart from the novelty of the physics results presented here, this
paper constitutes the first application of such a code to the 
fully-automated computation of cross sections that require the
subtraction of QED singularities. We stress, in particular, that
no aspect of the code has been designed or optimised in order to
deal specifically with $\ttV$ production, in keeping with the 
general strategy that underpins \aNLO. We also remark that the
calculation of EW corrections and their automation is receiving a
growing attention from the high-energy physics community (see 
e.g.~refs.~\cite{Actis:2012qn,Denner:2014ina,Denner:2014bna,Kallweit:2014xda}
for recent results).

This paper is organised as follows. In sect.~\ref{sec:setup} we 
introduce our notation, and briefly describe the calculation as
is performed by \aNLO. In sect.~\ref{sec:results} we present results 
for total cross sections and sample differential distributions. 
We summarise our findings in sect.~\ref{sec:outlook}.

\newpage
\section{Definitions and calculation details\label{sec:setup}}
We adopt, as far as it is possible, the definitions and notations
introduced in ref.~\cite{Frixione:2014qaa}. A generic observable
$\Sigma(\as,\aem)$ in $\ttv$ production can be written at the 
LO as follows:
\beqn
\Sigma_{\ttV}^{\rm (LO)}(\as,\aem)&=&
\as^2\aem\,\Sigma_{3,0}
+\as\aem^2\,\Sigma_{3,1}
+\aem^3\,\Sigma_{3,2}
\nonumber\\*
&\equiv&
\Sigma_{\LOo}+\Sigma_{\LOt}+\Sigma_{\LOth}\,.
\label{SigB}
\eeqn
This equation implicitly defines the leading, second-leading, and 
third-leading $\Sigma_{{\rm LO},q+1}$ contributions in terms of the
coefficients $\Sigma_{3,q}$ that appear naturally in a mixed-coupling
expansion (see sect.~2.4 of ref.~\cite{Alwall:2014hca} for more details).
Analogously, at the NLO one has:
\beqn
\Sigma_{\ttV}^{\rm (NLO)}(\as,\aem)&=&
\as^3\aem\,\Sigma_{4,0}
+\as^2\aem^2\,\Sigma_{4,1}
+\as\aem^3\,\Sigma_{4,2}
+\aem^4\,\Sigma_{4,3}\,,
\nonumber\\*
&\equiv&
\Sigma_{\NLOo}+\Sigma_{\NLOt}+\Sigma_{\NLOth}+\Sigma_{\NLOf}\,.
\label{SigNLO}
\eeqn
We shall retain in our computation the two dominant terms at the LO
and NLO, namely $\Sigma_{\LOo}$, $\Sigma_{\LOt}$, $\Sigma_{\NLOo}$, and
$\Sigma_{\NLOt}$.

The LO contribution $\Sigma_{\LOt}$ vanishes in the case of $\ttw$ 
production, and is numerically rather small in the cases of the cross
sections for electrically-neutral bosons, $\ttz$ and $\ttH$. As discussed
in ref.~\cite{Frixione:2014qaa}, the latter two final states stem
at this perturbative order from partonic processes with a $b\bb$ initial
state\footnote{Under the assumption that the CKM matrix be diagonal,
as is done here. Light-quark initial states $q\bq$ are possible when
this assumption is relaxed, but then the corresponding contributions
are CKM-suppressed.}, if only weak effects are considered. However, when 
one includes QED effects, diagrams with an initial-state photon contribute
as well. Partonic processes with one incoming photon also contribute
to the second-leading NLO term $\Sigma_{\NLOt}$. 
While this fact does not pose any problems at the level of short-distance
computations, it requires one to use PDFs that feature photon densities,
and that incorporate QED evolution. From this viewpoint, the situation
is less than satisfactory. The only modern such PDF set is 
NNPDF2.3QED~\cite{Ball:2013hta}, which has the disadvantage of
treating QED effects only at the leading order. While {\em formally} this
degrades the NLO accuracy of (part of) our computation, in practice
it does not constitute a major problem, given that the photon density
is anyhow rather poorly determined at present. In the following, we
shall be assessing carefully the impact of PDF uncertainties on our
predictions, both by employing the NNPDF prescription, and by artificially
setting the photon density equal to zero. We point out that another
consequence of having QED-LL-evolved PDFs is the possibility of
using an arbitrary scheme for the finite parts of the initial-state
QED subtractions. In this paper, we have adopted the $\overline{\rm MS}$
scheme, and have refrained from studying the dependence of our results
on the QED scheme choice for the PDFs.

As was already mentioned, all of our results are obtained by means
of the automatic code \aNLO, which contains all ingredients relevant
to the computations of LO and NLO cross sections, with or without
matching to parton showers. NLO results not matched to parton showers
are obtained by adopting the
FKS method~\cite{Frixione:1995ms,Frixione:1997np} for the subtraction 
of the singularities of the real-emission matrix elements (automated in 
the module \MadFKS~\cite{Frederix:2009yq}), and the OPP~\cite{Ossola:2006us} 
or Tensor Integral Reduction (TIR~\cite{Passarino:1978jh,Davydychev:1991va})
procedures for the computation of the one-loop matrix elements (automated 
in the module \ML~\cite{Hirschi:2011pa,Alwall:2014hca}, which makes use of
\CutTools~\cite{Ossola:2007ax} with OPP and of 
\IREGI~\cite{ShaoIREGI} with TIR, and of an in-house implementation 
of the representation proposed in ref.~\cite{Cascioli:2011va} (\OL)).
The automation of mixed-coupling expansions has now been fully
achieved also in \MadFKS, at variance with the situation of 
ref.~\cite{Frixione:2014qaa}, and the present paper is part of
the ongoing validation effort. We have performed all of the self-consistency
checks available in \aNLO, which are discussed in ref.~\cite{Alwall:2014hca}
(see in particular sect.~2.4.2 of that paper for what concerns one-loop
matrix elements). Here, we mention explicitly the {\em in}dependence
of the cross section of the values taken by the FKS subtraction
parameters~\cite{Frixione:1995ms} $\xi_{cut}$ and $\delta_{\sss I}$, 
which is directly relevant to the newly-implemented QED subtractions.
Furthermore, we have computed to high numerical accuracy (${\cal O}(0.1\%)$)
the LO and NLO contributions both separately and in the course of the same 
numerical simulation, and found full agreement between these two procedures.
We also remark that the second-leading NLO term $\Sigma_{\NLOt}$ can be
organised internally by \aNLO\ in two different ways, which correspond
to seeing it (in an {\em unphysical manner}~\cite{Frixione:2014qaa})
as either an EW correction to $\Sigma_{\LOo}$ or a QCD correction
to $\Sigma_{\LOt}$; we have verified that these two ways lead to
the same numerical results.

The notation of eqs.~(\ref{SigB}) and~(\ref{SigNLO})
may be unfamiliar to most readers. Given that in this paper
we restrict ourselves to the computation of the two dominant
terms at each perturbative order, one can introduce an alternative
notation, which is less precise (see ref.~\cite{Frixione:2014qaa}
for more details) but rather consistent with what has 
been used in the literature so far. Such an alternative
labeling scheme, which we shall adopt extensively in sect.~\ref{sec:results},
is summarised in table~\ref{tab:names}. We stress that the two lines
at the bottom of that table imply:
\beq
\Sigma_{\NLOt}=\Sigma_{\rm NLO~EW}+\Sigma_{\rm HBR}\,,
\label{sigNLOt}
\eeq
with:
\beqn
\Sigma_{\rm NLO~EW}(\ttV)&=&\as^2\aem^2\,
\sum_{X\ne W^\pm,Z,H}\Sigma_{4,1}(\ttV+X)\,,
\label{sigNLOEW}
\\*
\Sigma_{\rm HBR}(\ttV)&=&\as^2\aem^2\,
\sum_{X=W^\pm,Z,H}\Sigma_{4,1}(\ttV+X)\,.
\label{sigHBR}
\eeqn
The two terms on the r.h.s.~of eq.~(\ref{sigNLOt}) are both finite 
and theoretically well defined, and we shall present the corresponding
results separately (rather than only for their sum $\Sigma_{\NLOt}$).
In the vast majority of the results available in the literature,
the analogue of the HBR contribution is simply ignored, on the
basis of the fact that its final states are distinguishable from those
relevant to eq.~(\ref{sigNLOEW}). Unfortunately, such an argument is
rather unphysical, because it cannot be quantified {\em unless} a
proper study is made that uses the decay products of the vector bosons,
and suitable acceptance cuts are imposed on their momenta. A fully 
realistic simulation of this kind would be particularly important were 
the experimental results quoted for cross sections exclusive in exactly 
one heavy boson. Note, finally, that a very similar argument could be
made in the case of the radiation of a photon (of sufficient hardness),
whereas typically cross sections that include QED corrections are
computed fully inclusively in any extra photon (as we do 
in eq.~(\ref{sigNLOEW})).
%%%%%%%%%%%%%%%%%%%%%%%%%%%%%%%%%%%%%%%%%%%%%%%%%%%%%%%%%%%%%%%%%%%
\begin{table}%[h]%[ph]
\begin{center}
\begin{tabular}{l|l}
Label & Meaning \\
\hline
LO QCD & $\LOo$ \\
NLO QCD & $\NLOo$ \\
LO EW & $\LOt$ \\
NLO EW & $\NLOt$; no $pp\to t\bt V_1V_2$ \\
HBR & $\NLOt$; only $pp\to t\bt V_1V_2$\\
\end{tabular}
\end{center}
\caption{\label{tab:names}
Shorthand notation used in sect.~\protect\ref{sec:results}. $V_1$ and 
$V_2$ stand for a Higgs, a $W^\pm$, or a $Z$ boson. HBR is an acronym 
for Heavy Boson Radiation, and for a given $V_1$ understands the sum
over $V_2$. The reader is encouraged to check sect.~\protect\ref{sec:setup} 
for the precise definitions of all the quantities involved.
}
\end{table}
%%%%%%%%%%%%%%%%%%%%%%%%%%%%%%%%%%%%%%%%%%%%%%%%%%%%%%%%%%%%%%%%%%%

\section{Results\label{sec:results}}
In this section we present our predictions for inclusive rates relevant 
to the production of $\ttH$, $\ttz$, $\ttwp$, and $\ttwm$ at a $pp$ collider
with a c.m.~energy of 8~TeV (LHC Run I), 13~TeV (LHC Run II),
and 100~TeV. In the case of the LHC Run II, we shall also study the
four production processes at the level of several differential
distributions. Furthermore, for such a c.m.~energy we shall consider
the implications of a ``boosted'' regime, effectively obtained
by imposing the following final-state cuts:
\beq
\pt(t)\ge 200~\gev\,,\;\;\;\;\;\;
\pt(\bt)\ge 200~\gev\,,\;\;\;\;\;\;
\pt(V)\ge 200~\gev\,.
\label{eq:boosted}
\eeq
In HBR processes, the transverse momentum of the vector boson denoted 
by $X$ in eq.~(\ref{sigHBR}) is not constrained; this implies that, in
the case of identical particles ($X=V$), a single vector boson fulfilling
the last condition in eq.~(\ref{eq:boosted}) is sufficient for the
corresponding event to contribute to the cross section.
While a high-$\pt$ regime might be advocated in the context of
Higgs searches~\cite{Butterworth:2008iy,Plehn:2009rk,Buckley:2013auc}
to increase the relative strength of the signal, in the present case
it is interesting regardless of the nature of the associated heavy
boson, because it is known to enhance the impact of EW effects through
large Sudakov logarithms~\cite{Ciafaloni:1998xg,Ciafaloni:2000df,
Denner:2000jv,Denner:2001gw}. Thus, it allows one to gauge directly
the impact of EW corrections where they should matter most, and hence 
to assess the reliability of predictions that include only NLO QCD 
effects.

We have chosen the particle masses as follows:
\beqn
&&
m_t=173.3~\gev\,,\;\;\;\;\;\;\phantom{aa}
m_H=125~\gev\,,\;\;\;\;\;\;
\\*&&
m_W=80.385~\gev\,,\;\;\;\;\;\;
m_Z=91.188~\gev\,.
\eeqn
All widths are set equal to zero, and the massive modes and Yukawas
are renormalised on-shell. We point out that these settings are 
not hard-coded in \aNLO, but are inherited~\cite{Alwall:2014hca} 
from the adopted \UFO~\cite{Degrande:2011ua} model.
We have chosen the NNPDF2.3QED PDF set~\cite{Ball:2013hta}
(particularly for the reasons discussed in sect.~\ref{sec:setup}) 
that is associated with the value
\beq
\as(m_Z)=0.118\,.
\eeq
Our default EW scheme is the $\aem(m_Z)$ scheme, where we set:
\beq
\frac{1}{\aem(m_Z)}=128.93\,.
\label{aem}
\eeq
We shall also present results in the $G_\mu$ scheme, where:
\beq
G_\mu = 1.16639\cdot 10^{-5}\;\;\;\;\longrightarrow\;\;\;\;
\frac{1}{\aem}=132.23\,.
\label{aemGmu}
\eeq
The central values of the renormalisation ($\muR$) and factorisation 
($\muF$) scales have been taken equal to the reference scale:
\beq
\mu=\frac{\Ht}{2}\equiv 
\frac{1}{2}\sum_i\sqrt{m_i^2+\pt^2(i)}\,,
\label{scref}
\eeq
where the sum runs over all final-state particles. The theoretical 
uncertainties due to the $\muR$ and $\muF$ dependencies have been 
evaluated by varying these scales independently in the range:
\beq
\frac{1}{2}\mu\le\muR,\muF\le 2\mu\,,
\label{scalevar}
\eeq
and by taking the envelope of the resulting predictions;
the value of $\aem$ is kept fixed. In this work, we have
limited ourselves to considering the scale dependence of
$\Sigma_{\LOo}$ and $\Sigma_{\NLOo}$, which corresponds to what
is usually identified with the scale uncertainty of the
QCD cross section. We point out that the calculation of
this theory systematics does not entail any independent runs, being
performed through the exact reweighting technique introduced in
ref.~\cite{Frederix:2011ss}, which is fully automated in \aNLO.
The PDF uncertainties are computed, again through reweighting, 
by following the NNPDF methodology~\cite{Ball:2008by}; we report
the 68\% CL symmetric interval 
(that is the one that contains only 68 replicas out of a total
of a hundred; this is done in order to avoid the problem of outliers, 
which is severe in this case owing to the photon PDF~\cite{Ball:2013hta})

We stress that, because of the choice of PDFs made in this paper,
the present results for $\ttH$ production would not be exactly identical 
to those of ref.~\cite{Frixione:2014qaa} even if QED effects were ignored.
However, the differences are tiny, so that a direct comparison between
the $\ttH$ results of this paper and those of ref.~\cite{Frixione:2014qaa} 
is possible, which allows one to assess the impact of QED-only corrections.

\subsection{Inclusive rates\label{sec:rates}}
We begin by reporting, in table~\ref{tab:hbrtot}, the results relevant
to the individual contributions that enter the definition of a given
HBR cross section. As is implied by eq.~(\ref{sigHBR}), by summing the
relevant entries of table~\ref{tab:hbrtot} one obtains the desired
HBR rate. For example, in the case of $\ttH$ production:
\beq
\sigma_{\HBR}(\ttH)=\sigma(t\bt HH)+\sigma(t\bt HZ)+
\sigma(t\bt HW^+)+\sigma(t\bt HW^-)\,,
\eeq
and analogously for the other processes. Note that HBR cross sections
are inclusive by definition, and cannot be summed; this is evident if one
considers that one given contribution may enter in more than one HBR rate
(e.g.~$\sigma(t\bt HZ)$ contributes to the HBR's of both $\ttH$ and $\ttz$). 
The entries of table~\ref{tab:hbrtot} have a relative integration error 
of about $0.1\%$.
%%%%%%%%%%%%%%%%%%%%%%%%%%%%%%%%%%%%%%%%%%%%%%%%%%%%%%%%%%%%%%%%%%%%%%%
\begin{table}
    \centering
    \begin{tabular}{l|ccc}
          $\sigma (\pb)$  & $8\tev$  & $13\tev$  & $100\tev$ \\
        \hline
        $ \tthh  $ &  $  1.640 \cdot 10^{-4} $   &  $  6.947 \cdot 10^{-4} $   &  $  6.078 \cdot 10^{-2} $ \\
        $ \tthz  $ &  $  2.831 \cdot 10^{-4} $   &  $  1.214 \cdot 10^{-3} $   &  $  1.212 \cdot 10^{-1} $ \\
        $ \tthwp $ &  $  2.918 \cdot 10^{-4} $   &  $  8.996 \cdot 10^{-4} $   &  $  1.982 \cdot 10^{-2} $ \\
        $ \tthwm $ &  $  1.139 \cdot 10^{-4} $   &  $  4.074 \cdot 10^{-4} $   &  $  1.366 \cdot 10^{-2} $ \\
        $ \ttzz  $ &  $  3.373 \cdot 10^{-4} $   &  $  1.385 \cdot 10^{-3} $   &  $  1.209 \cdot 10^{-1} $ \\
        $ \ttzwp $ &  $  5.036 \cdot 10^{-4} $   &  $  1.711 \cdot 10^{-3} $   &  $  4.634 \cdot 10^{-2} $ \\
        $ \ttzwm $ &  $  1.919 \cdot 10^{-4} $   &  $  7.455 \cdot 10^{-4} $   &  $  3.084 \cdot 10^{-2} $ \\
        $ \ttww  $ &  $  1.618 \cdot 10^{-3} $   &  $  7.066 \cdot 10^{-3} $   &  $  7.747 \cdot 10^{-1} $ 
    \end{tabular}
\caption{Total rates for the individual contributions to HBR 
cross sections.}
    \label{tab:hbrtot}
\end{table}
%%%%%%%%%%%%%%%%%%%%%%%%%%%%%%%%%%%%%%%%%%%%%%%%%%%%%%%%%%%%%%%%%%%%%%%

We now present, in turn, the results for the total rates relevant to 
$\ttH$, $\ttz$, $\ttwp$, and $\ttwm$ production. Each of these processes
corresponds to a set of two tables: tables~\ref{tab:ttH_abs} 
and~\ref{tab:ttH_rel} for $\ttH$, tables~\ref{tab:ttZ_abs} 
and~\ref{tab:ttZ_rel} for $\ttz$, tables~\ref{tab:ttWp_abs} 
and~\ref{tab:ttWp_rel} for $\ttwp$, and tables~\ref{tab:ttWm_abs}
and~\ref{tab:ttWm_rel} for $\ttwm$. In the first table of each
set we give the values, in pb, of the various contributions 
to the total cross section, namely LO QCD, NLO QCD, LO EW, NLO EW, and HBR;
at a given c.m.~energy, these results have an integration error which is 
{\em at most} $0.1\%$ times the LO QCD cross section\footnote{The typical 
errors are such that the statistical uncertainties affect the last digit
of the results quoted in the tables.} relevant to that energy.
The two contributions labelled with ``EW'' are also computed by setting the
photon density equal to zero, as explained in sect.~\ref{sec:setup}. 
In the case of the 13~TeV LHC, we also give (in parentheses) the rates 
within the cuts of eq.~(\ref{eq:boosted}). The second table of each set
displays the value of the ratios:
\beq
\delta_{\rm X}=\frac{\sigma_{\rm X}}{\sigma_{\xLOo}}\,,
\label{deltadef}
\eeq
with X equal to NLO QCD, LO EW, NLO EW, and HBR. In other words, for any
given column the entry in the $n^{th}$ row of the second table is equal 
to the ratio of the entry in the $(n+1)^{th}$ row  of the first table 
over the entry in the first row of that table. Except for HBR, 
the results for the ratios $\delta$ are associated with uncertainties. 
These fractional uncertainties are computed by using eq.~(\ref{deltadef}),
with the numerator set equal to the maximum and minimum of either the scale
or the PDF envelope, and the denominator always computed with central scales
and PDFs. Note that the denominator is a LO quantity, at variance with what
is done usually in QCD where the central NLO cross section is used; the 
present choice allows one to treat QCD and EW effects on a 
more equal footing in the context of a mixed-coupling expansion.
In the case of NLO QCD, the uncertainties quoted in the tables are due 
to scale variations (leftmost errors) and PDF variations (rightmost errors); 
in the case of the LO and NLO EW contributions, to PDF variations. 
%%%%%%%%%%%%%%%%%%%%%%%%%%%%%%%%%%%%%%%%%%%%%%%%%%%%%%%%%%%%%%%%%%%%%%%
\begin{table}
    %%% RESULTS FOR TTH: ABSOLUTE CORRECTIONS
    \centering
    \begin{tabular}{c|cccc}
        $\tth$ : $\sigma(\pb)$ & 8 TeV & 13 TeV & 100 TeV \\
        \hline
        $\xLOo$         & $  9.685 \cdot 10^{-2} $  & $  3.617 \cdot 10^{-1} $ ($  1.338 \cdot 10^{-2} $)  & $ 23.57 $  \\  
        $\xNLOo$        & $  2.507 \cdot 10^{-2} $  & $  1.073 \cdot 10^{-1} $ ($  3.230 \cdot 10^{-3} $)  & $ 9.61 $  \\  
        \hline
        $\xLOt$         & $  1.719 \cdot 10^{-3} $  & $  4.437 \cdot 10^{-3} $ ($  3.758 \cdot 10^{-4} $)  & $  1.123 \cdot 10^{-2} $  \\  
        $\xLOt$ 
                            no $\gamma$  & $ -2.652 \cdot 10^{-4} $  & $ -1.390 \cdot 10^{-3} $ ($ -2.452 \cdot 10^{-5} $)  & $ -1.356 \cdot 10^{-1} $  \\  
        \hline
        $\xNLOt$        & $ -5.367 \cdot 10^{-4} $  & $ -4.408 \cdot 10^{-3} $ ($ -1.097 \cdot 10^{-3} $)  & $ -6.261 \cdot 10^{-1} $  \\  
        $\xNLOt$ 
                            no $\gamma$  & $ -7.039 \cdot 10^{-4} $  & $ -4.919 \cdot 10^{-3} $ ($ -1.131 \cdot 10^{-3} $)  & $ -6.367 \cdot 10^{-1} $  \\  
        \hline
        $\HBR$            & $  8.529 \cdot 10^{-4} $  & $  3.216 \cdot 10^{-3} $ ($  2.496 \cdot 10^{-4} $)  & $  2.154 \cdot 10^{-1} $\\
    \end{tabular}
    \caption{Contributions, as defined in table~\ref{tab:names}, to the 
total rate (in pb) of $\ttH$ production, for three different collider 
energies. The results in parentheses are relevant to the boosted scenario,
eq.~(\ref{eq:boosted}).}
    \label{tab:ttH_abs}
\end{table}
%%%%%%%%%%%%%%%%%%%%%%%%%%%%%%%%%%%%%%%%%%%%%%%%%%%%%%%%%%%%%%%%%%%%%%%
%%%%%%%%%%%%%%%%%%%%%%%%%%%%%%%%%%%%%%%%%%%%%%%%%%%%%%%%%%%%%%%%%%%%%%%
\begin{table}
    %%% RESULTS FOR TTH: RELATIVE CORRECTIONS
    \centering
    \begin{tabular}{c|cccc}
        $\tth$ : $\delta(\%)$ & 8 TeV & 13 TeV & 100 TeV \\
        \hline
        $\xNLOo$       & $   25.9  ^{+5.4}_{-11.1}\pm 3.5 $  &  $ 29.7  ^{+
	6.8}  _{-11.1}\pm 2.8 $ ($   24.2  ^{+ 4.8}  _{- 10.6}\pm 4.5$) &  $ 40.8
	^{+ 9.3}  _{- 9.1}\pm 1.0 $ \\  
        \hline
        $\xLOt$        & $    1.8 \pm    1.3 $  & $    1.2 \pm    0.9 $ ($    2.8 \pm    2.0 $)  & $    0.0 \pm    0.2 $  \\  
        $\xLOt$  
                            no $\gamma$ & $   -0.3 \pm    0.0 $  & $   -0.4 \pm    0.0 $ ($   -0.2 \pm    0.0 $)  & $   -0.6 \pm    0.0 $  \\  
        \hline
        $\xNLOt$       & $   -0.6 \pm    0.1 $  & $   -1.2 \pm    0.1 $ ($   -8.2 \pm    0.3 $)  & $   -2.7 \pm    0.0 $  \\  
        $\xNLOt$  
                            no $\gamma$ & $   -0.7 \pm    0.0 $  & $   -1.4 \pm    0.0 $ ($   -8.5 \pm    0.2 $)  & $   -2.7 \pm    0.0 $  \\  
        \hline
        $\HBR$           &    0.88   &    0.89 (1.87)  &    0.91   \\
    \end{tabular}
    \caption{Same as in table~\ref{tab:ttH_abs}, but given as fractions of
corresponding LO QCD cross sections. Scale (for NLO QCD) and PDF uncertainties
are also shown.}
    \label{tab:ttH_rel}
\end{table}
%%%%%%%%%%%%%%%%%%%%%%%%%%%%%%%%%%%%%%%%%%%%%%%%%%%%%%%%%%%%%%%%%%%%%%%

The results for the total cross sections exhibit a few features common
to all four processes considered here. Firstly, the leading NLO term
(NLO QCD) is very large, and grows with the collider energy. Its 
impact is particularly striking in the case of $\ttw$ production, owing
to the opening at the NLO of partonic channels ($qg$) that feature a 
gluon PDF, while no initial-state gluon is present at the LO -- 
in the case of $\ttH$ and $\ttz$ production, one has 
$gg$-initiated partonic processes already at the Born level. As a
consequence of this, the scale uncertainty, which is relatively large
for all processes, becomes extremely significant in $\ttw$ production of
increasing hardness (large c.m.~energy or boosted regime), where it is
predominantly of LO-type because of the growing contributions of 
$qg$-initiated partonic processes. In all cases, the PDF uncertainties
of the NLO QCD term are smaller than those due to the hard scales, 
and decrease with the c.m.~energy.
Secondly, the contributions due to processes with initial-state photons 
are quite large at the LO (except for $\ttw$ production, which has a LO EW
cross section identically equal to zero), but constitute only a small
fraction of the total at the NLO. This is due to the fact that LO EW 
processes proceed only through two types of initial states, namely $\gamma g$ 
and $b\bb$, whereas NLO EW ones have richer incoming-parton luminosities.
Thirdly, as a consequence of the previous point, the uncertainty of the
photon density only marginally increases (if at all) the total PDF 
uncertainty that affects the NLO EW term, while it constitutes a dominant
factor at the LO EW level (for $\ttH$ and $\ttz$). 
%%%%%%%%%%%%%%%%%%%%%%%%%%%%%%%%%%%%%%%%%%%%%%%%%%%%%%%%%%%%%%%%%%%%%%%
\begin{table}
    %%% RESULTS FOR TTZ: ABSOLUTE CORRECTIONS
    \centering
    \begin{tabular}{c|cccc}
        $\ttz$ : $\sigma(\pb)$ & 8 TeV & 13 TeV & 100 TeV \\
        \hline
        $\xLOo$         & $  1.379 \cdot 10^{-1} $  & $  5.282 \cdot 10^{-1} $ ($  1.955 \cdot 10^{-2} $)  & $ 37.69 $  \\  
        $\xNLOo$        & $  5.956 \cdot 10^{-2} $  & $  2.426 \cdot 10^{-1} $ ($  7.856 \cdot 10^{-3} $)  & $ 18.99 $  \\  
        \hline
        $\xLOt$         & $  6.552 \cdot 10^{-4} $  & $ -2.172 \cdot 10^{-4} $ ($  4.039 \cdot 10^{-4} $)  & $ -4.278 \cdot 10^{-1} $  \\  
        $\xLOt$   
                            no $\gamma$  & $ -1.105 \cdot 10^{-3} $  & $ -5.771 \cdot 10^{-3} $ ($ -6.179 \cdot 10^{-5} $)  & $ -5.931 \cdot 10^{-1} $  \\  
        \hline
        $\xNLOt$        & $ -4.540 \cdot 10^{-3} $  & $ -2.017 \cdot 10^{-2} $ ($ -2.172 \cdot 10^{-3} $)  & $ -1.974 $  \\  
        $\xNLOt$   
                            no $\gamma$  & $ -5.069 \cdot 10^{-3} $  & $ -2.158 \cdot 10^{-2} $ ($ -2.252 \cdot 10^{-3} $)  & $ -2.036 $  \\  
        \hline
        $\HBR$            & $  1.316 \cdot 10^{-3} $  & $  5.056 \cdot 10^{-3} $ ($  4.162 \cdot 10^{-4} $)  & $  3.192 \cdot 10^{-1} $\\
    \end{tabular}
    \caption{Same as in table~\ref{tab:ttH_abs}, for $\ttz$ production.}
    \label{tab:ttZ_abs}
\end{table}
%%%%%%%%%%%%%%%%%%%%%%%%%%%%%%%%%%%%%%%%%%%%%%%%%%%%%%%%%%%%%%%%%%%%%%%
%%%%%%%%%%%%%%%%%%%%%%%%%%%%%%%%%%%%%%%%%%%%%%%%%%%%%%%%%%%%%%%%%%%%%%%
\begin{table}
    %%% RESULTS FOR TTZ: RELATIVE CORRECTIONS
    \centering
    \begin{tabular}{c|cccc}
        $\ttz$ : $\delta(\%)$ & 8 TeV & 13 TeV & 100 TeV \\
        \hline
        $\xNLOo$       & $   43.2  ^{+ 12.8}  _{- 15.9}\pm 3.6 $  & $   45.9
	^{+ 13.2}  _{- 15.5}\pm 2.9 $ ($   40.2  ^{+ 11.1}  _{- 15.0}\pm 4.7
	$)  & $   50.4  ^{+ 11.4}  _{- 10.9}\pm 1.1 $  \\  
        \hline
        $\xLOt$        & $    0.5 \pm    0.9 $  & $   0.0 \pm    0.7 $  ($    2.1 \pm    1.6 $) & $   -1.1 \pm    0.2 $  \\  
        $\xLOt$  
                            no $\gamma$ & $   -0.8 \pm    0.1 $  & $   -1.1 \pm    0.0 $  ($   -0.3 \pm    0.0 $) & $   -1.6 \pm    0.0 $  \\  
        \hline
        $\xNLOt$       & $   -3.3 \pm    0.3 $  & $   -3.8 \pm    0.2 $  ($  -11.1 \pm    0.5 $) & $   -5.2 \pm    0.1 $  \\  
        $\xNLOt$  
                            no $\gamma$ & $   -3.7 \pm    0.1 $  & $   -4.1 \pm    0.1 $  ($  -11.5 \pm    0.3 $) & $   -5.4 \pm    0.0 $  \\  
        \hline
        $\HBR$           &    0.95   &    0.96 (2.13)  &    0.85   \\
    \end{tabular}
    \caption{Same as in table~\ref{tab:ttH_rel}, for $\ttz$ production.}
    \label{tab:ttZ_rel}
\end{table}
%%%%%%%%%%%%%%%%%%%%%%%%%%%%%%%%%%%%%%%%%%%%%%%%%%%%%%%%%%%%%%%%%%%%%%%

Other aspects characterise differently the four $\ttV$ processes.
The relative importance of NLO EW contributions w.r.t.~the NLO QCD
ones increases with energy in the cases of $\ttH$ and $\ttz$ production,
while it decreases for $\ttw$ production. At the 8-TeV LHC, NLO EW
terms have the largest impact on $\ttwp$ (about 17\% of the NLO QCD ones),
and the smallest on $\tth$ (2.7\%). This is reflected in the fact that
for $\ttw$ production the NLO EW effects are barely
within the NLO QCD scale uncertainty band; conversely, 
for $\ttH$ and $\ttz$ production NLO EW contributions are amply within
the NLO QCD uncertainties. By imposing at the NLO EW level and at the 13-TeV 
LHC the boosted conditions enforced by eq.~(\ref{eq:boosted}), the change 
w.r.t.~the non-boosted scenario is largest in the case of $\tth$ 
production (by a factor equal to about 6.8); $\ttz$ and $\ttw$ 
behave similarly, with enhancement factors in the range $2.5-3$.
However, for all processes the boosted kinematics are such that 
the NLO EW terms are equal or larger than the scale uncertainties that 
affect the corresponding NLO QCD terms. For both of the processes 
which have a non-trivial LO EW cross section ($\ttH$ and $\ttz$),
the $b\bb$- and $\gamma g$-initiated contributions tend to cancel each other. 
In the case of $\tth$, an almost complete (and accidental) cancellation
(relative to the LO QCD term) occurs at a c.m.~energy of 100~TeV,
while for $\ttz$ it so does at the much lower LHC Run II energy. This 
implies that the impact of EW effects at the 13-TeV LHC is more important 
in the case of $\ttz$ than for $\tth$ production, given that for the
latter process the LO and NLO contributions tend to cancel in the sum 
at this collider energy. However, it is necessary to keep in mind the
observation about the uncertainties induced on the LO EW cross section
by the photon density: a better determination of such a PDF would
be desirable, in order to render the statement above quantitatively
more precise. Finally for $\ttH$ production, 
by comparing the results of table~\ref{tab:ttH_rel}
relevant to the NLO EW terms with those of table~6 of 
ref.~\cite{Frixione:2014qaa} relevant to the weak-only contributions
to the NLO cross section, one sees that the relative impact of QED 
effects decreases with the c.m.~energy and is rather negligible in
the boosted scenario, as expected. These QED effects have the opposite
sign w.r.t.~those of weak origin, and can be as large as half of the
latter at the LHC Run I.
%%%%%%%%%%%%%%%%%%%%%%%%%%%%%%%%%%%%%%%%%%%%%%%%%%%%%%%%%%%%%%%%%%%%%%%
\begin{table}
    %%% RESULTS FOR TTW+: ABSOLUTE CORRECTIONS
    \centering
    \begin{tabular}{c|cccc}
        $\ttwp$ : $\sigma(\pb)$ & 8 TeV & 13 TeV & 100 TeV \\
        \hline
        $\xLOo$         & $  1.003 \cdot 10^{-1} $  & $  2.496 \cdot 10^{-1} $  ($  7.749 \cdot 10^{-3} $) & $ 3.908 $  \\  
        $\xNLOo$        & $  4.089 \cdot 10^{-2} $  & $  1.250 \cdot 10^{-1} $  ($  4.624 \cdot 10^{-3} $) & $ 6.114 $  \\  
        \hline
        $\xLOt$         & 0  & 0  & 0  \\  
        $\xLOt$   
                            no $\gamma$  & 0  & 0  & 0  \\  
        \hline
        $\xNLOt$        & $ -6.899 \cdot 10^{-3} $  & $ -1.931 \cdot 10^{-2} $  ($ -1.490 \cdot 10^{-3} $) & $ -3.650 \cdot 10^{-1} $  \\  
        $\xNLOt$   
                            no $\gamma$  & $ -7.103 \cdot 10^{-3} $  & $ -1.988 \cdot 10^{-2} $  ($ -1.546 \cdot 10^{-3} $) & $ -3.762 \cdot 10^{-1} $  \\  
        \hline
        $\HBR$            & $  2.414 \cdot 10^{-3} $  & $  9.677 \cdot 10^{-3} $  ($  5.743 \cdot 10^{-4} $) & $  8.409 \cdot 10^{-1} $\\
    \end{tabular}
    \caption{Same as in table~\ref{tab:ttH_abs}, for $\ttwp$ production.}
    \label{tab:ttWp_abs}
\end{table}
%%%%%%%%%%%%%%%%%%%%%%%%%%%%%%%%%%%%%%%%%%%%%%%%%%%%%%%%%%%%%%%%%%%%%%%
%%%%%%%%%%%%%%%%%%%%%%%%%%%%%%%%%%%%%%%%%%%%%%%%%%%%%%%%%%%%%%%%%%%%%%%
\begin{table}
    %%% RESULTS FOR TTW+: RELATIVE CORRECTIONS
    \centering
    \begin{tabular}{c|cccc}
        $\ttwp$ : $\delta(\%)$ & 8 TeV & 13 TeV & 100 TeV \\
        \hline
        $\xNLOo$       & $   40.8  ^{+ 11.2}  _{- 12.3}\pm 2.9 $  &  $   50.1
	^{+ 14.2}  _{- 13.5}\pm 2.4 $  ($   59.7  ^{+ 18.9}  _{- 17.7}\pm 3.1
	$) &  $ 156.4  ^{+ 38.3}  _{- 35.0}\pm 2.4 $  \\  
        \hline
        $\xLOt$        & 0  & 0  & 0  \\  
        $\xLOt$   
                            no $\gamma$ & 0  & 0  & 0  \\  
        \hline
        $\xNLOt$       & $   -6.9 \pm    0.2 $  & $   -7.7 \pm    0.2 $  ($  -19.2 \pm    0.7 $) & $   -9.3 \pm    0.2 $  \\  
        $\xNLOt$  
                            no $\gamma$ & $   -7.1 \pm    0.2 $  & $   -8.0 \pm    0.2 $  ($  -20.0 \pm    0.5 $) & $   -9.6 \pm    0.1 $  \\  
        \hline
        $\HBR$           &    2.41   &    3.88 (7.41)  &   21.52   \\
    \end{tabular}
    \caption{Same as in table~\ref{tab:ttH_rel}, for $\ttwp$ production.}
    \label{tab:ttWp_rel}
\end{table}
%%%%%%%%%%%%%%%%%%%%%%%%%%%%%%%%%%%%%%%%%%%%%%%%%%%%%%%%%%%%%%%%%%%%%%%

As far as the HBR cross sections are concerned, some general considerations
about the various mechanisms that govern the (partial) compensation between
these terms and the one-loop contributions of weak origin have already
been given in ref.~\cite{Frixione:2014qaa}; they are not $\ttH$-specific,
and hence will not be repeated here. We limit ourselves to observing,
by inspection of tables~\ref{tab:ttH_rel}, \ref{tab:ttZ_rel}, 
\ref{tab:ttWp_rel}, and~\ref{tab:ttWm_rel}, that relative to the LO
QCD cross sections the $\ttH$ and $\ttz$ HBR contributions have a 
mild dependence on the c.m.~energy (slightly increasing for the
former process and decreasing for the latter one); the NLO EW contribution
tend to become clearly dominant over HBR by increasing the collider  
energy and especially in a boosted scenario. The situation is quite the
opposite for $\ttw$ production, where the growth of the HBR rates is
not matched by that of the NLO EW terms, so that the HBR cross section
is largely dominant over the latter at a 100~TeV collider (but not quite 
so in a boosted configuration at the LHC Run II). The origin of this fact is 
the same as that responsible for the growth of the NLO QCD contributions,
namely partonic luminosities; in particular, the $t\bt W^+ W^-$ final 
state can be obtained from a $gg$-initiated partonic process.
While the above statement must be carefully reconsidered in the context
of fully-realistic simulations, where acceptance cuts are imposed on
the decay products of the tops and of the vector bosons, it does say
that, in such simulations, HBR contributions cannot simply be neglected.
Note that the behaviour with the c.m.~energy of the $\ttwp$ and $\ttwm$
cross sections is not identical, mainly owing to the fact that the former 
(latter) process is more sensitive to valence (sea) quark densities.
%%%%%%%%%%%%%%%%%%%%%%%%%%%%%%%%%%%%%%%%%%%%%%%%%%%%%%%%%%%%%%%%%%%%%%%
\begin{table}
    %%% RESULTS FOR TTW-: ABSOLUTE CORRECTIONS
    \centering
    \begin{tabular}{c|cccc}
        $\ttwm$ : $\sigma(\pb)$ & 8 TeV & 13 TeV & 100 TeV \\
        \hline
        $\xLOo$        & $  4.427 \cdot 10^{-2} $  & $  1.265 \cdot 10^{-1} $  ($  3.186 \cdot 10^{-3} $) & $ 2.833 $  \\  
        $\xNLOo$       & $  1.870 \cdot 10^{-2} $  & $  6.515 \cdot 10^{-2} $  ($  2.111 \cdot 10^{-3} $) & $ 4.351 $  \\  
        \hline
        $\xLOt$        & 0  & 0  & 0  \\  
        $\xLOt$   
                            no $\gamma$ & 0  & 0  & 0  \\  
        \hline
        $\xNLOt$       & $ -2.634 \cdot 10^{-3} $  & $ -8.502 \cdot 10^{-3} $  ($ -5.838 \cdot 10^{-4} $) & $ -2.400 \cdot 10^{-1} $  \\  
        $\xNLOt$  
                            no $\gamma$ & $ -2.761 \cdot 10^{-3} $  & $ -8.912 \cdot 10^{-3} $  ($ -6.094 \cdot 10^{-4} $) & $ -2.484 \cdot 10^{-1} $  \\  
        \hline
        $\HBR$           & $  1.924 \cdot 10^{-3} $  & $  8.219 \cdot 10^{-3} $  ($  4.781 \cdot 10^{-4} $) &  $  8.192 \cdot 10^{-1} $\\
    \end{tabular}
    \caption{Same as in table~\ref{tab:ttH_abs}, for $\ttwm$ production.}
    \label{tab:ttWm_abs}
\end{table}
%%%%%%%%%%%%%%%%%%%%%%%%%%%%%%%%%%%%%%%%%%%%%%%%%%%%%%%%%%%%%%%%%%%%%%%
%%%%%%%%%%%%%%%%%%%%%%%%%%%%%%%%%%%%%%%%%%%%%%%%%%%%%%%%%%%%%%%%%%%%%%%
\begin{table}
    %%% RESULTS FOR TTW-: RELATIVE CORRECTIONS
    \centering
    \begin{tabular}{c|cccc}
        $\ttwm$ : $\delta(\%)$ & 8 TeV & 13 TeV & 100 TeV \\
        \hline
        $\xNLOo$        & $   42.2  ^{+ 11.9}  _{- 12.7}\pm 3.3 $  & $   51.5
	^{+ 14.8}  _{- 13.8}\pm 2.8 $  ($   66.3  ^{+ 21.7}  _{- 19.6}\pm 3.9
	$) & $  153.6  ^{+ 37.7}  _{- 34.9}\pm 2.2 $  \\  
        \hline
        $\xLOt$         & 0  & 0  & 0  \\  
        $\xLOt$   
                            no $\gamma$  & 0  & 0  & 0  \\  
        \hline
        $\xNLOt$        & $   -6.0 \pm    0.3 $  & $   -6.7 \pm    0.2 $  ($  -18.3 \pm    0.8 $) &  $   -8.5 \pm    0.2 $  \\  
        $\xNLOt$  
                            no $\gamma$  & $   -6.2 \pm    0.2 $  & $   -7.0 \pm    0.2 $  ($  -19.1 \pm    0.6 $) &  $   -8.8 \pm    0.1 $  \\  
        \hline
        $\HBR$            &    4.35   &    6.50 (15.01)  &   28.91   \\
    \end{tabular}
    \caption{Same as in table~\ref{tab:ttH_rel}, for $\ttwm$ production.}
    \label{tab:ttWm_rel}
\end{table}
%%%%%%%%%%%%%%%%%%%%%%%%%%%%%%%%%%%%%%%%%%%%%%%%%%%%%%%%%%%%%%%%%%%%%%%

We now turn to discussing how the results presented so far might be
affected by a change of EW scheme. We thus give predictions obtained
in the $G_\mu$ scheme, with the parameters set as in eq.~(\ref{aemGmu});
we limit ourselves to considering the 13-TeV LHC, and do not include HBR
cross sections in this study. We define a quantity analogous to that of
eq.~(\ref{deltadef}) in the $G_\mu$ scheme:
\beq
\delta_{\rm X}^{G_\mu}=\frac{\sigma_{\rm X}^{G_\mu}}{\sigma_{\xLOo}^{G_\mu}}\,.
\label{deltadefGmu}
\eeq
We also introduce the following ratios, that help measure
the differences between analogous results in the two schemes:
\beqn
\Delta^{G_\mu}_{\xLOo} &=& 
\frac{\sigma_{\xLOo}-\sigma_{\xLOo}^{G_\mu}}
{\sigma_{\xLOo}}\,, 
\label{Del1}
\eeqn
\beqn
\Delta^{G_\mu}_{\xLOt} & = & 
\frac{\sigma_{\xLOo} + \sigma_{\xLOt} - 
(\sigma_{\xLOo}^{G_\mu} + \sigma_{\xLOt}^{G_\mu})}
{\sigma_{\xLOo} + \sigma_{\xLOt}}\,, 
\label{Del2}
\\
\Delta^{G_\mu}_{\xNLOt} & = & 
\frac{\sigma_{\xLOo} + \sigma_{\xLOt} + 
\sigma_{\xNLOt} - (\sigma_{\xLOo}^{G_\mu} + 
\sigma_{\xLOt}^{G_\mu} + \sigma_{\xNLOt}^{G_\mu})}
{\sigma_{\xLOo} + \sigma_{\xLOt} + \sigma_{\xNLOt}}\,.\phantom{aaaa}
\label{Del3}
\eeqn
The results are collected in table~\ref{tab:amzvsgmu}, where for ease of
comparison we also report the relevant predictions given previously in
the $\alpha(m_Z)$ scheme (see tables~\ref{tab:ttH_abs}--\ref{tab:ttWm_rel}).

%%%%%%%%%%%%%%%%%%%%%%%%%%%%%%%%%%%%%%%%%%%%%%%%%%%%%%%%%%%%%%%%%%%%%%%
\begin{table}
    %%% COMPARISON OF RESULTS IN THE ALPHA(MZ) / GMU SCHEME, AT 13 TEV
    \centering
    \begin{tabular}{c|cccc}
            & $\tth$  & $\ttz$  & $\ttwp$  & $\ttwm$ \\
        \hline
  $\sigma_{\xLOo}(\pb)$               & $ 3.617 \cdot 10^{-1} $  & $  5.282 \cdot 10^{-1} $  & $  2.496 \cdot 10^{-1} $ & $  1.265 \cdot 10^{-1} $ \\ 
  $\sigma_{\xLOo}^{G_\mu}(\pb)$       & $ 3.527 \cdot 10^{-1} $  & $  5.152 \cdot 10^{-1} $  & $  2.433 \cdot 10^{-1} $ & $  1.234 \cdot 10^{-1} $ \\
  \hline
  $\Delta^{G_\mu}_{\xLOo}(\perc)$     & $2.5 $ & $2.5 $ & $2.5 $ & $2.5 $  \\
  \hline\hline				      	       	        	  
  $\delta_{\xLOt}(\perc)$             & $1.2 $ & $0.0 $ & $0   $ & $0   $  \\ 
  $\delta_{\xLOt}^{G_\mu}(\perc)$     & $1.2 $ & $0.0 $ & $0   $ & $0   $  \\
  \hline				      	       	        	  
  $\Delta^{G_\mu}_{\xLOt}(\perc)$     & $2.5 $ & $2.5 $ & $2.5 $ & $2.5 $  \\
  \hline\hline				      	       	        	  
  $\delta_{\xNLOt}(\perc)$            & $-1.2$ & $-3.8$ & $-7.7$ & $-6.7$  \\ 
  $\delta_{\xNLOt}^{G_\mu}(\perc)$    & $1.8 $ & $-0.7$ & $-4.5$ & $-3.5$  \\
  \hline				      	       	        	  
  $\Delta^{G_\mu}_{\xNLOt}(\perc)$    & $-0.5$ & $-0.7$ & $-0.9$ & $-0.9$  \\
    \end{tabular}
    \caption{Comparison between results in the $\alpha(m_Z)$ and $G_\mu$ 
scheme, at 13 TeV.}
    \label{tab:amzvsgmu}
\end{table}
%%%%%%%%%%%%%%%%%%%%%%%%%%%%%%%%%%%%%%%%%%%%%%%%%%%%%%%%%%%%%%%%%%%%%%%
The scheme dependence of the dominant LO term, $\sigma_{\xLOo}$, is
solely due to the value of $\aem$; thus, the 2.5\% reported in the
third row of table~\ref{tab:amzvsgmu} is simply the relative difference
between the two values of $\aem$ given in eqs.~(\ref{aem}) and~(\ref{aemGmu}),
since this LO term factorises a single power of $\aem$. The smallness
of $\sigma_{\xLOt}$ is such that $\Delta^{G_\mu}_{\xLOt}$,
defined in eq.~(\ref{Del2}), is largely dominated by $\sigma_{\xLOo}$.
Hence its values are also equal to 2.5\% within the numerical accuracy
of our results; by increasing the statistics, one would observe tiny
differences w.r.t.~the predictions for $\Delta^{G_\mu}_{\xLOo}$.
The predictions for the relative differences at the LO imply that
a change of EW scheme may be significant, being of the same order as
the NLO EW relative contributions, in particular in the case of $\ttH$ 
and $\ttz$ production, and slightly less so for $\ttw$ production 
(compare $\Delta^{G_\mu}_{\xLOt}$ with $\delta_{\xNLOt}$).
These higher-order EW results are also affected by a change of EW
scheme, as one can see by comparing the results for $\delta_{\xNLOt}$
and for $\delta_{\xNLOt}^{G_\mu}$ in table~\ref{tab:amzvsgmu}, with the
$G_\mu$ scheme responsible for a systematic shift towards positive
cross sections. However, the most relevant figure of merit is actually
$\Delta^{G_\mu}_{\xNLOt}$, defined in eq.~(\ref{Del3}), which must 
be compared with its LO counterparts, $\Delta^{G_\mu}_{\xLOo}$
and $\Delta^{G_\mu}_{\xLOt}$; the values of the former ratio
are significantly smaller than those of the latter two ratios,
as a result of the stabilisation against changes of scheme that
is characteristic of higher-order computations. 

We conclude this section by mentioning that we have also computed
the LO contributions of ${\cal O}(\aem^3)$ to the total rates, since 
they factor the same power of $\lambda^6$ as the ${\cal O}(\as^2\aem^2)$
NLO terms, according to the naive scaling law $\as\to\lambda\as$
and $\aem\to\lambda^2\aem$. We find that these third-leading LO
rates are smaller (for $\ttH$ and $\ttz$), or much smaller (for $\ttw$,
by a factor of about ten), than the $\xNLOt$ ones; furthermore, they
are not enhanced by any Sudakov logarithms at large hardness. We finally
remark that photon-initiated contributions of ${\cal O}(\aem^3)$ are 
negligibly small. For these reasons, we have not reported any 
${\cal O}(\aem^3)$ results in the tables above, and have ignored
their contributions to differential distributions.

\subsection{Differential distributions\label{sec:distr}}
\begin{figure}[t]
    \centering
    \includegraphics[width=0.42\textwidth]{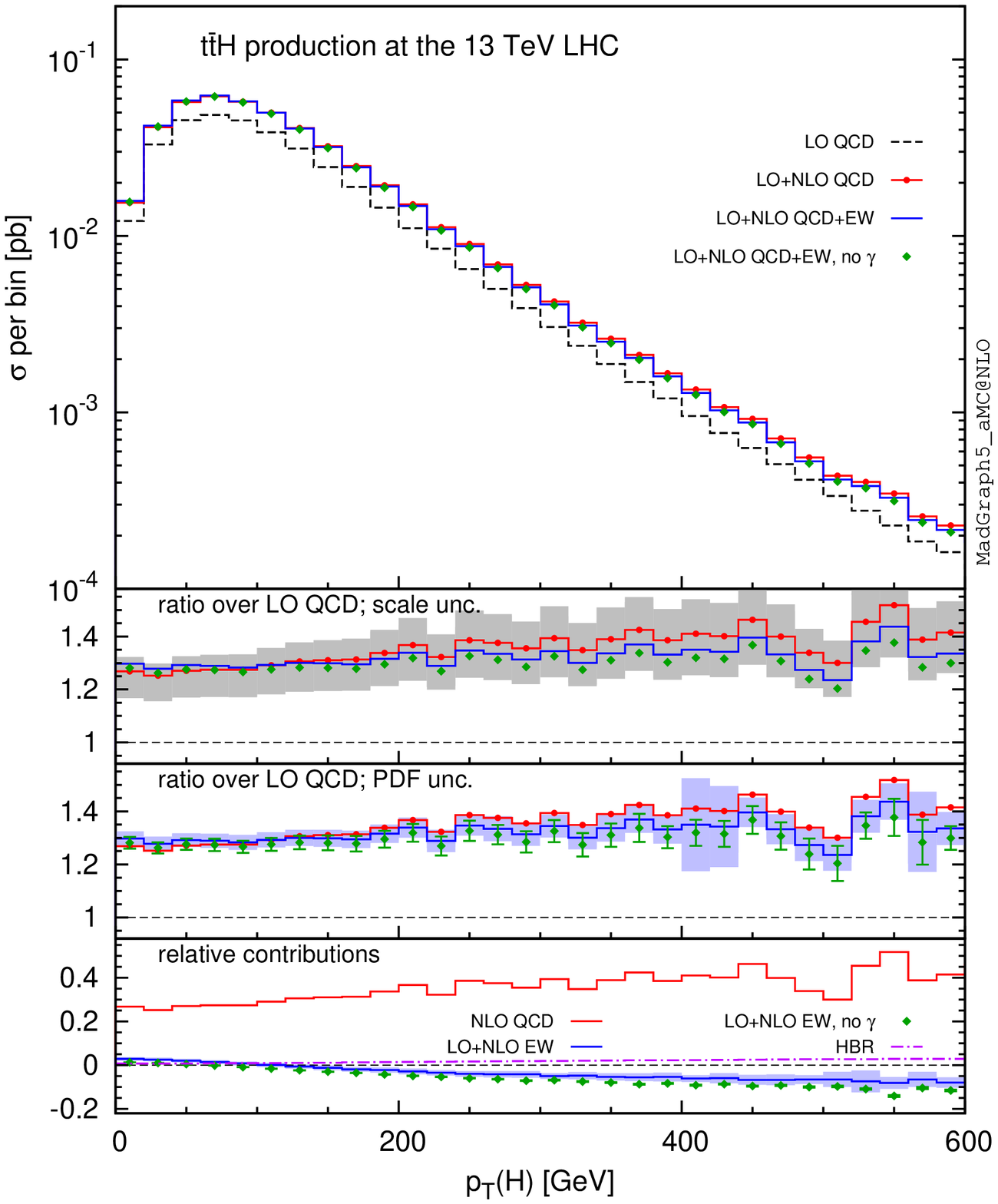}
    \includegraphics[width=0.42\textwidth]{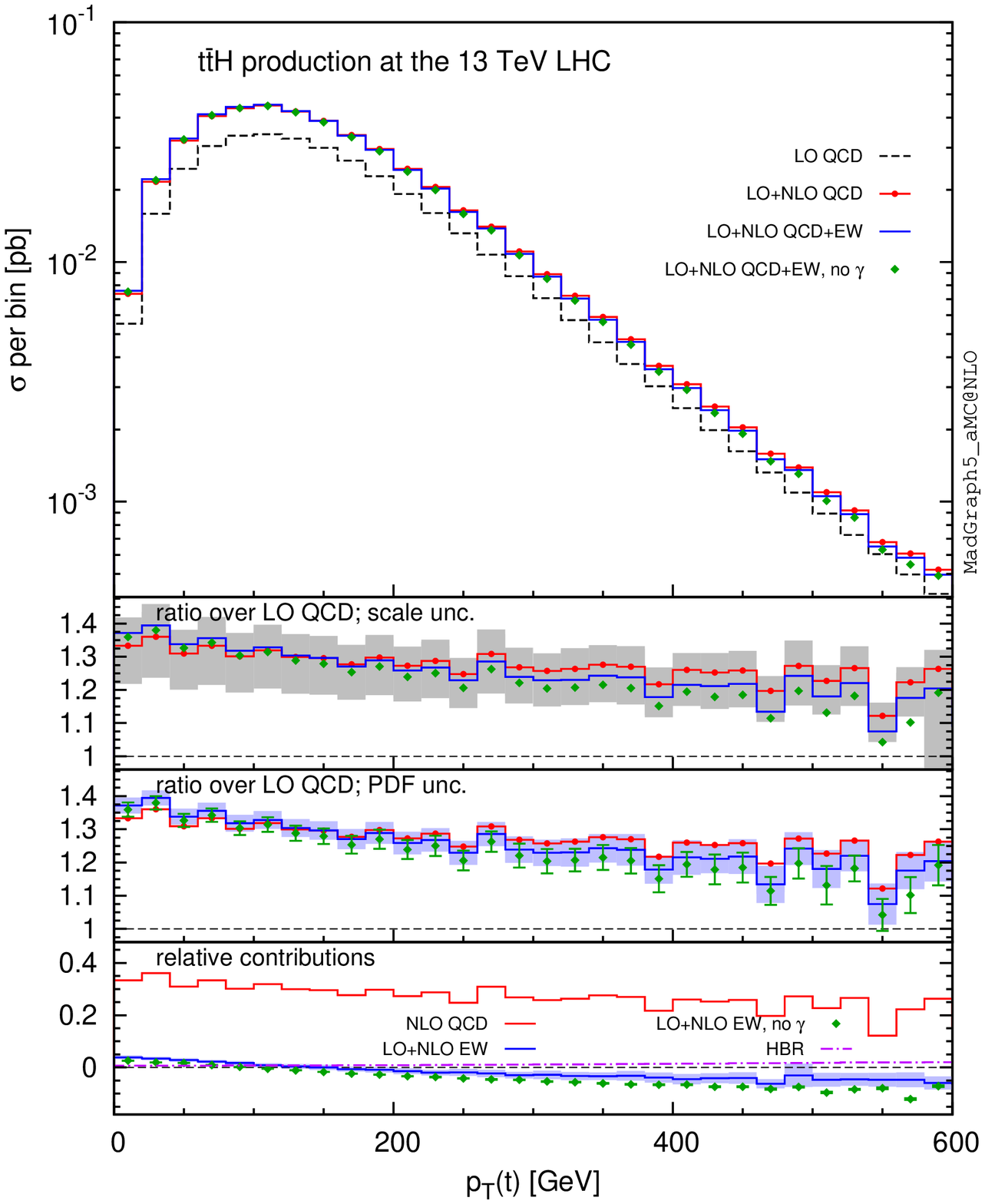}\\
    \includegraphics[width=0.42\textwidth]{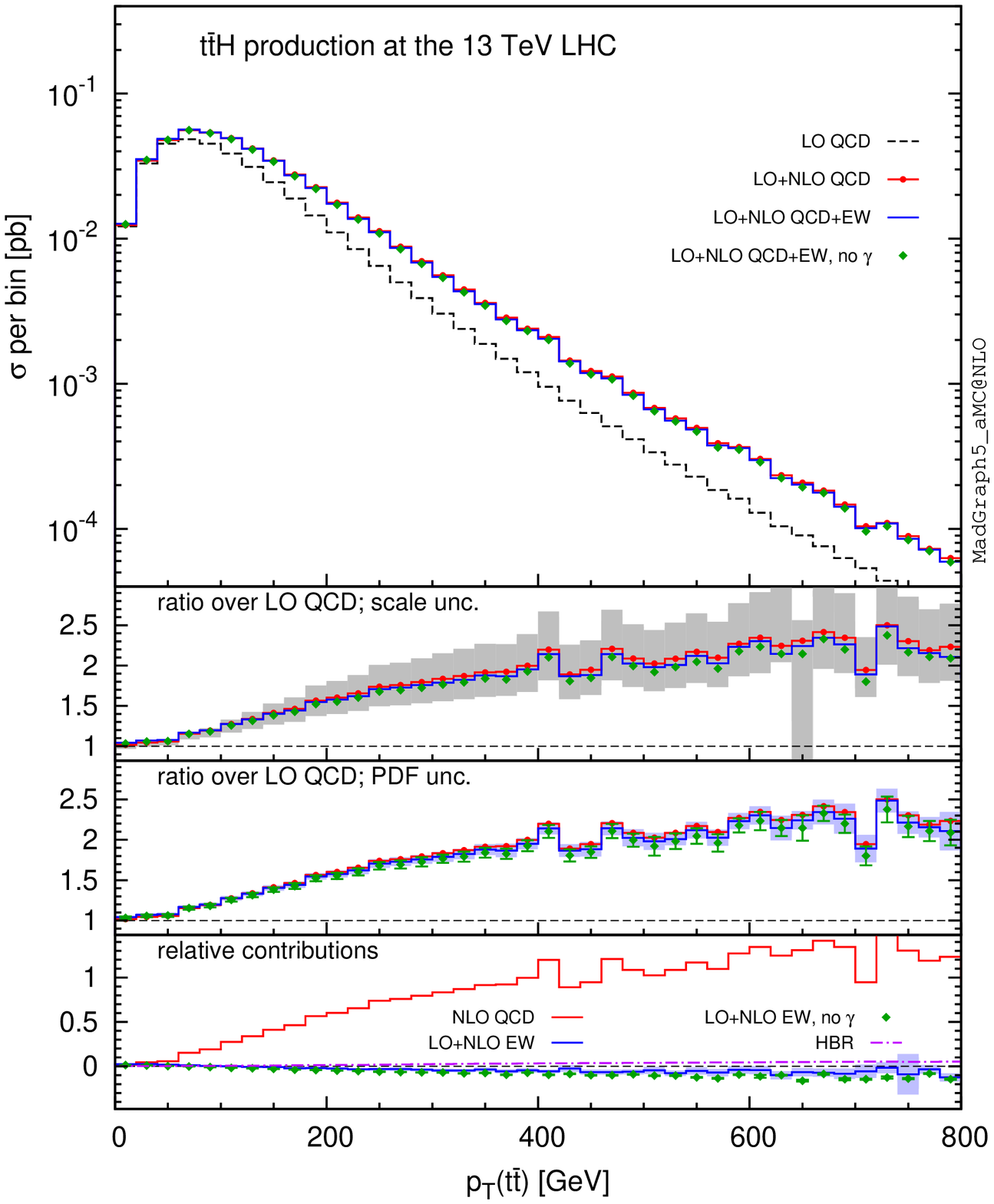}
    \includegraphics[width=0.42\textwidth]{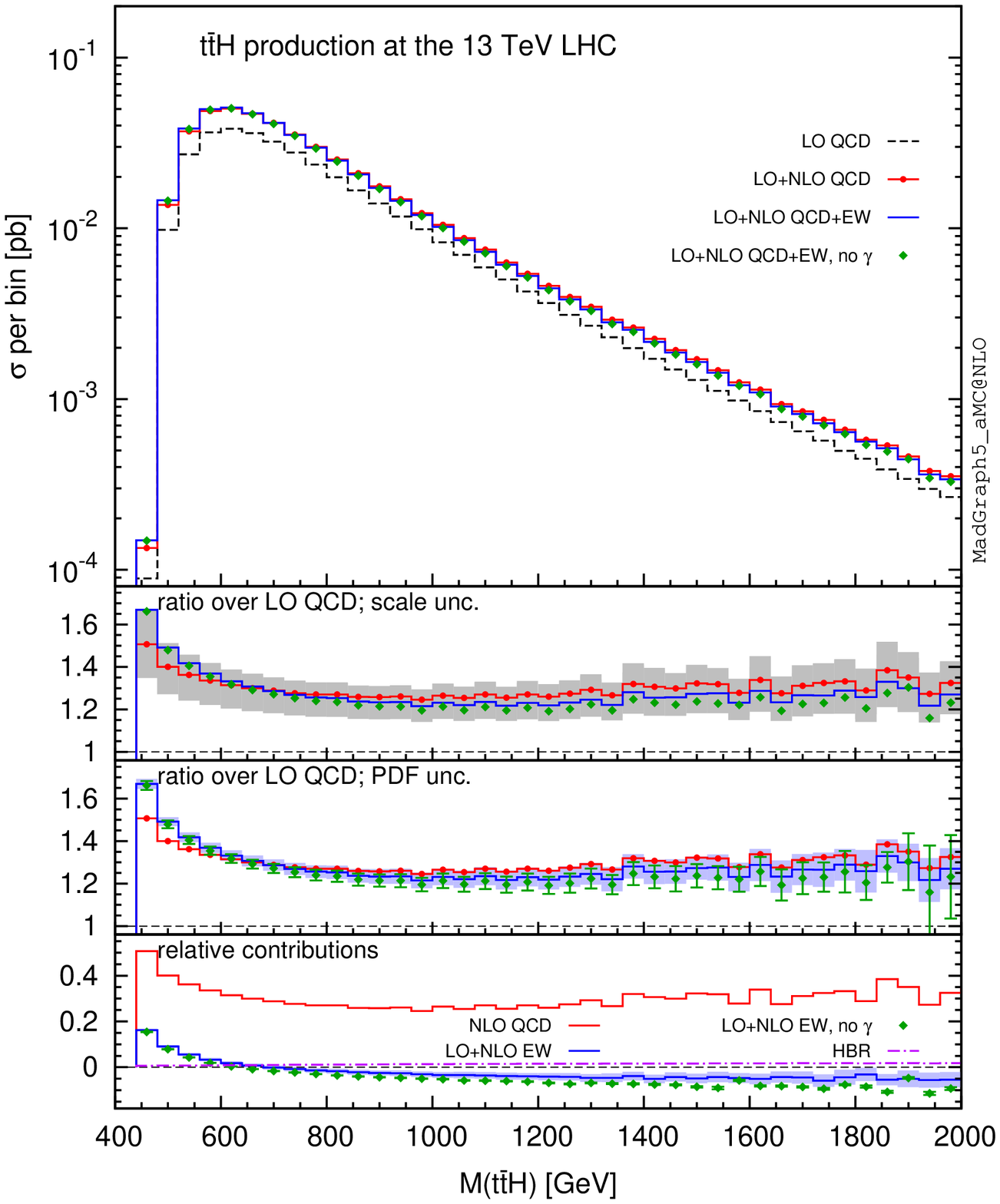}\\
    \includegraphics[width=0.42\textwidth]{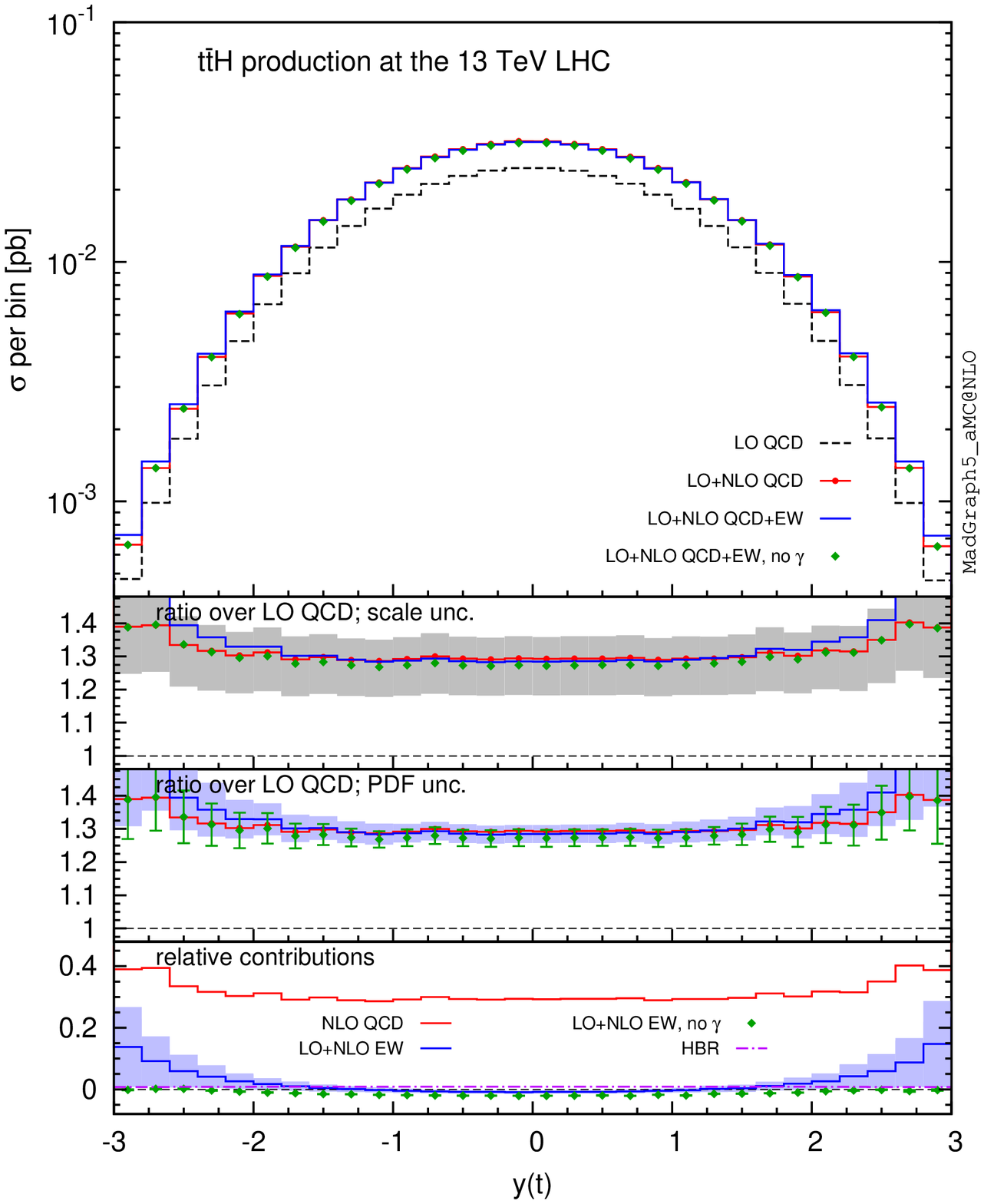}
    \includegraphics[width=0.42\textwidth]{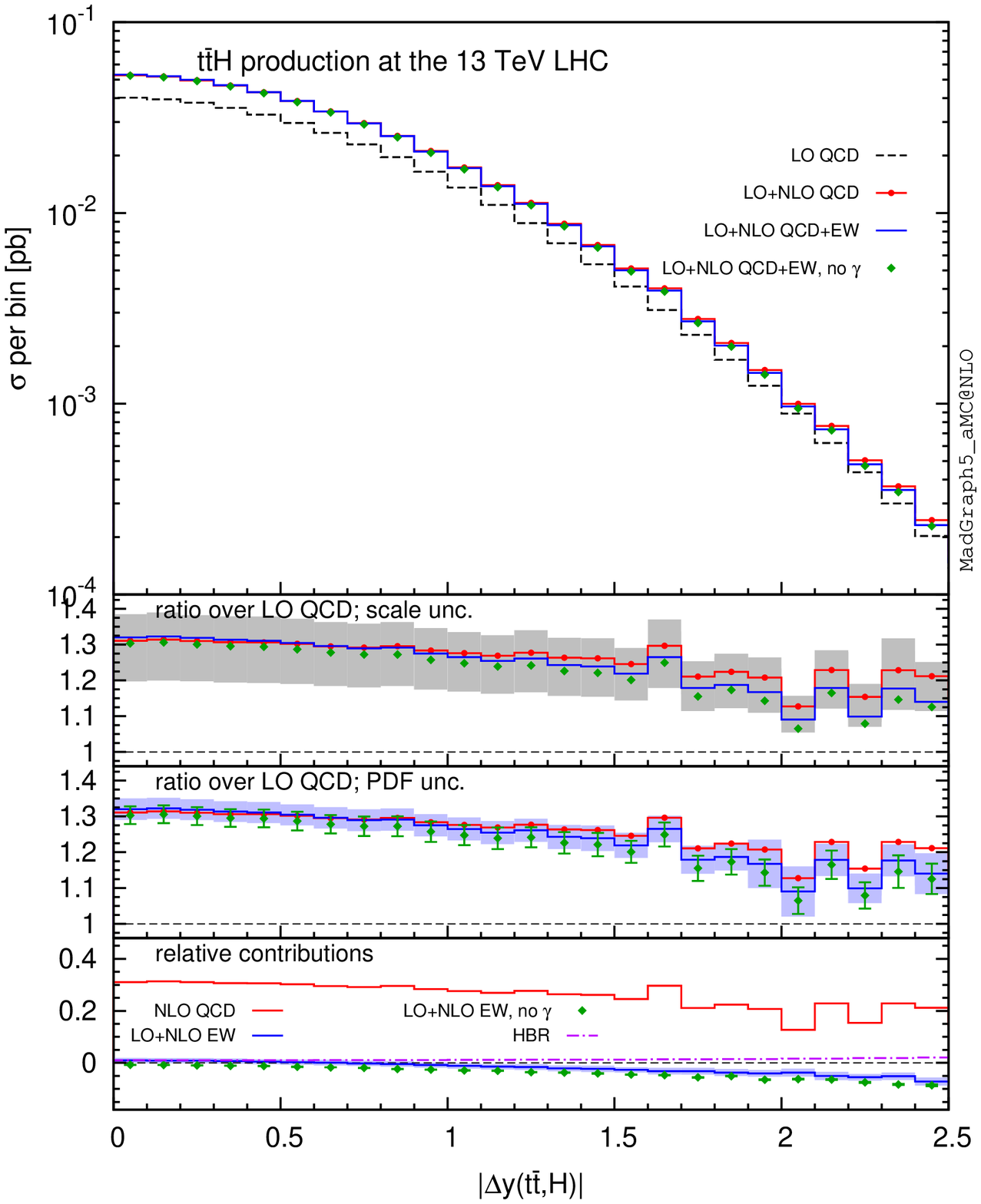}
    \caption{LO- and NLO-accurate results for $\tth$ production at 13 TeV.}
    \label{fig:tth13-nocuts}
\end{figure}

\begin{figure}[t]
    \centering
    \includegraphics[width=0.42\textwidth]{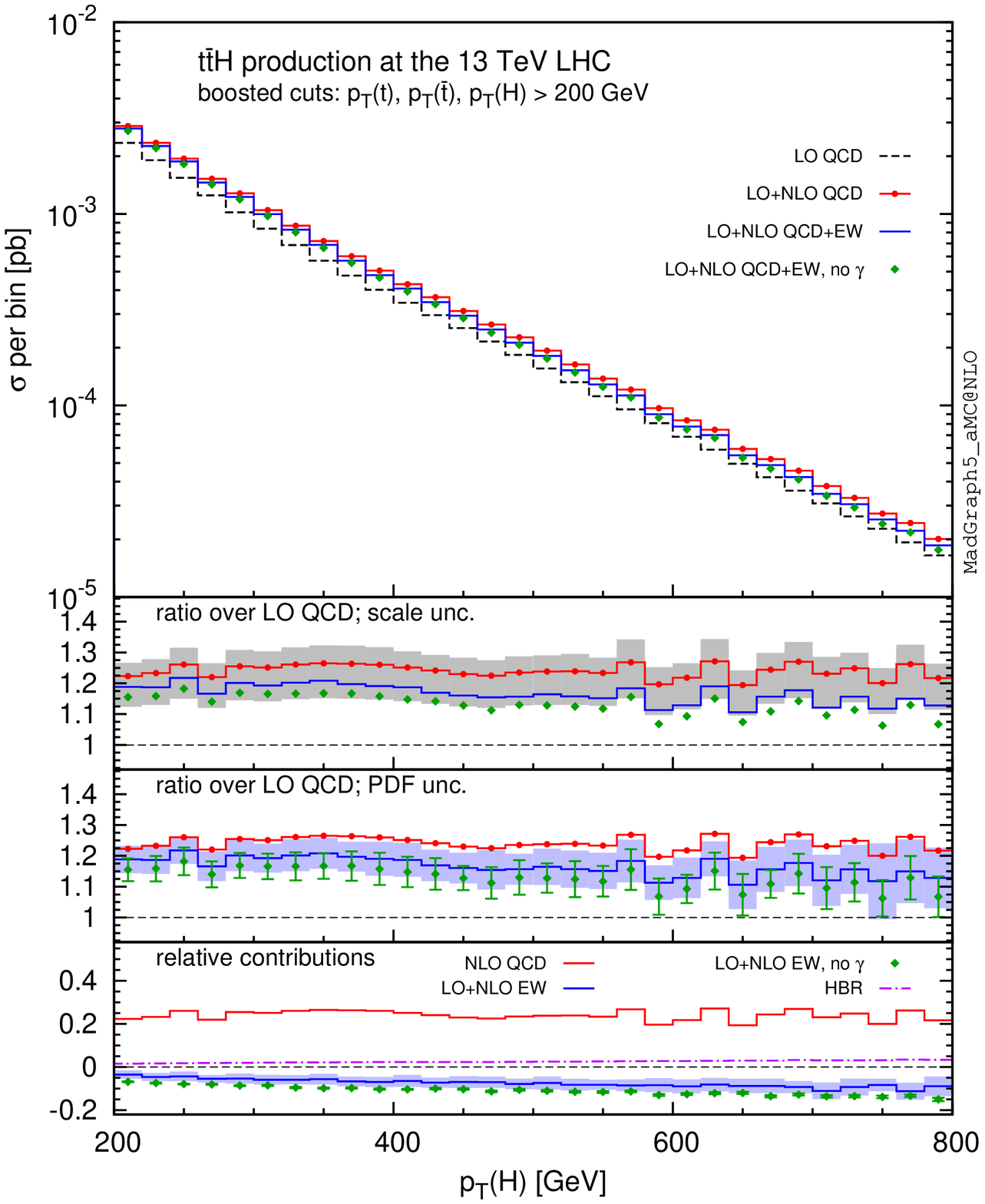}
    \includegraphics[width=0.42\textwidth]{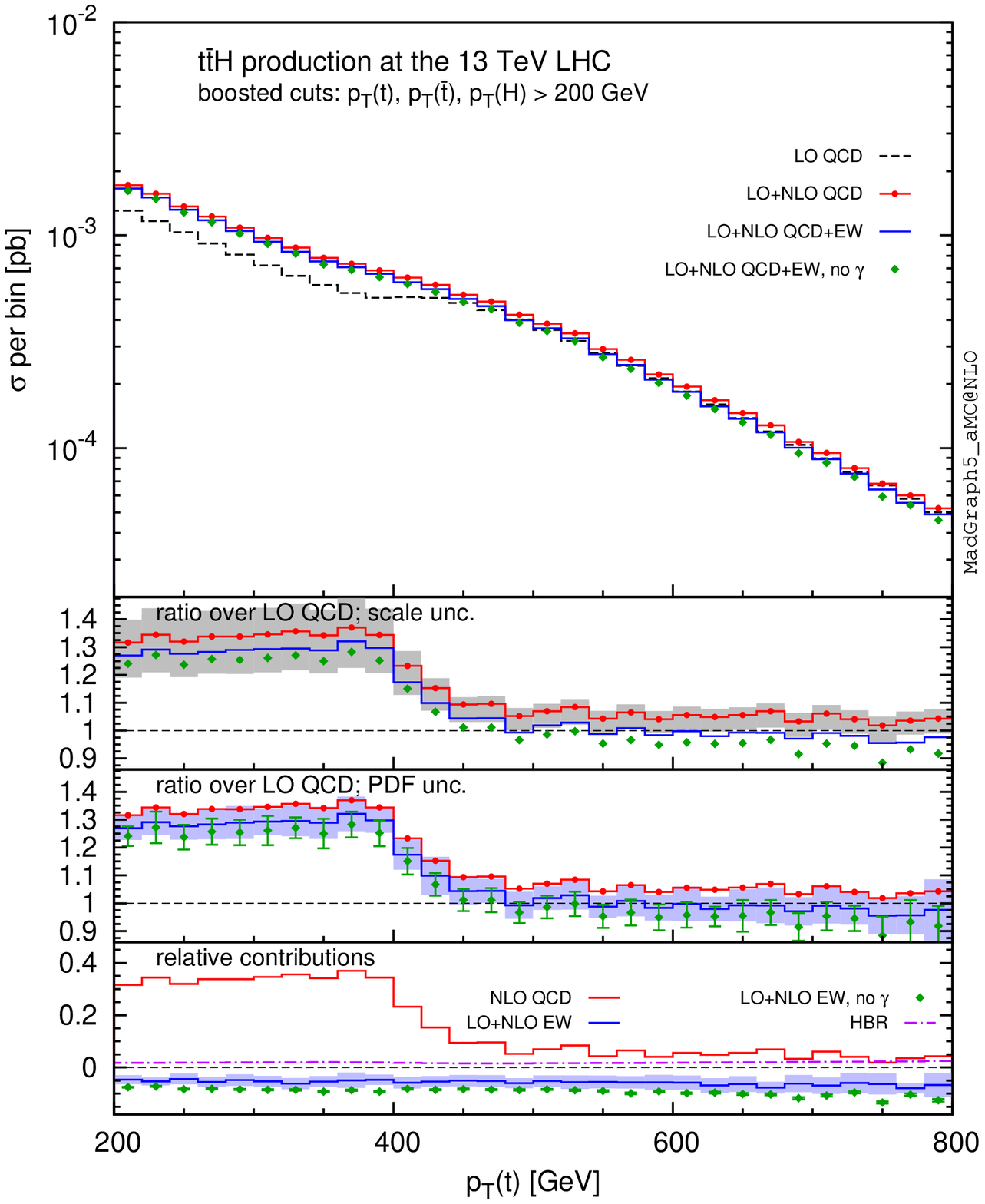}\\
    \includegraphics[width=0.42\textwidth]{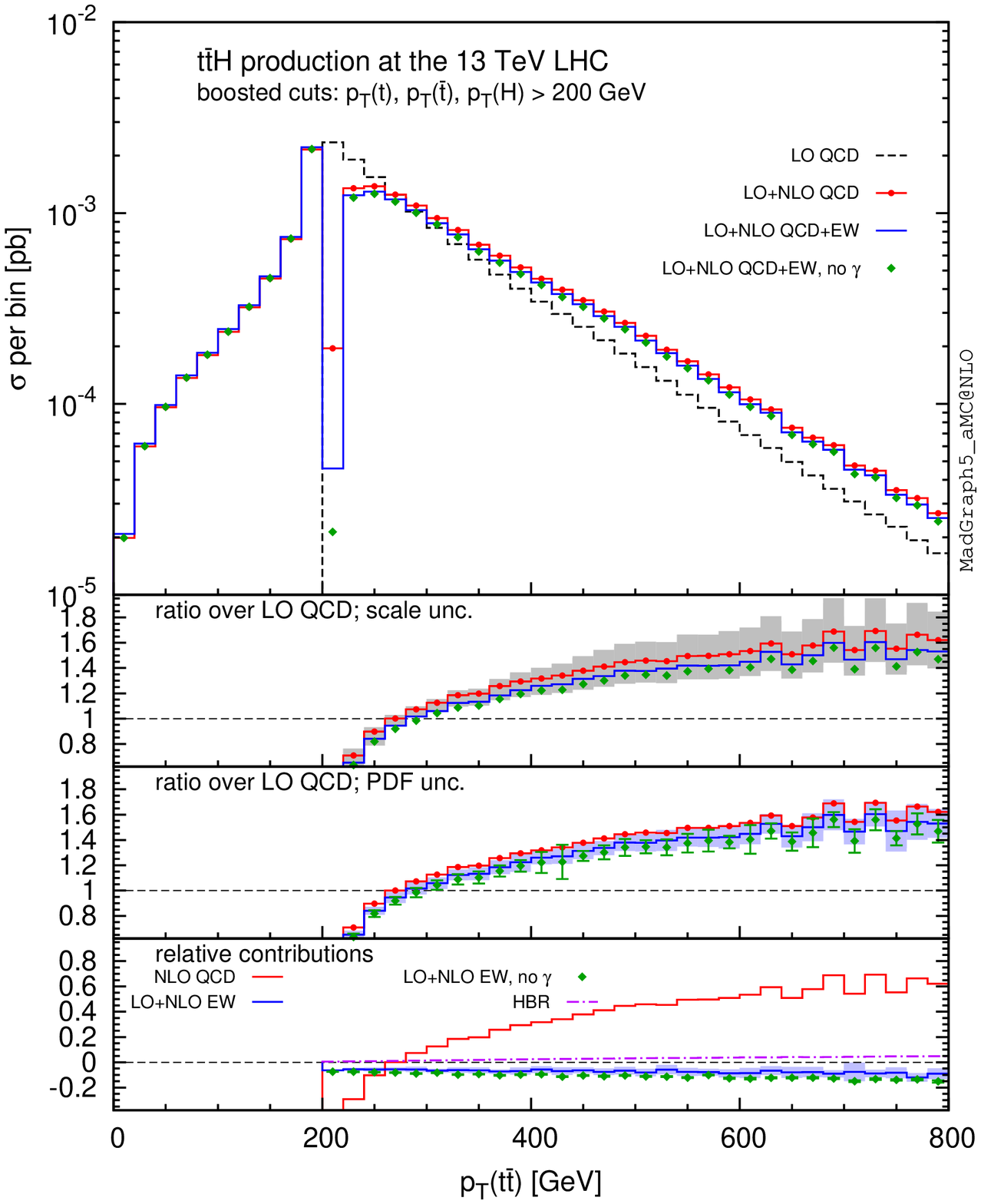}
    \includegraphics[width=0.42\textwidth]{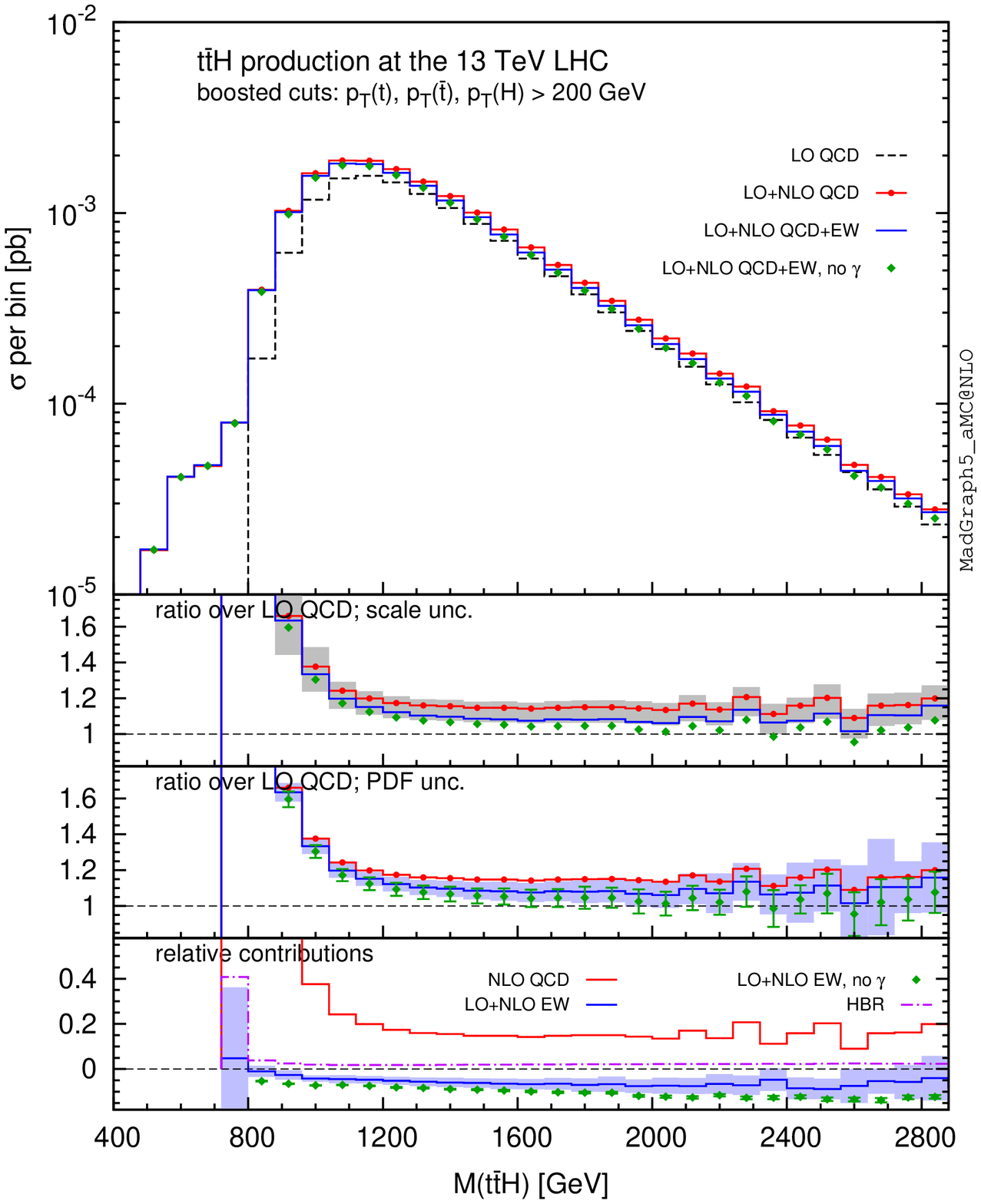}\\
    \includegraphics[width=0.42\textwidth]{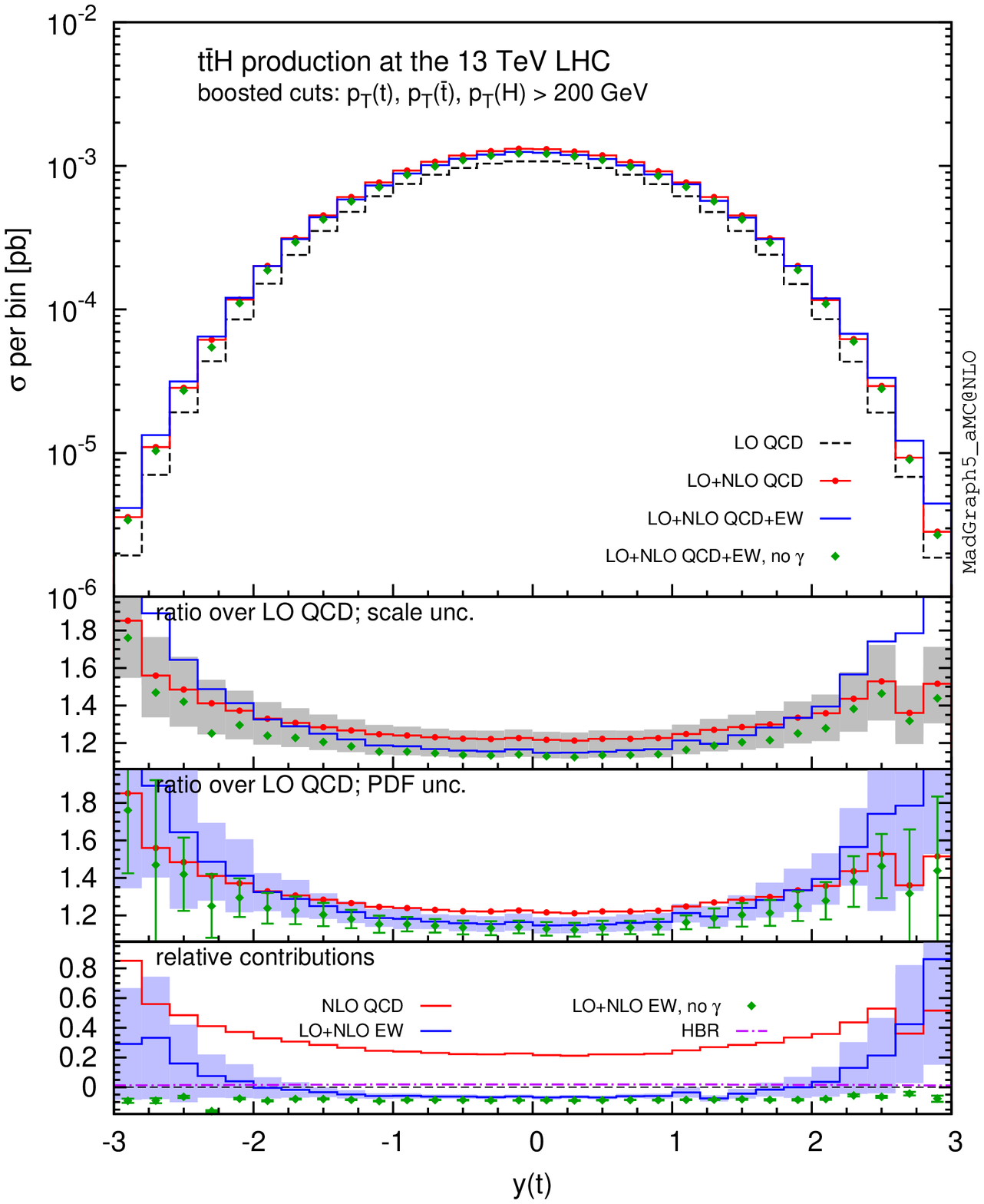}
    \includegraphics[width=0.42\textwidth]{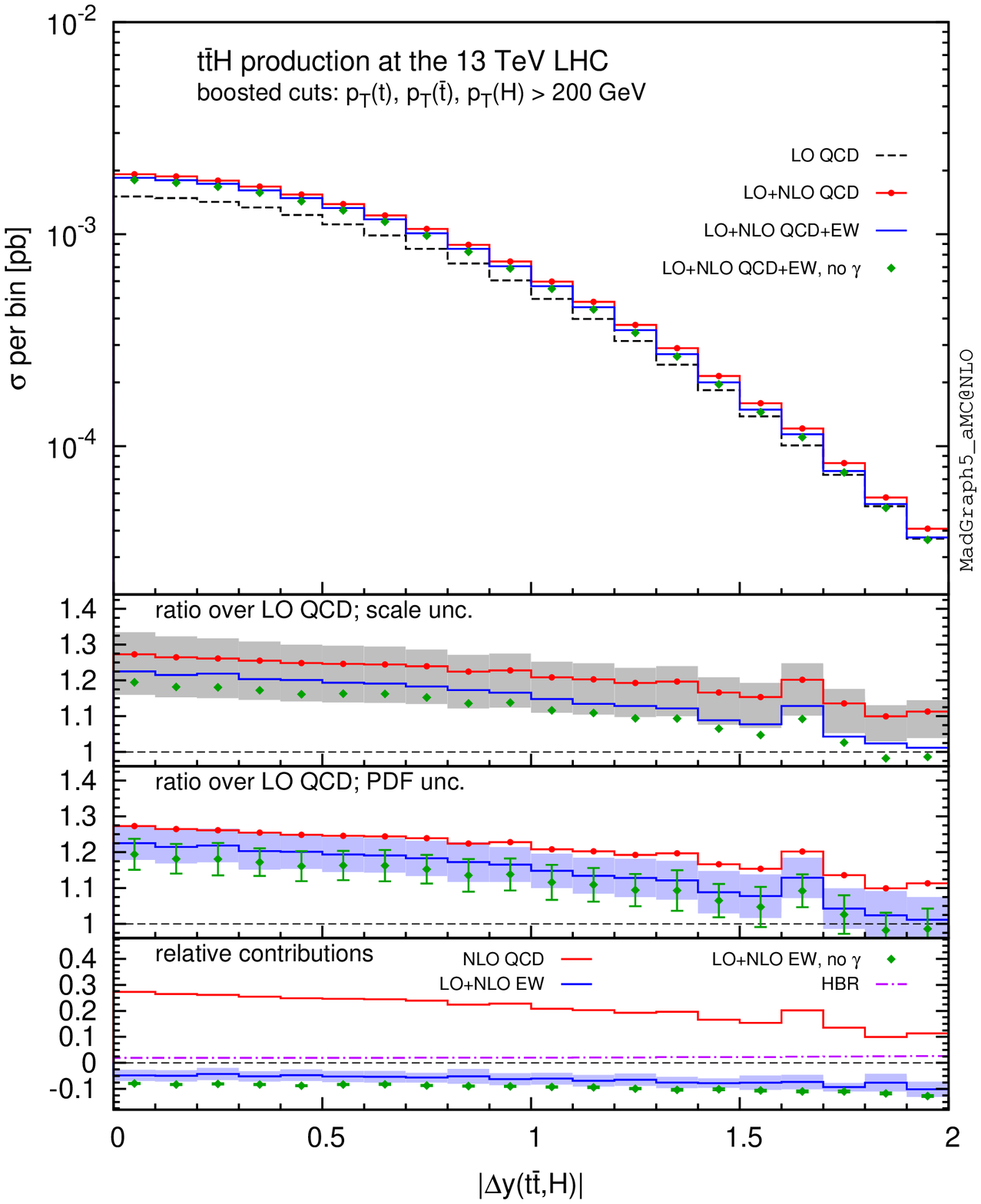}
    \caption{Same as in fig.~\ref{fig:tth13-nocuts}, with the cuts
of eq.~(\ref{eq:boosted}).}
    \label{fig:tth13-cuts}
\end{figure}

\begin{figure}[t]
    \centering
    \includegraphics[width=0.42\textwidth]{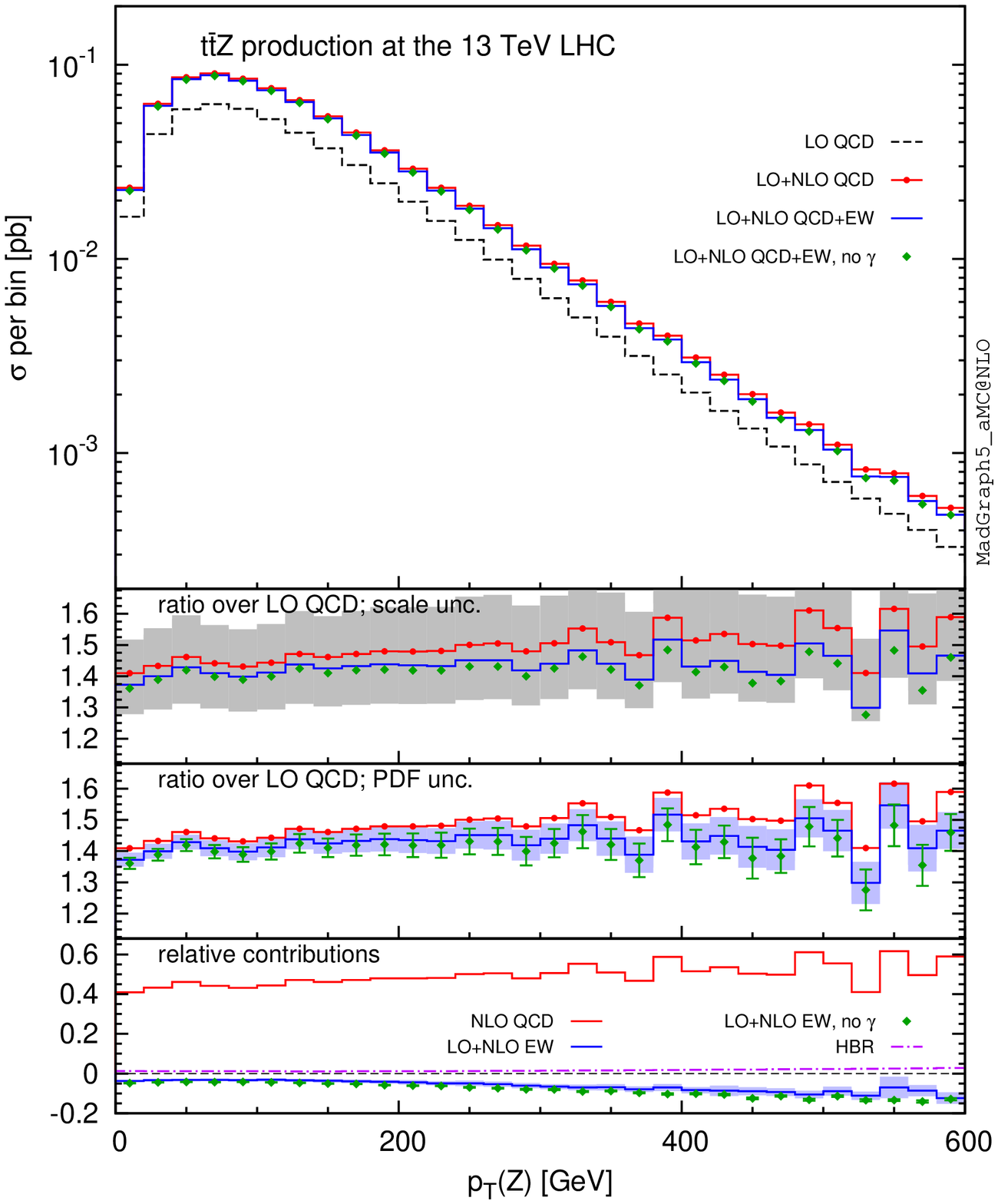}
    \includegraphics[width=0.42\textwidth]{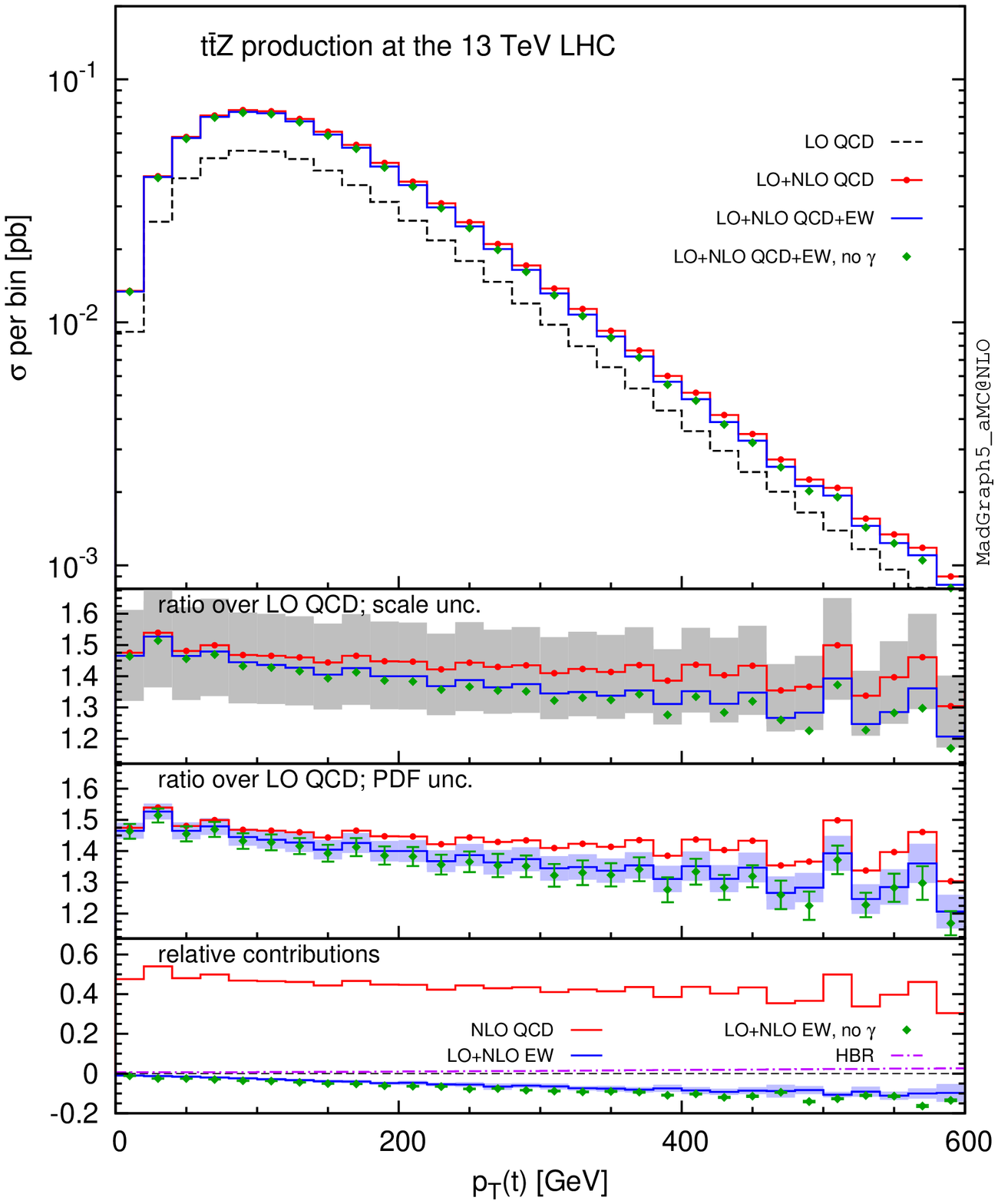}\\
    \includegraphics[width=0.42\textwidth]{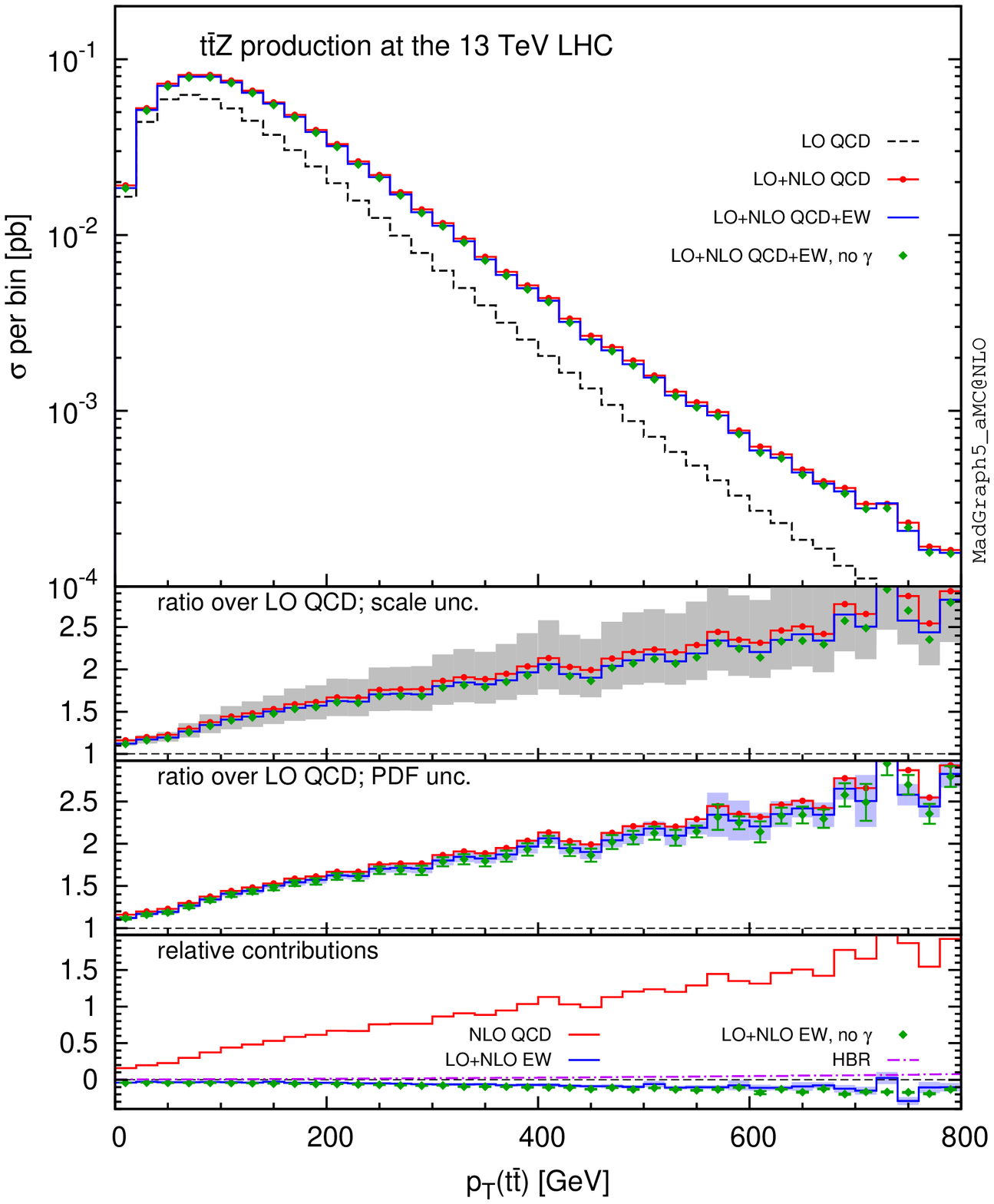}
    \includegraphics[width=0.42\textwidth]{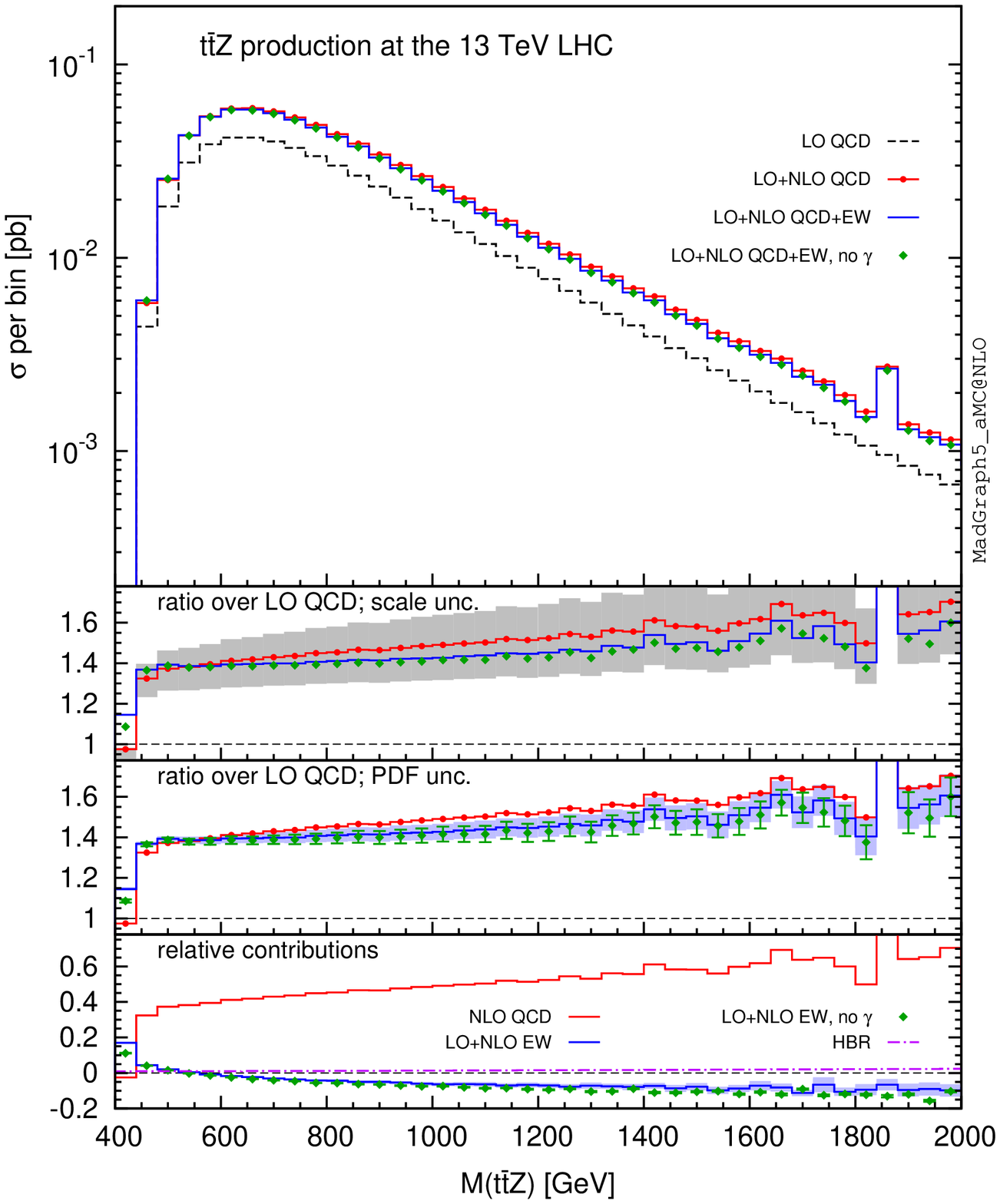}\\
    \includegraphics[width=0.42\textwidth]{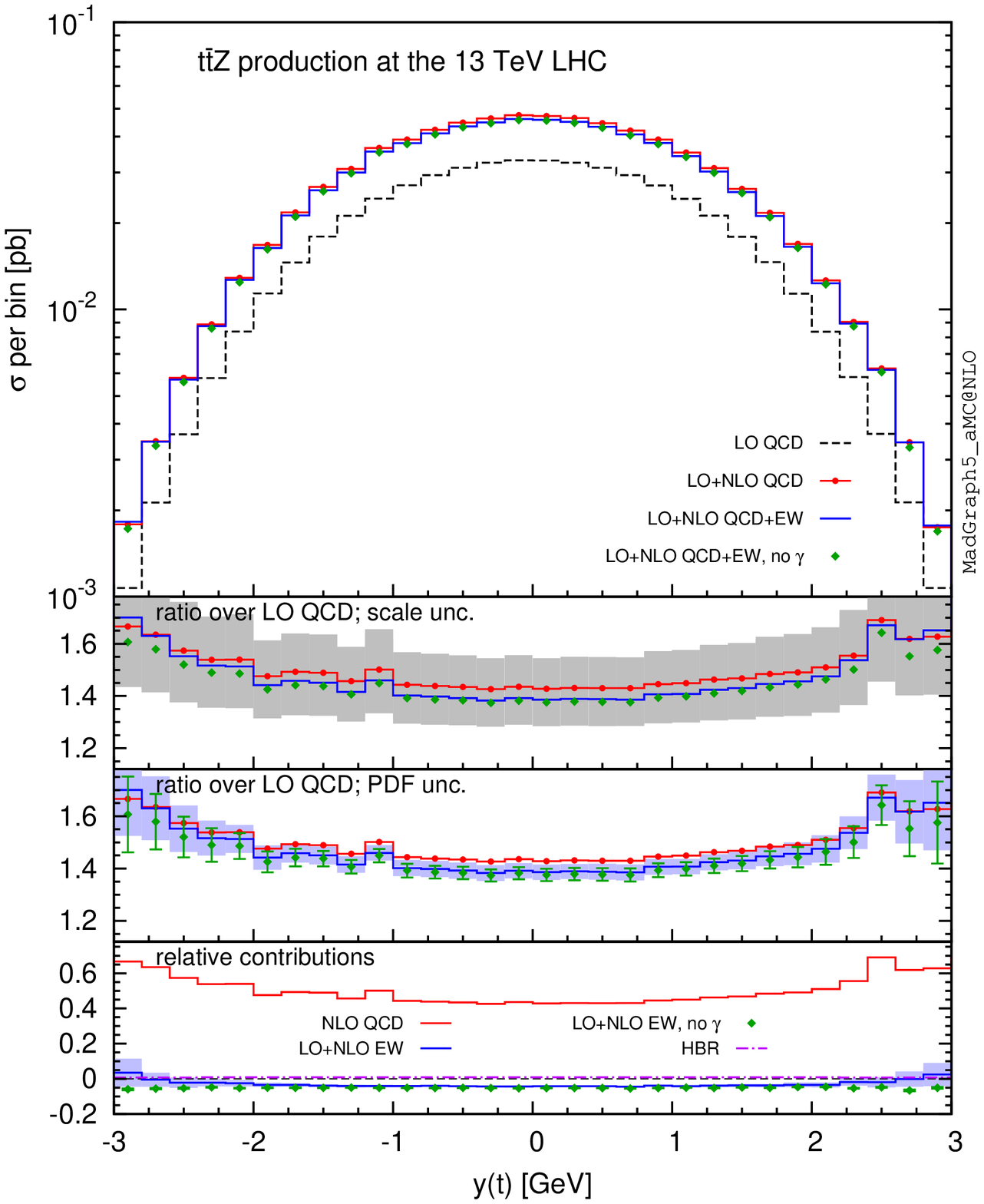}
    \includegraphics[width=0.42\textwidth]{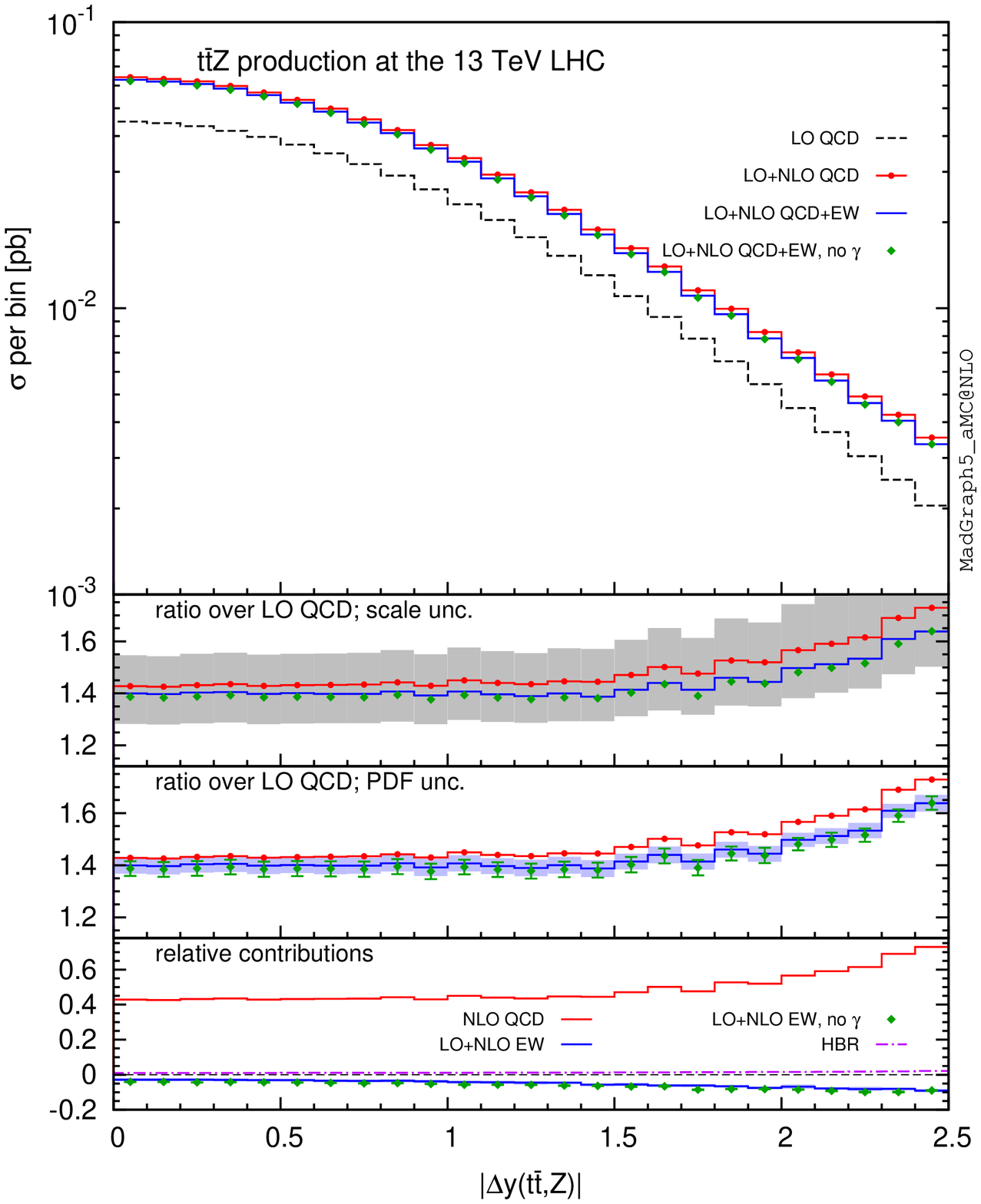}
    \caption{Same as in fig.~\ref{fig:tth13-nocuts}, for $\ttz$ production.}
    \label{fig:ttz13-nocuts}
\end{figure}

\begin{figure}[t]
    \centering
    \includegraphics[width=0.42\textwidth]{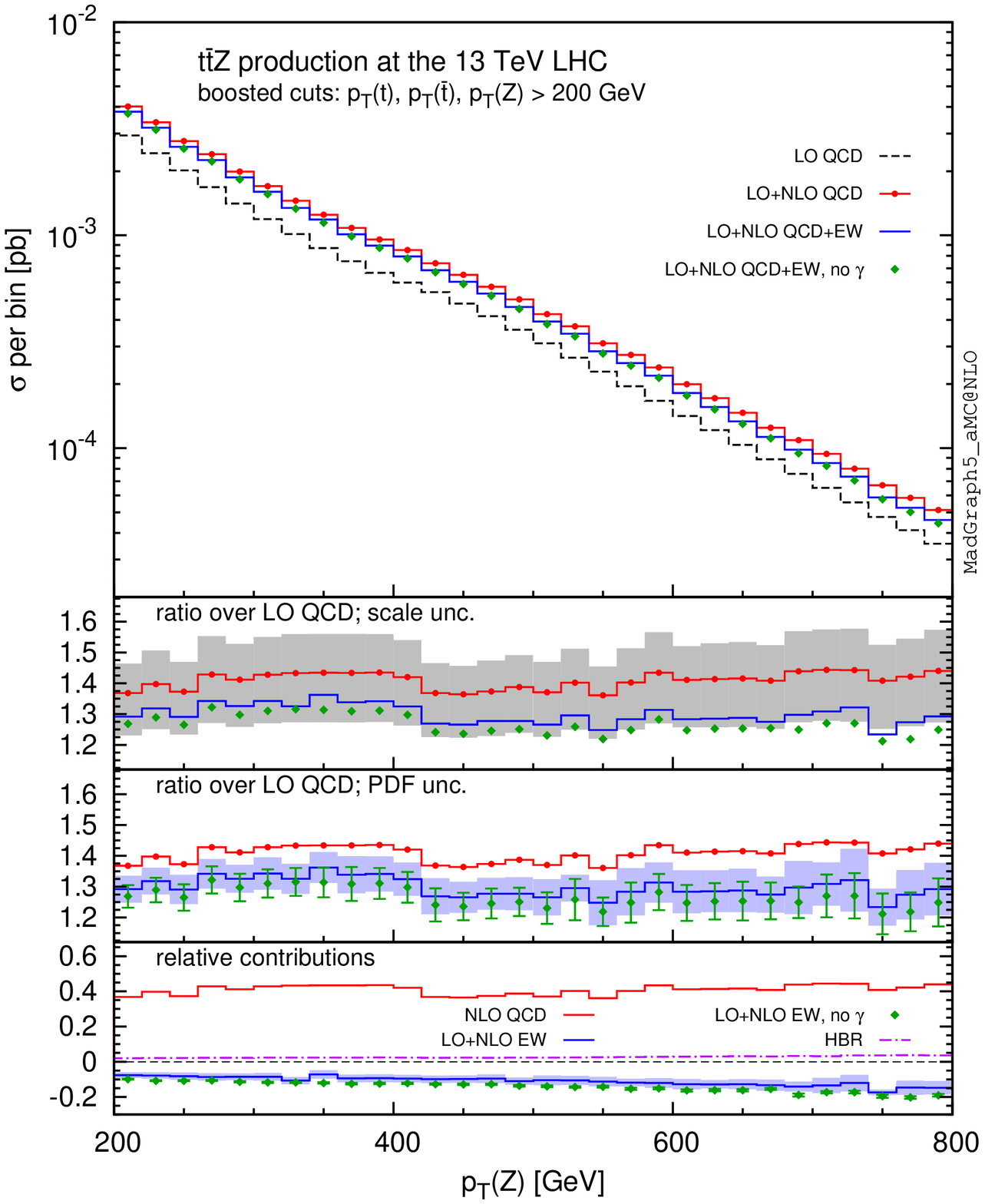}
    \includegraphics[width=0.42\textwidth]{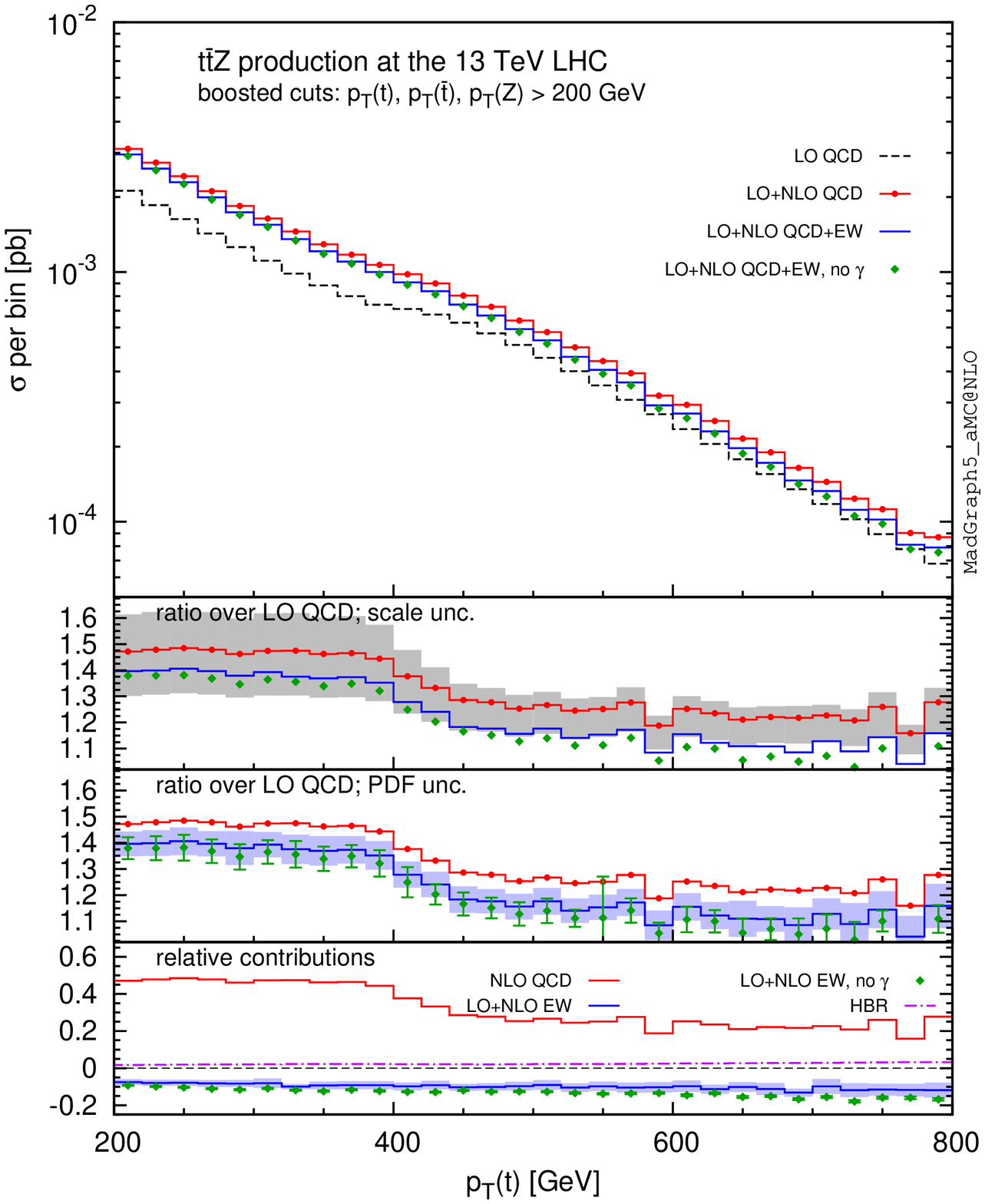}\\
    \includegraphics[width=0.42\textwidth]{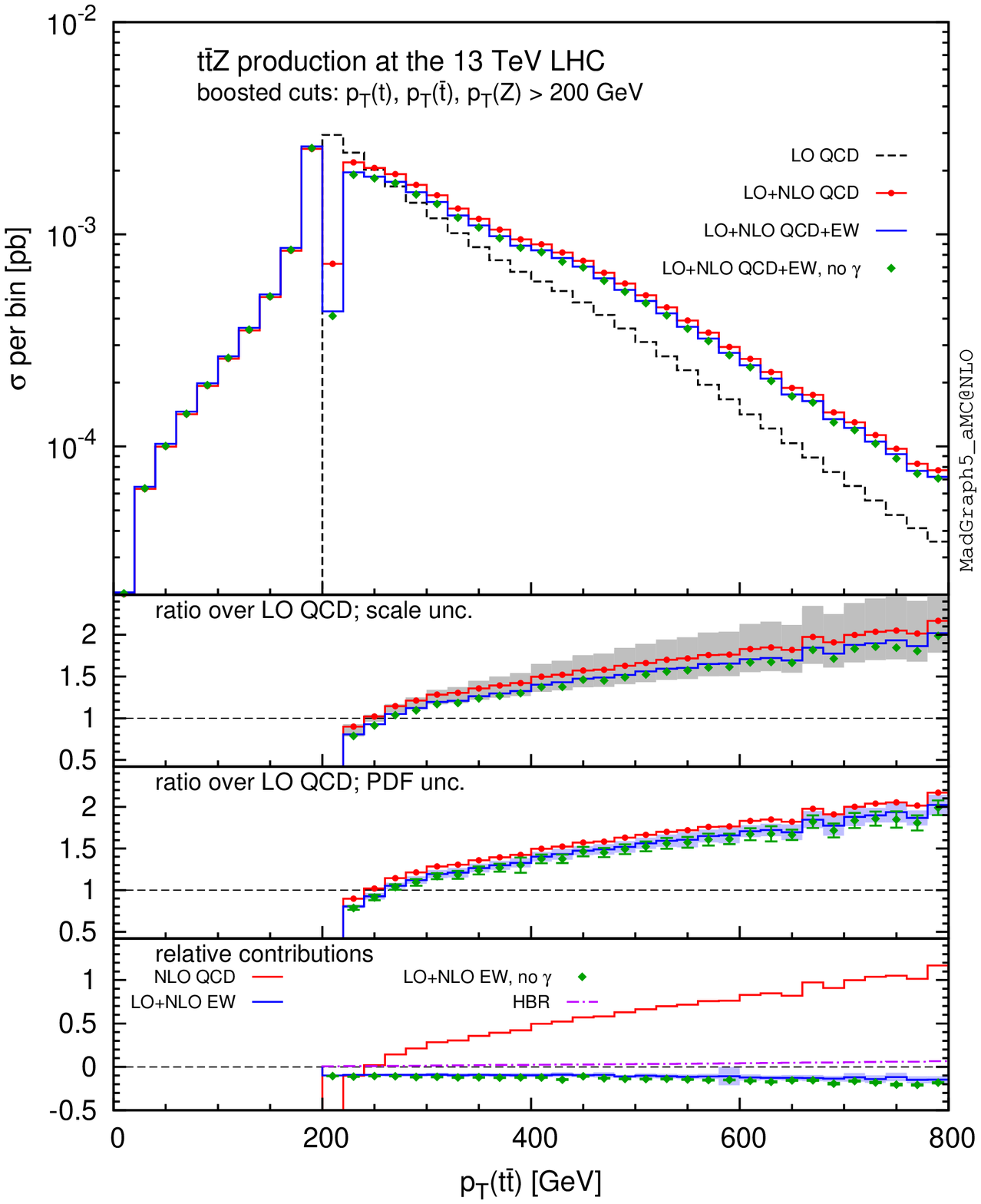}
    \includegraphics[width=0.42\textwidth]{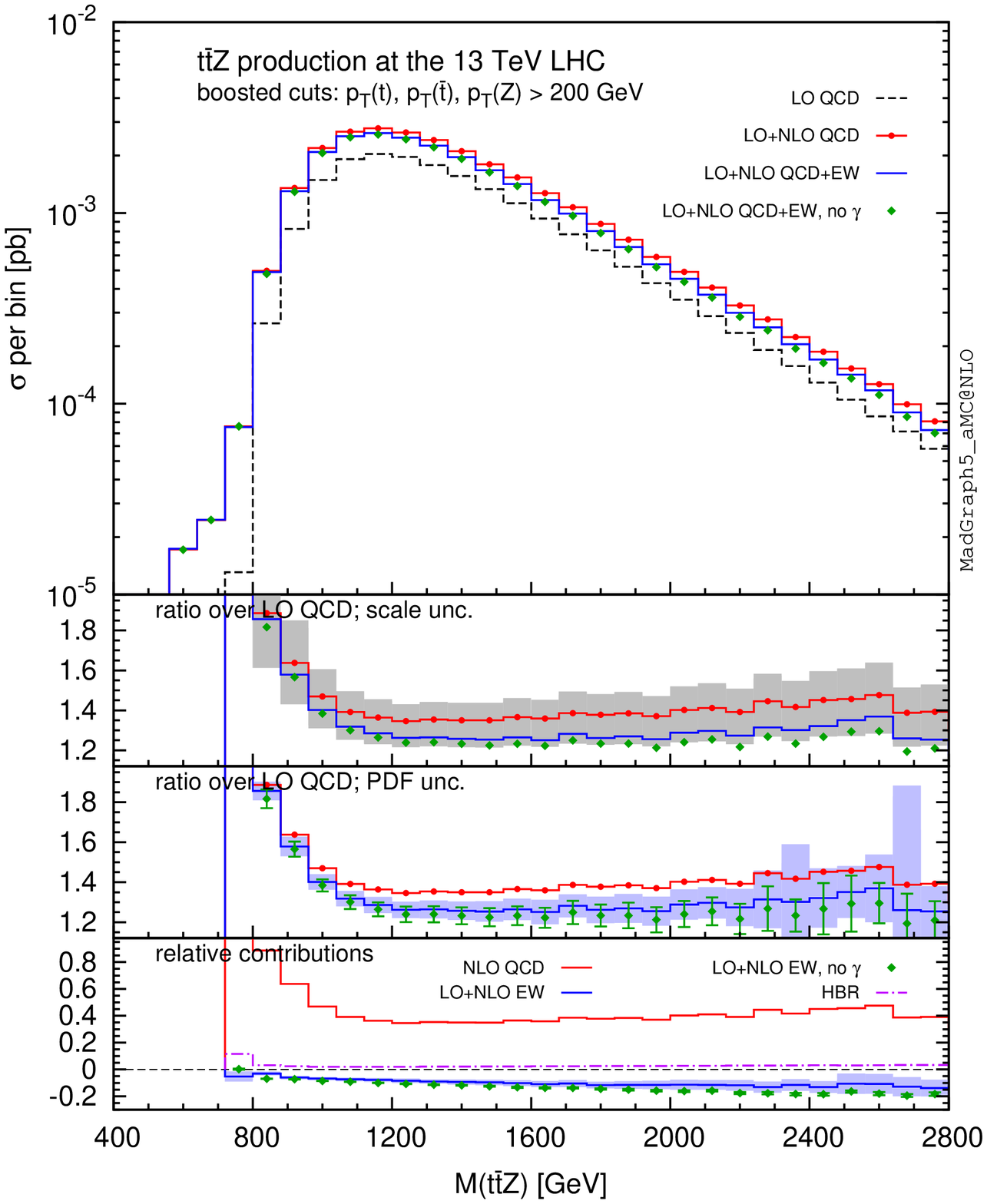}\\
    \includegraphics[width=0.42\textwidth]{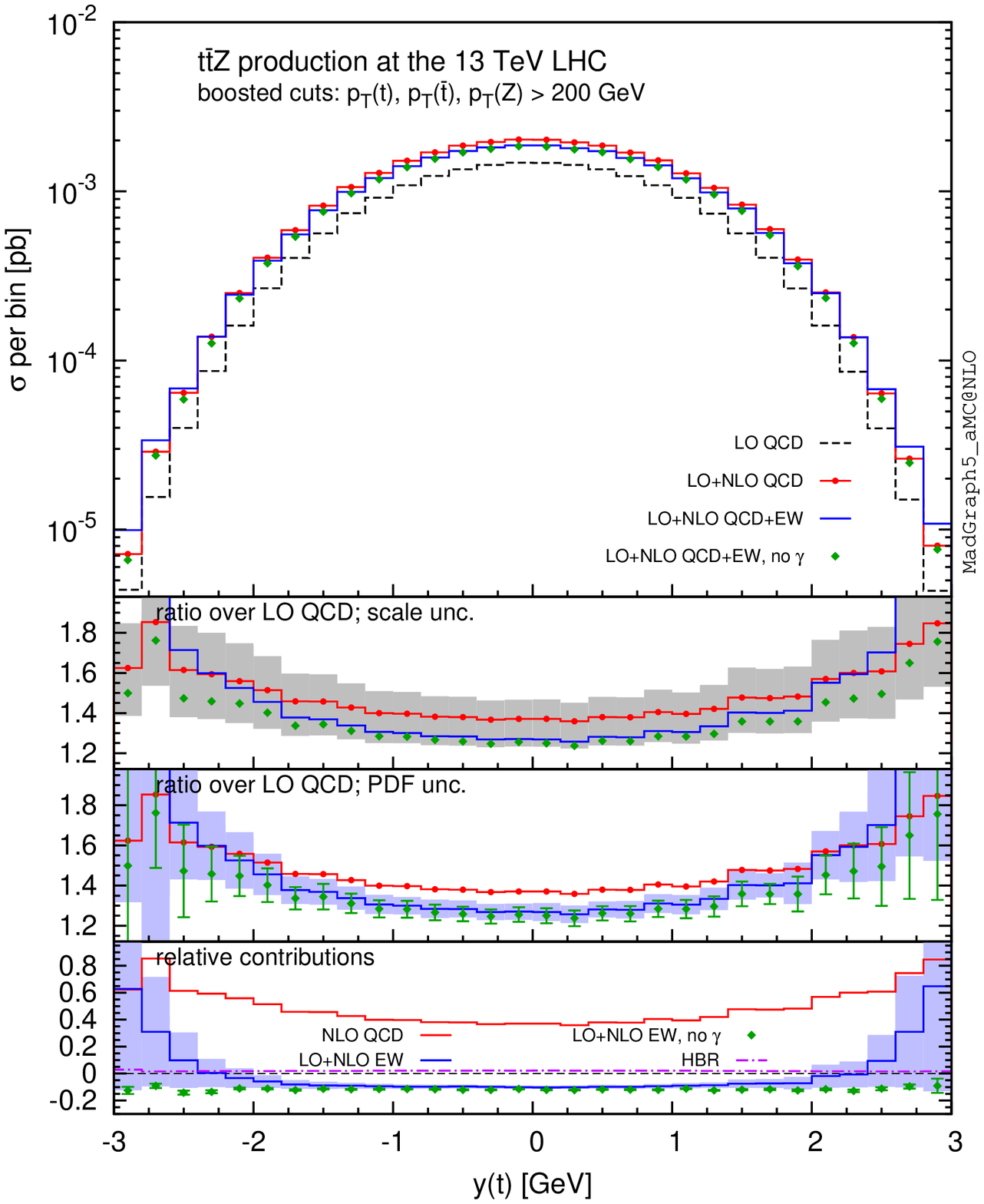}
    \includegraphics[width=0.42\textwidth]{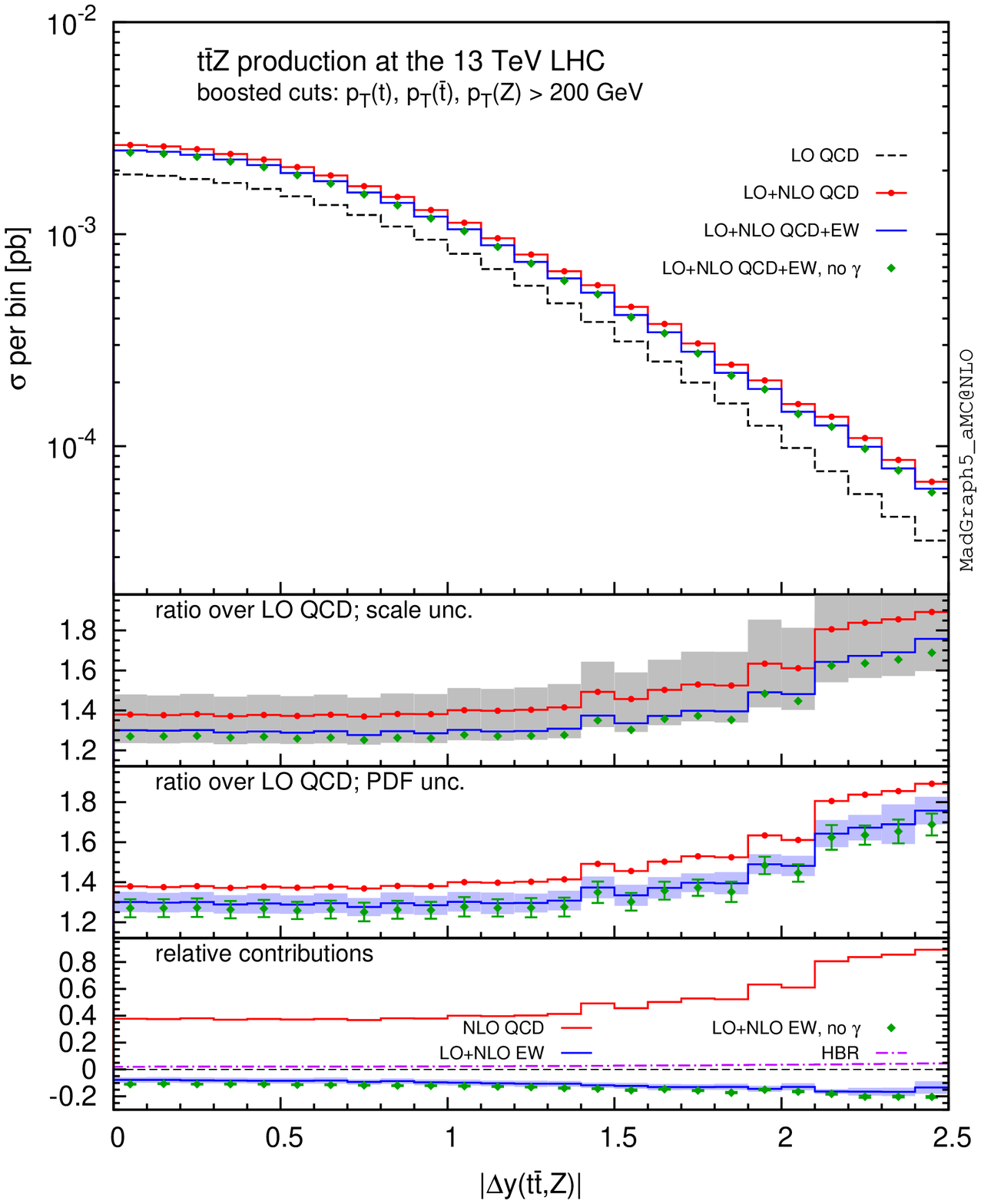}
    \caption{Same as in fig.~\ref{fig:tth13-cuts}, for $\ttz$ production.}
    \label{fig:ttz13-cuts}
\end{figure}

\begin{figure}[t]
    \centering
    \includegraphics[width=0.42\textwidth]{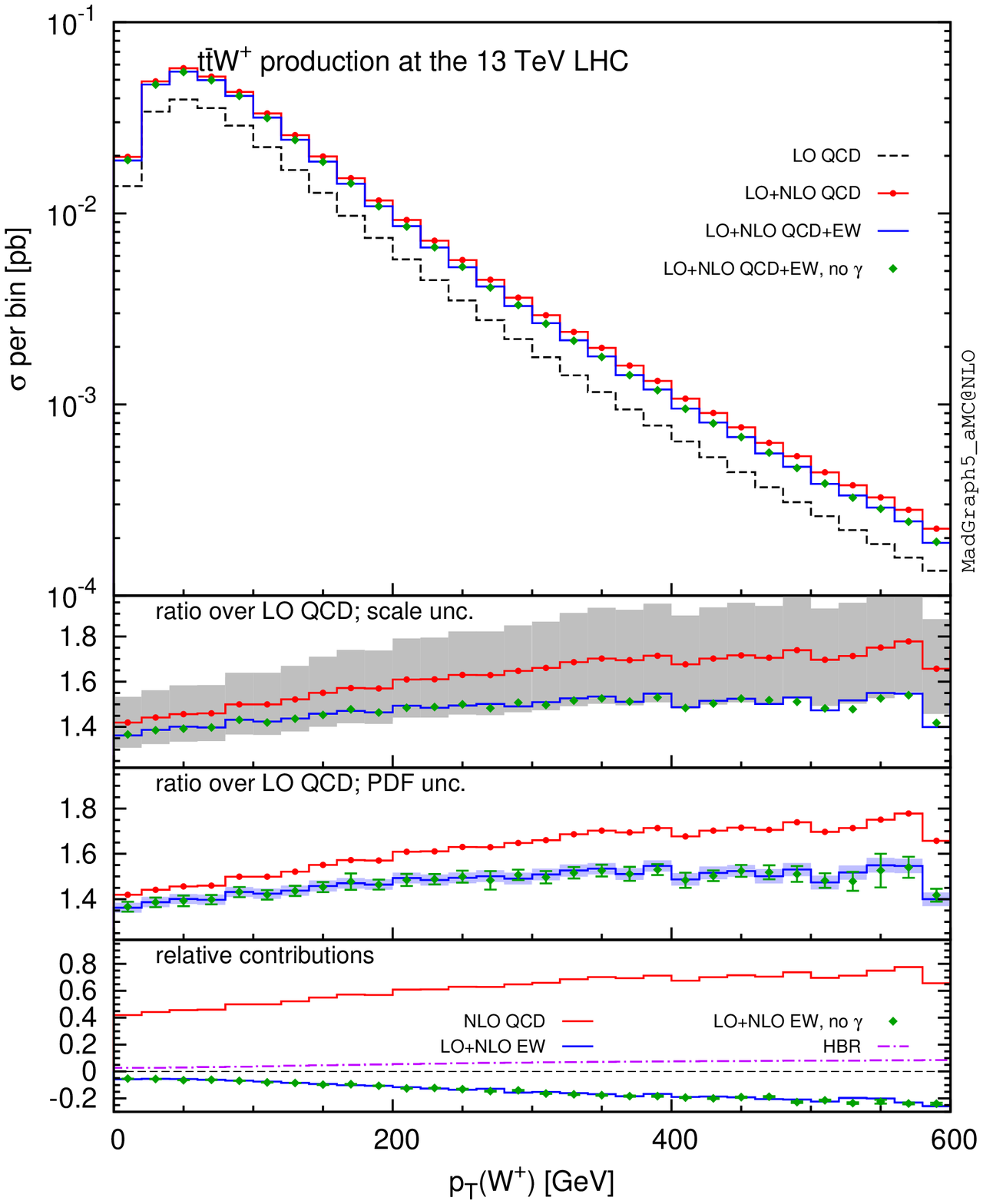}
    \includegraphics[width=0.42\textwidth]{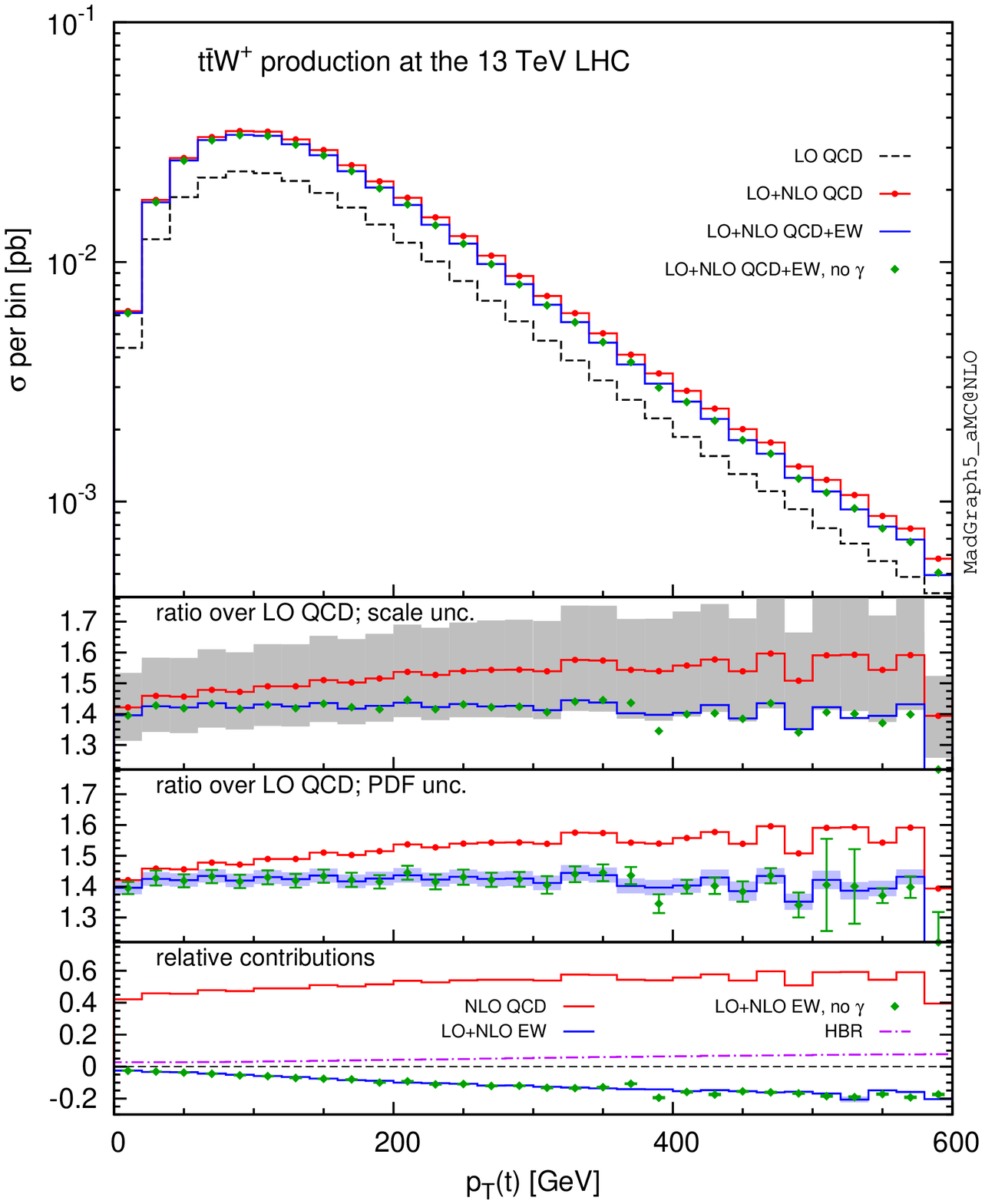}\\
    \includegraphics[width=0.42\textwidth]{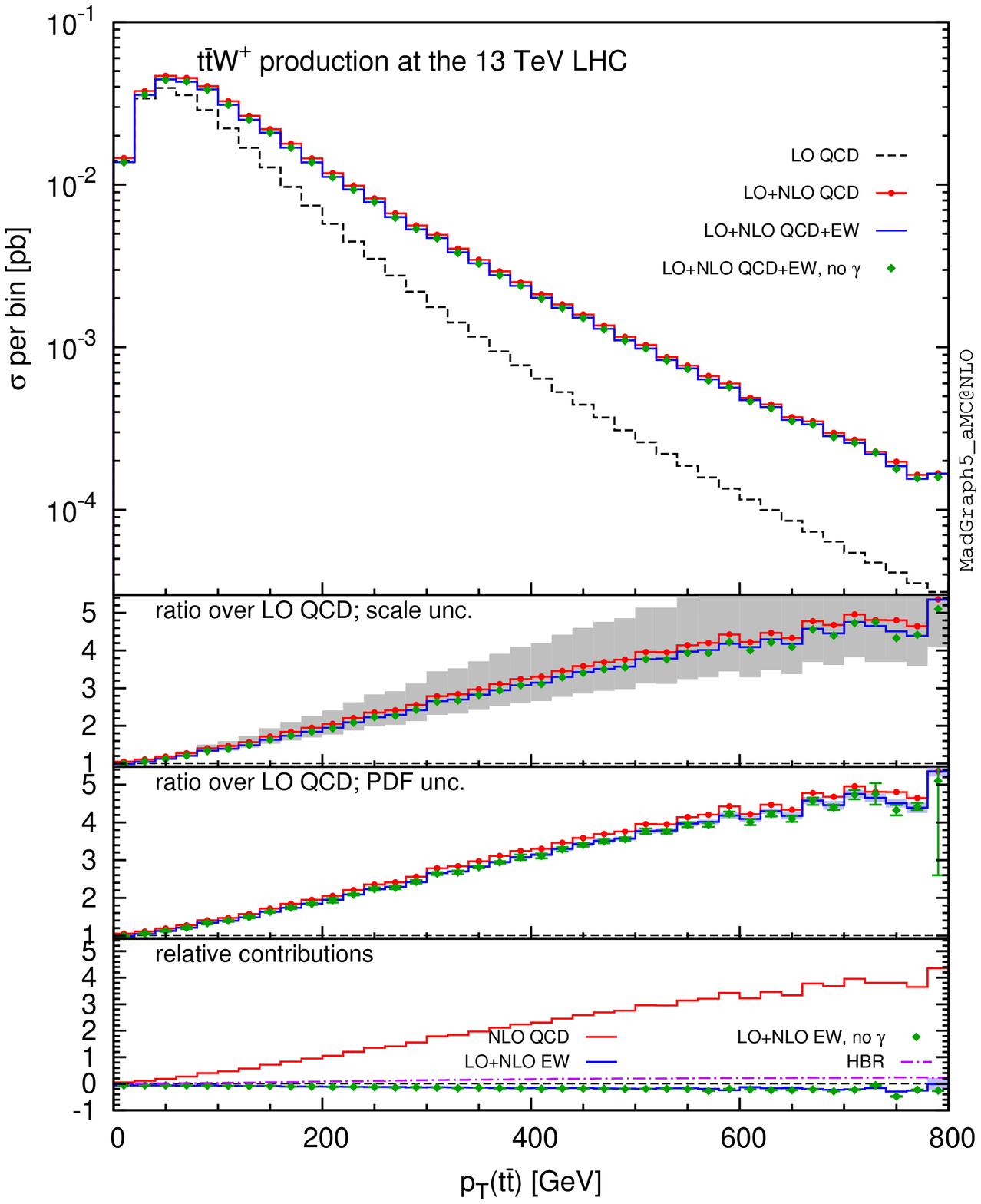}
    \includegraphics[width=0.42\textwidth]{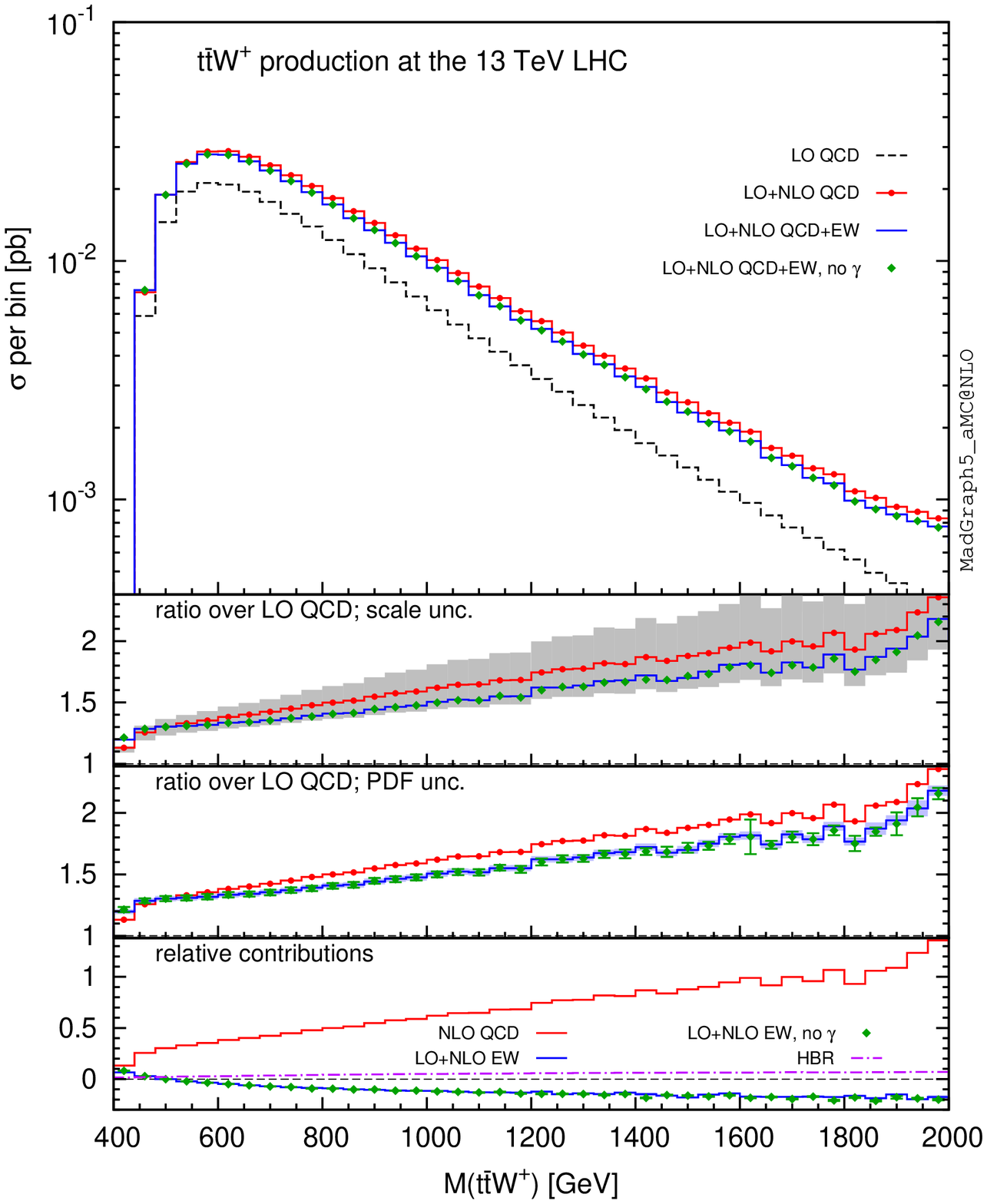}\\
    \includegraphics[width=0.42\textwidth]{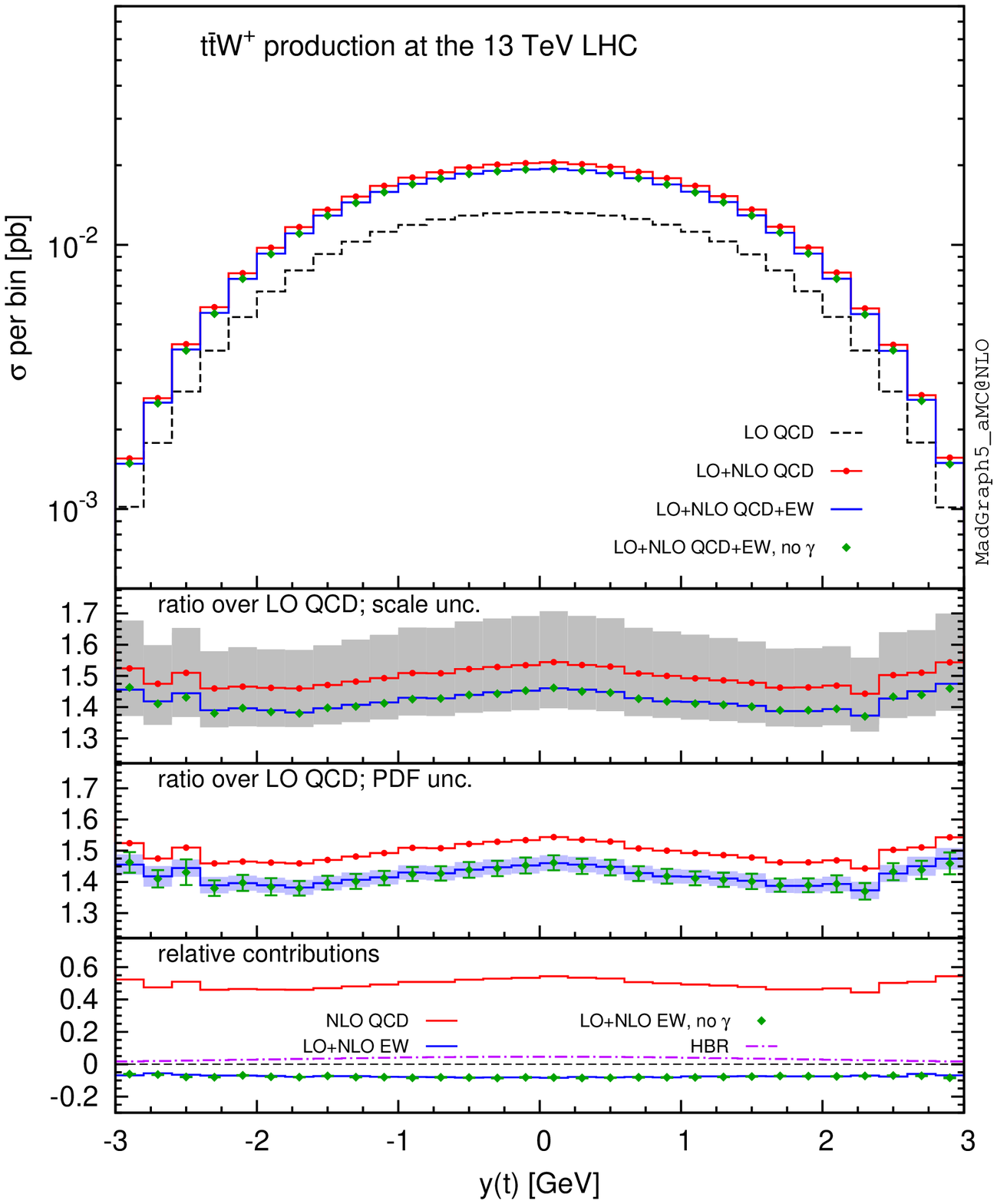}
    \includegraphics[width=0.42\textwidth]{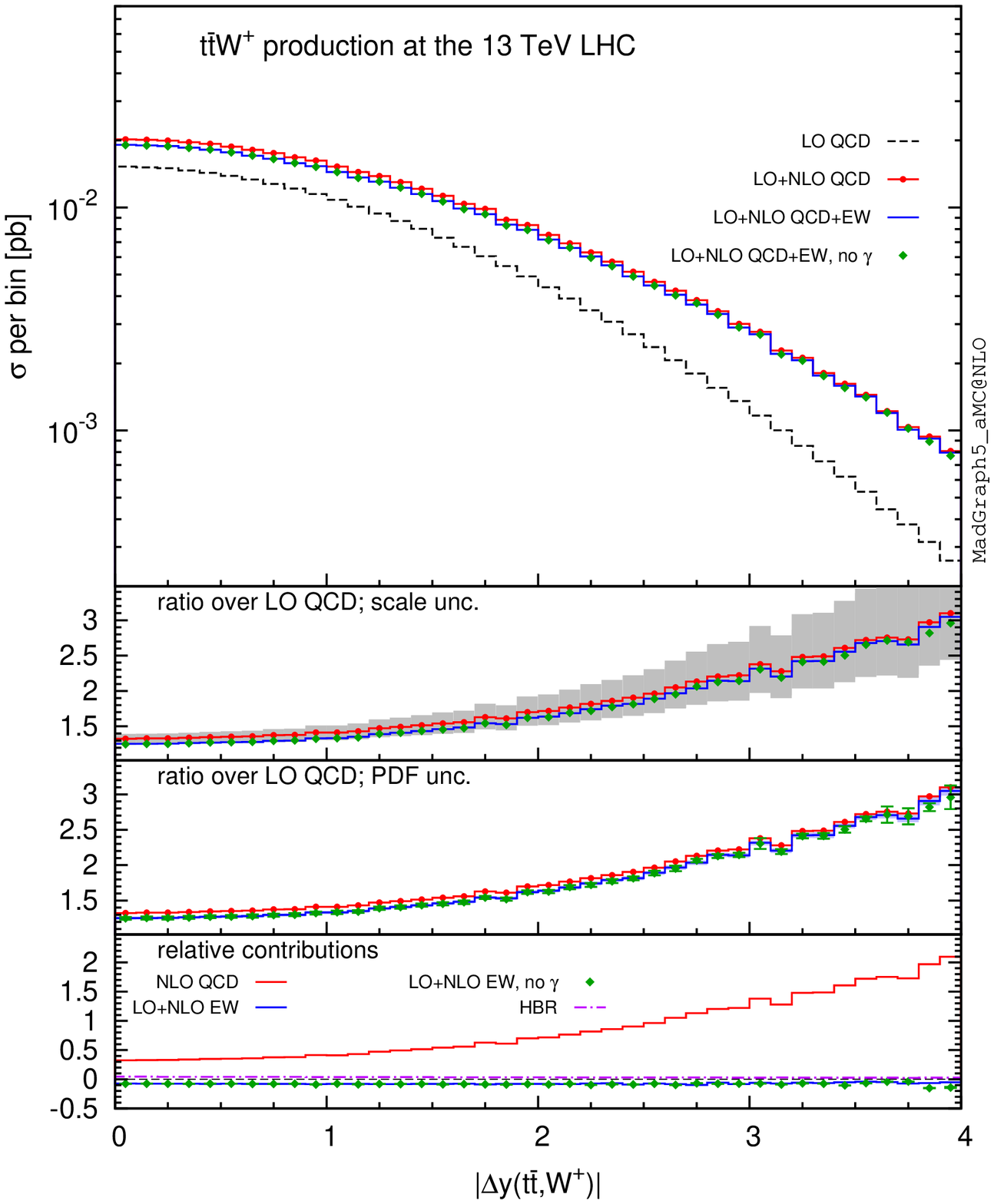}
    \caption{Same as in fig.~\ref{fig:tth13-nocuts}, for $\ttwp$ production.}
    \label{fig:ttwp13-nocuts}
\end{figure}

\begin{figure}[t]
    \centering
    \includegraphics[width=0.42\textwidth]{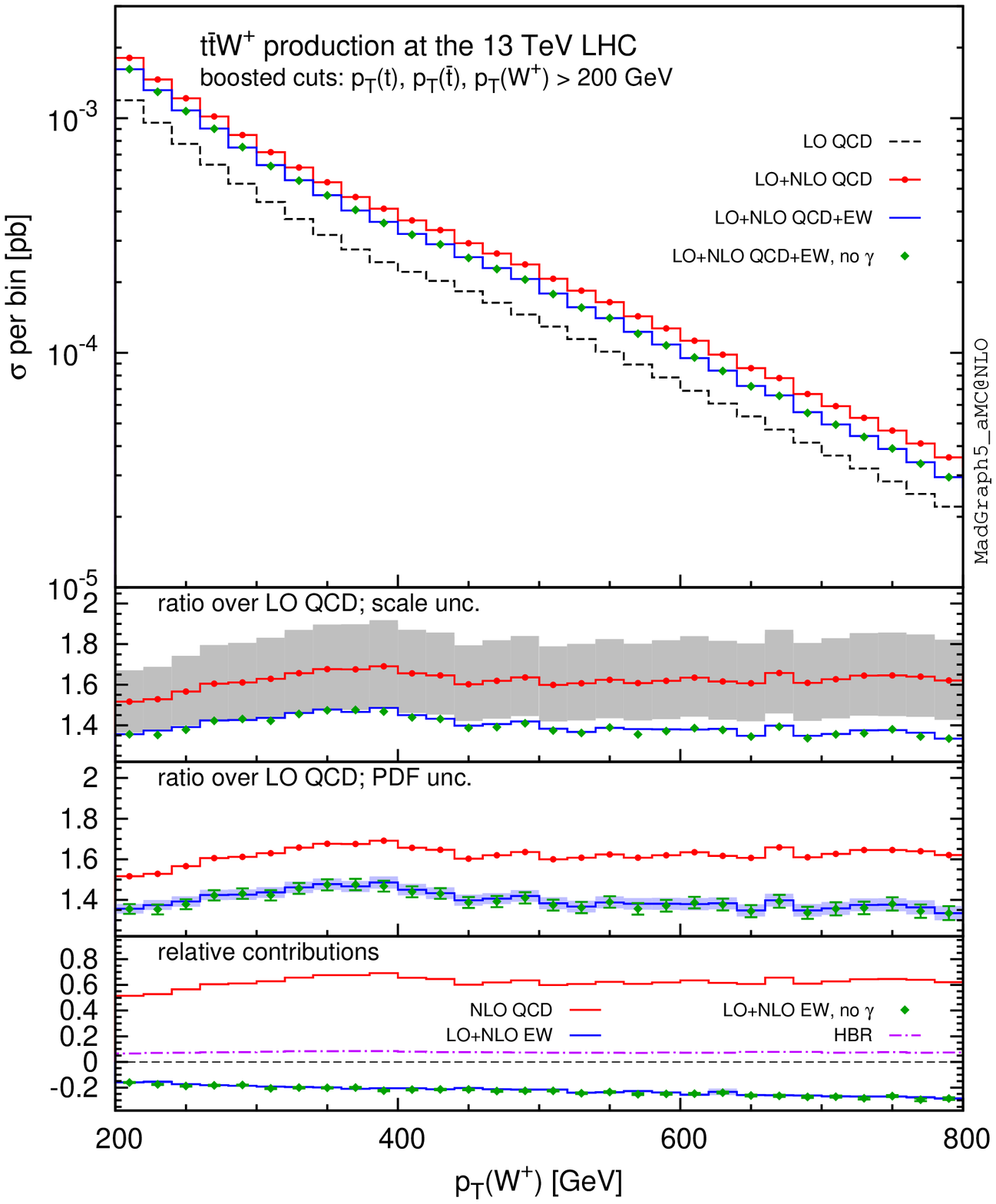}
    \includegraphics[width=0.42\textwidth]{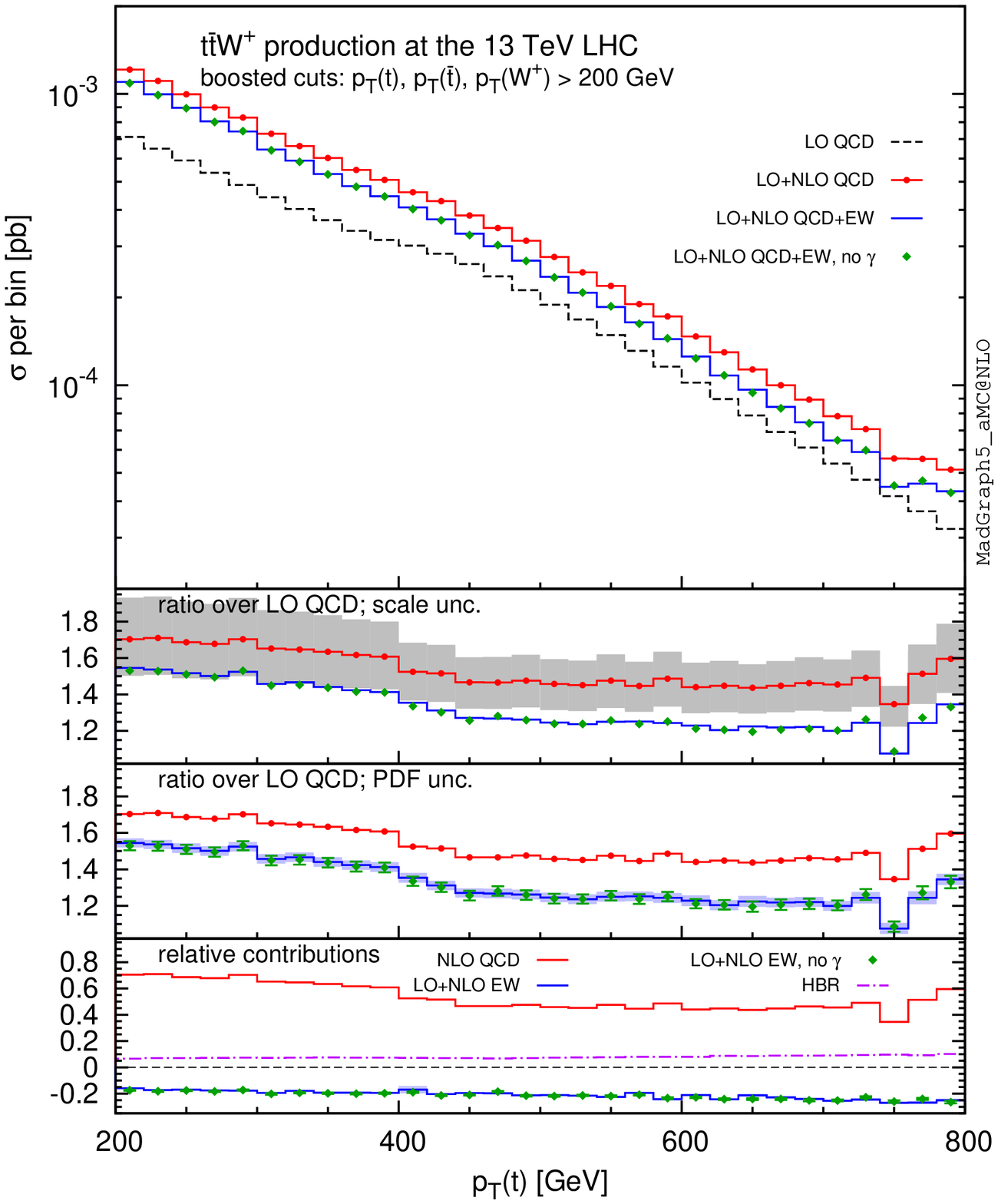}\\
    \includegraphics[width=0.42\textwidth]{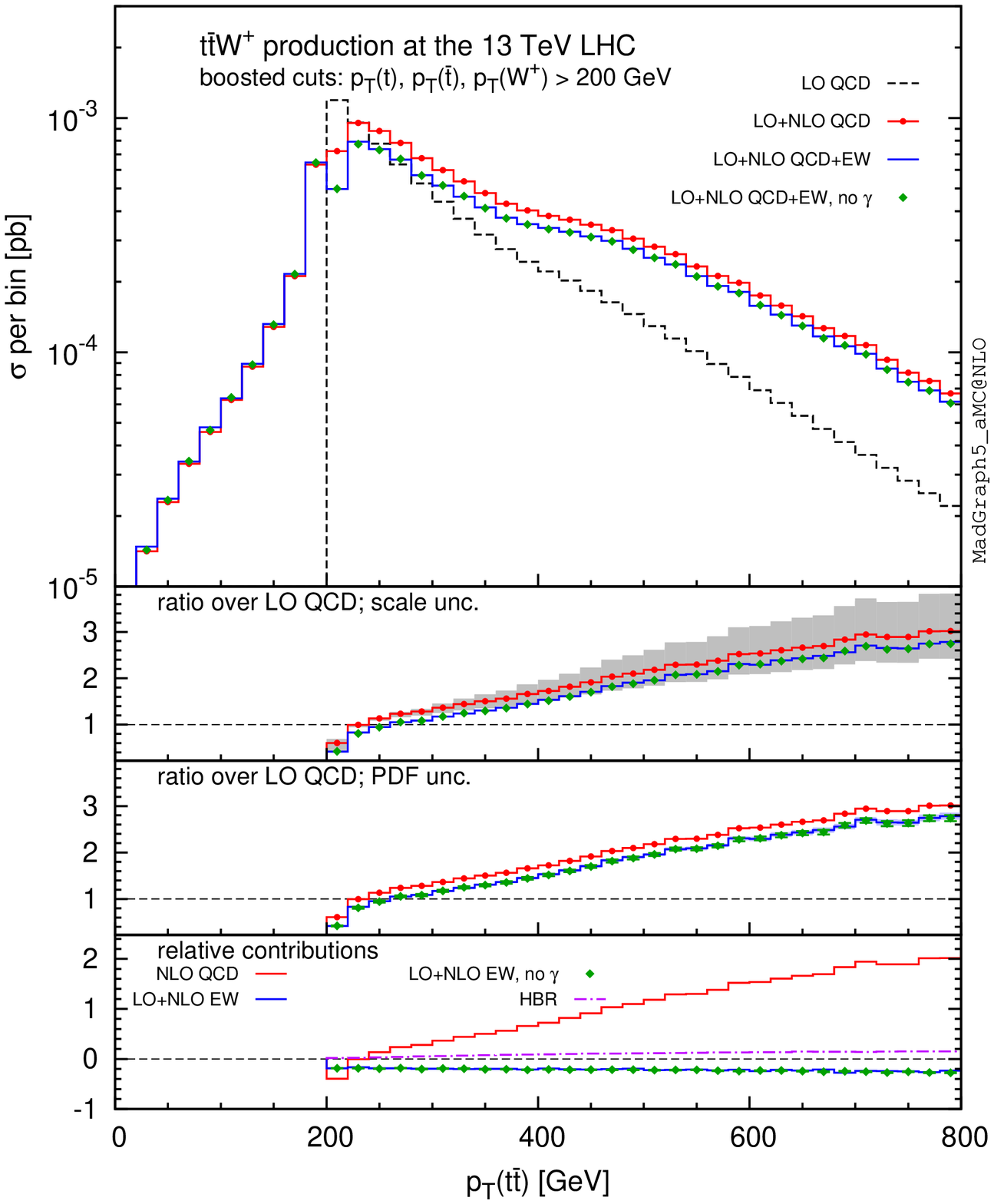}
    \includegraphics[width=0.42\textwidth]{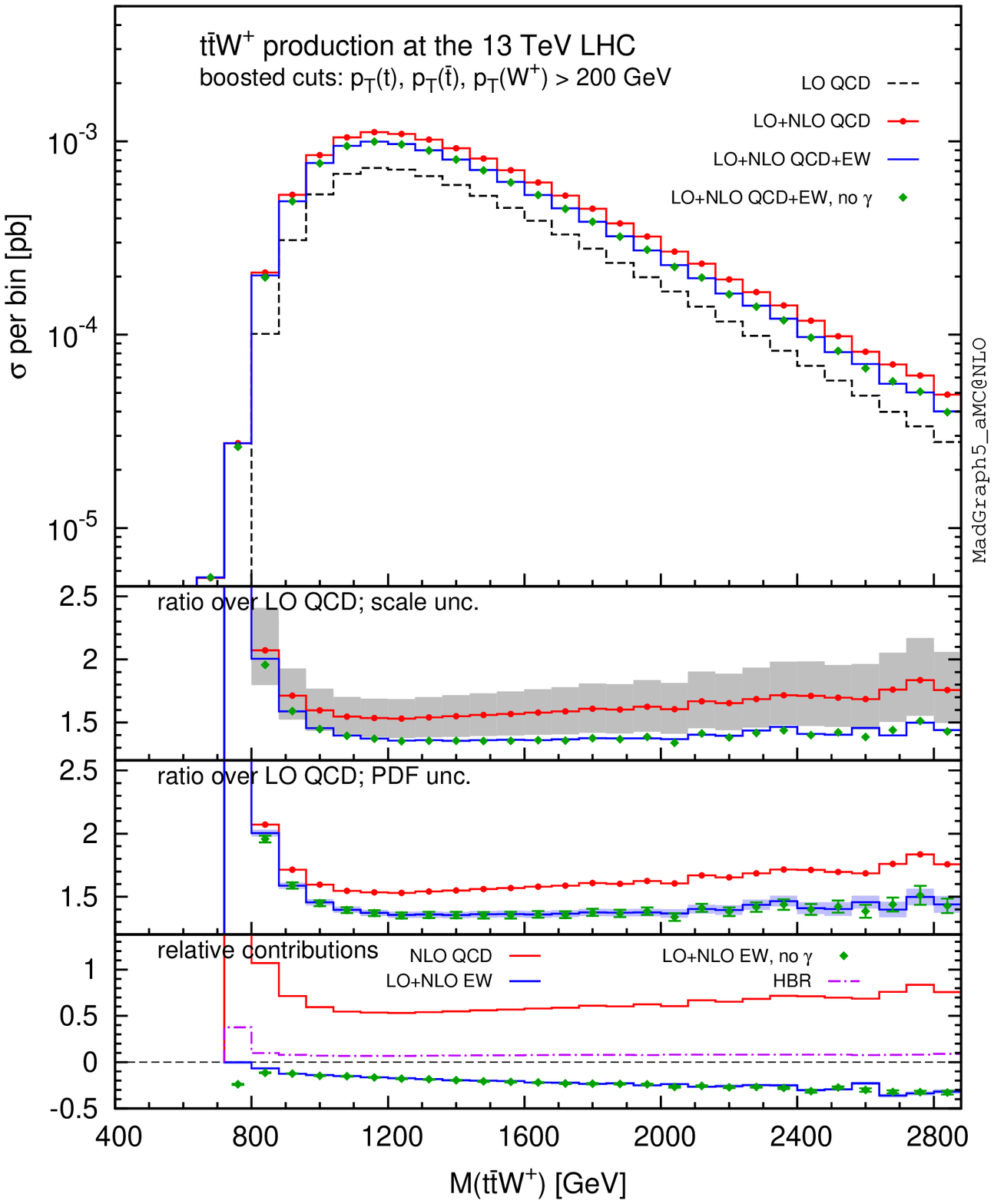}\\
    \includegraphics[width=0.42\textwidth]{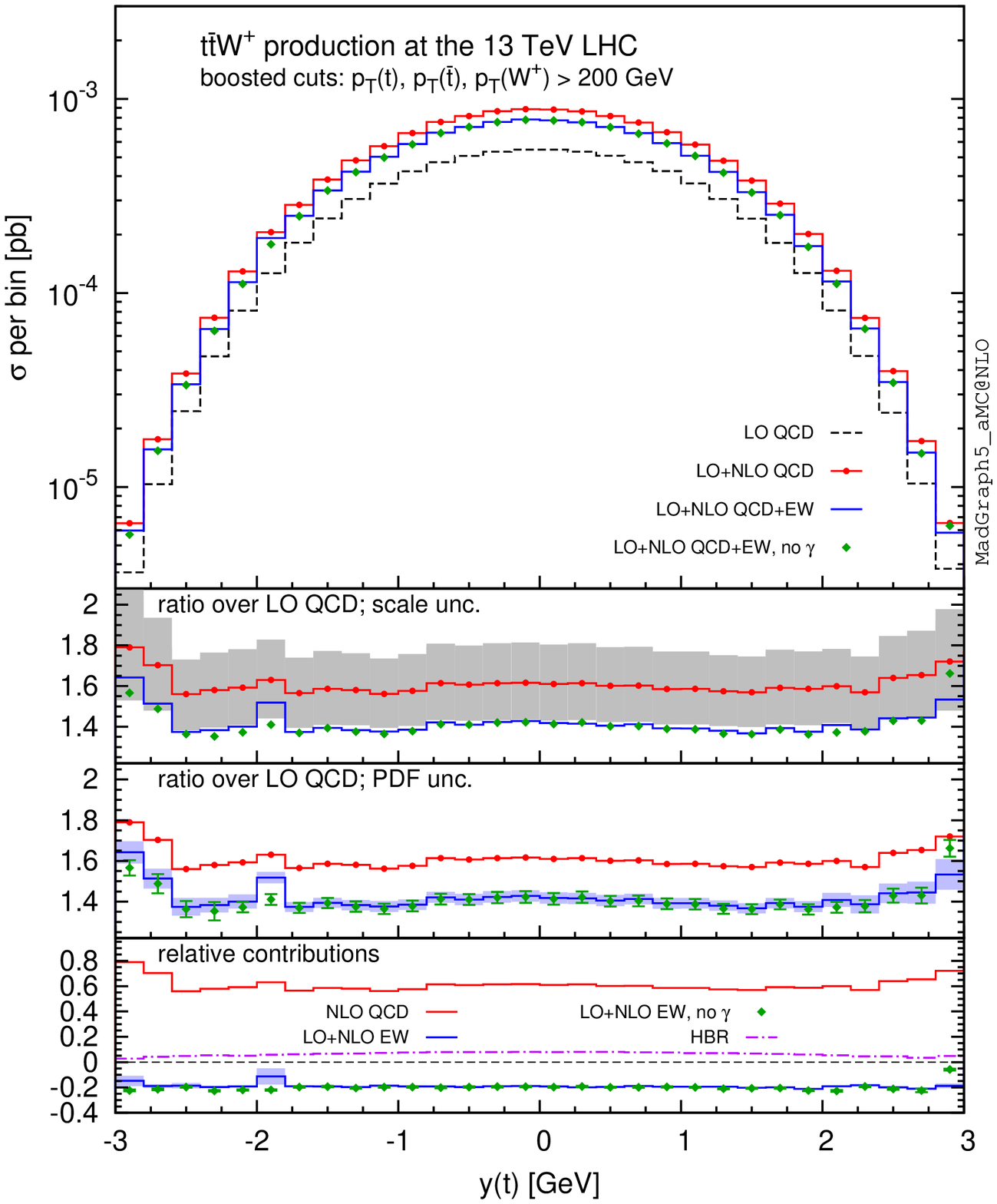}
    \includegraphics[width=0.42\textwidth]{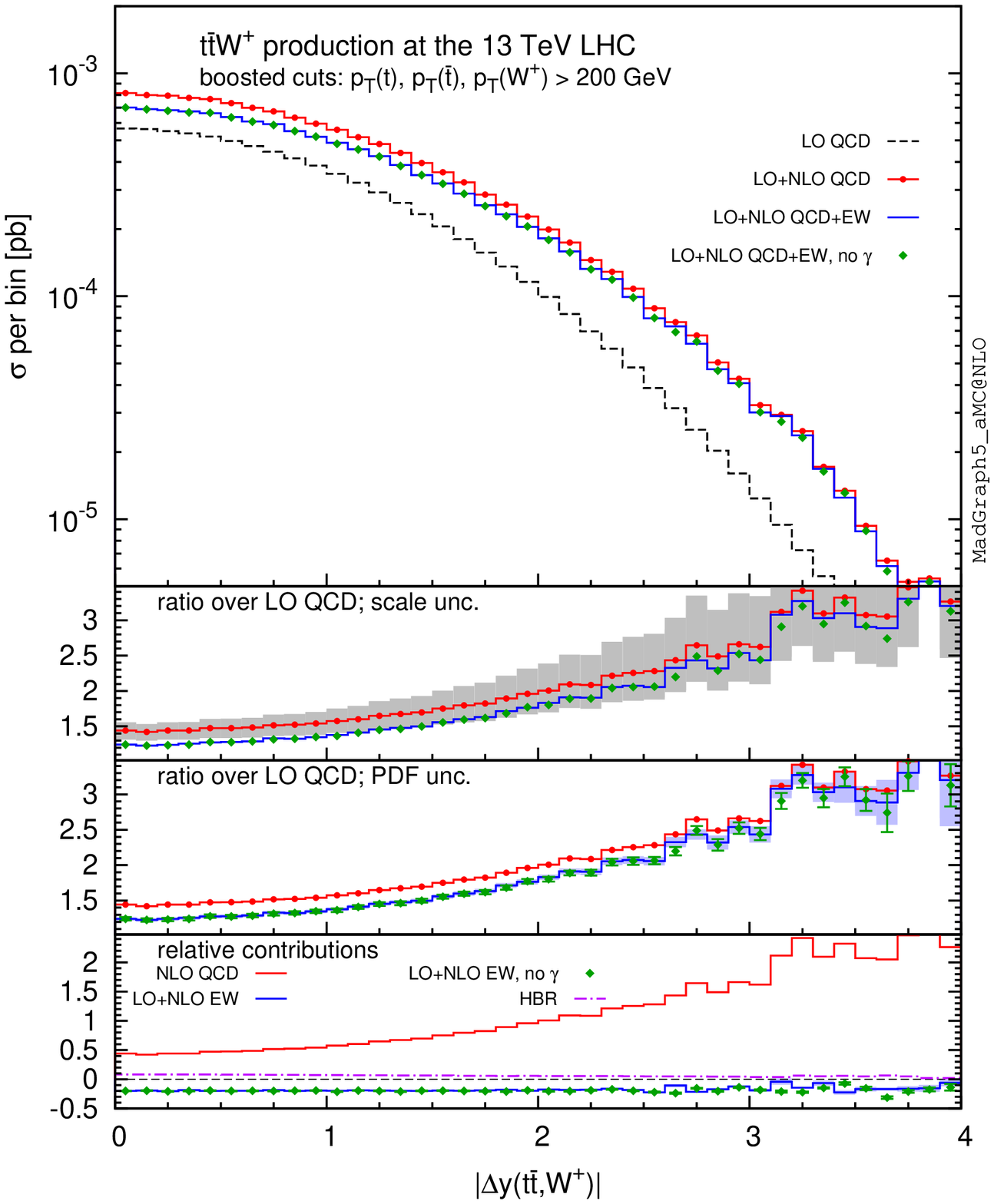}
    \caption{Same as in fig.~\ref{fig:tth13-cuts}, for $\ttwp$ production.}
    \label{fig:ttwp13-cuts}
\end{figure}

\begin{figure}[t]
    \centering
    \includegraphics[width=0.42\textwidth]{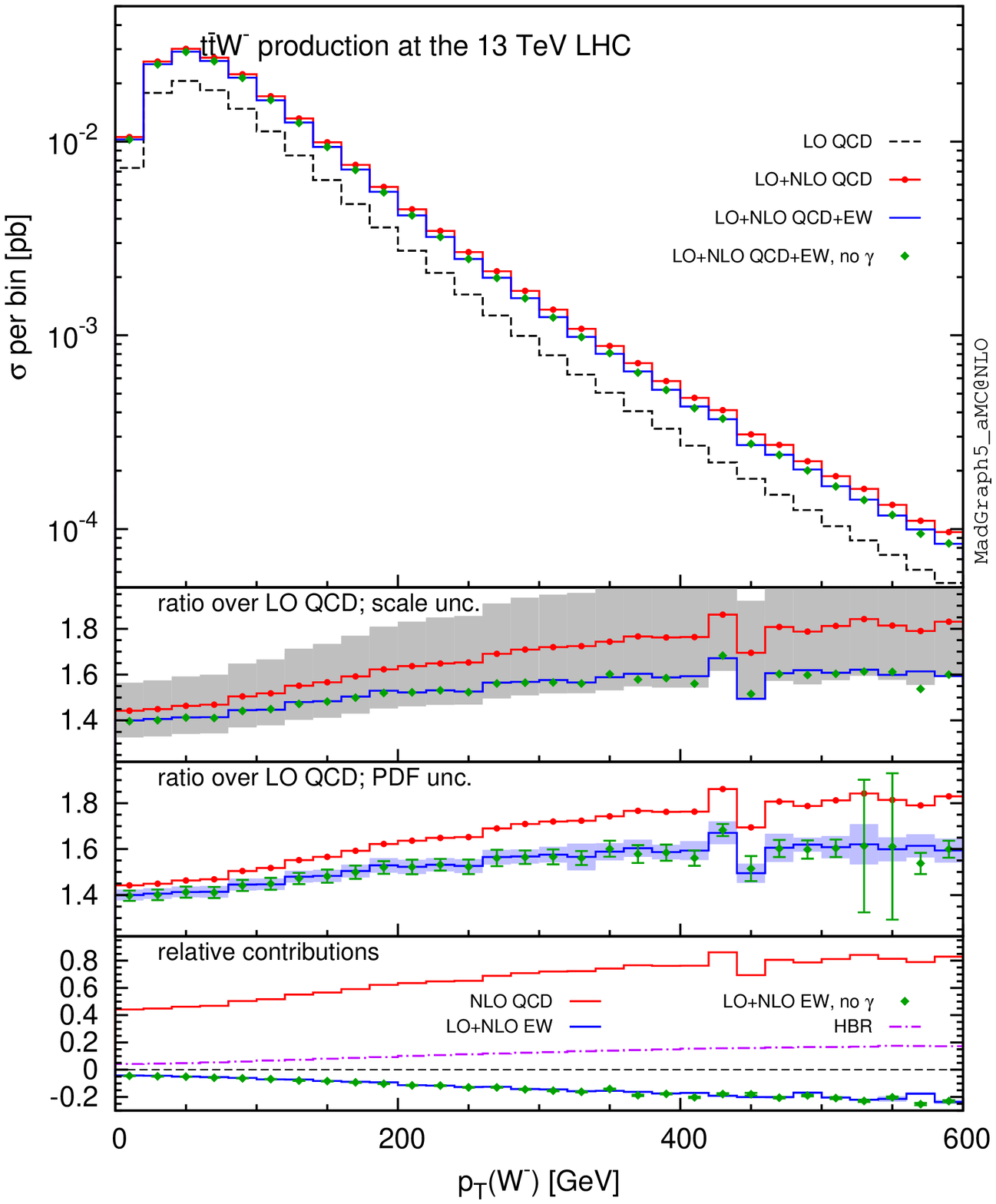}
    \includegraphics[width=0.42\textwidth]{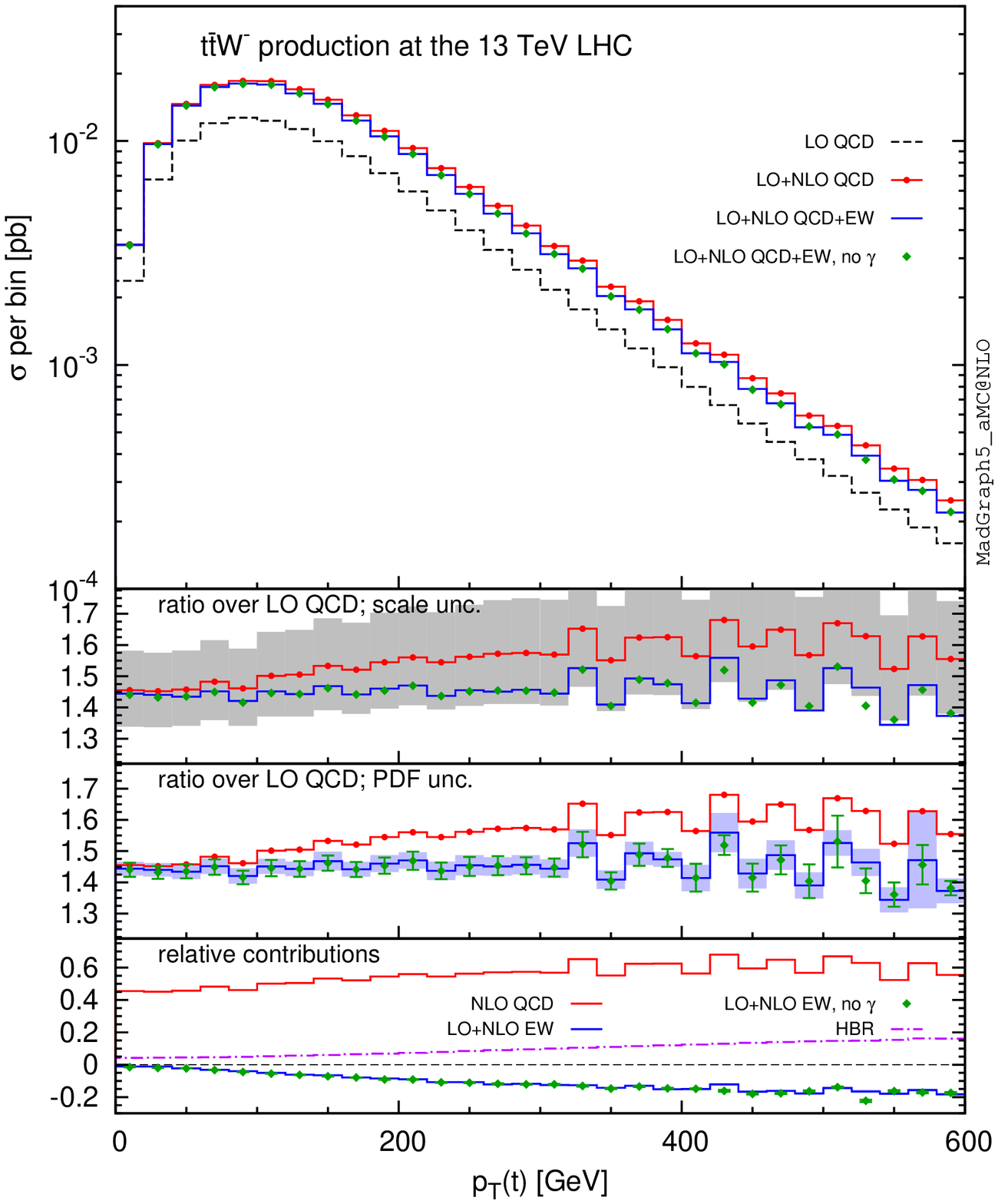}\\
    \includegraphics[width=0.42\textwidth]{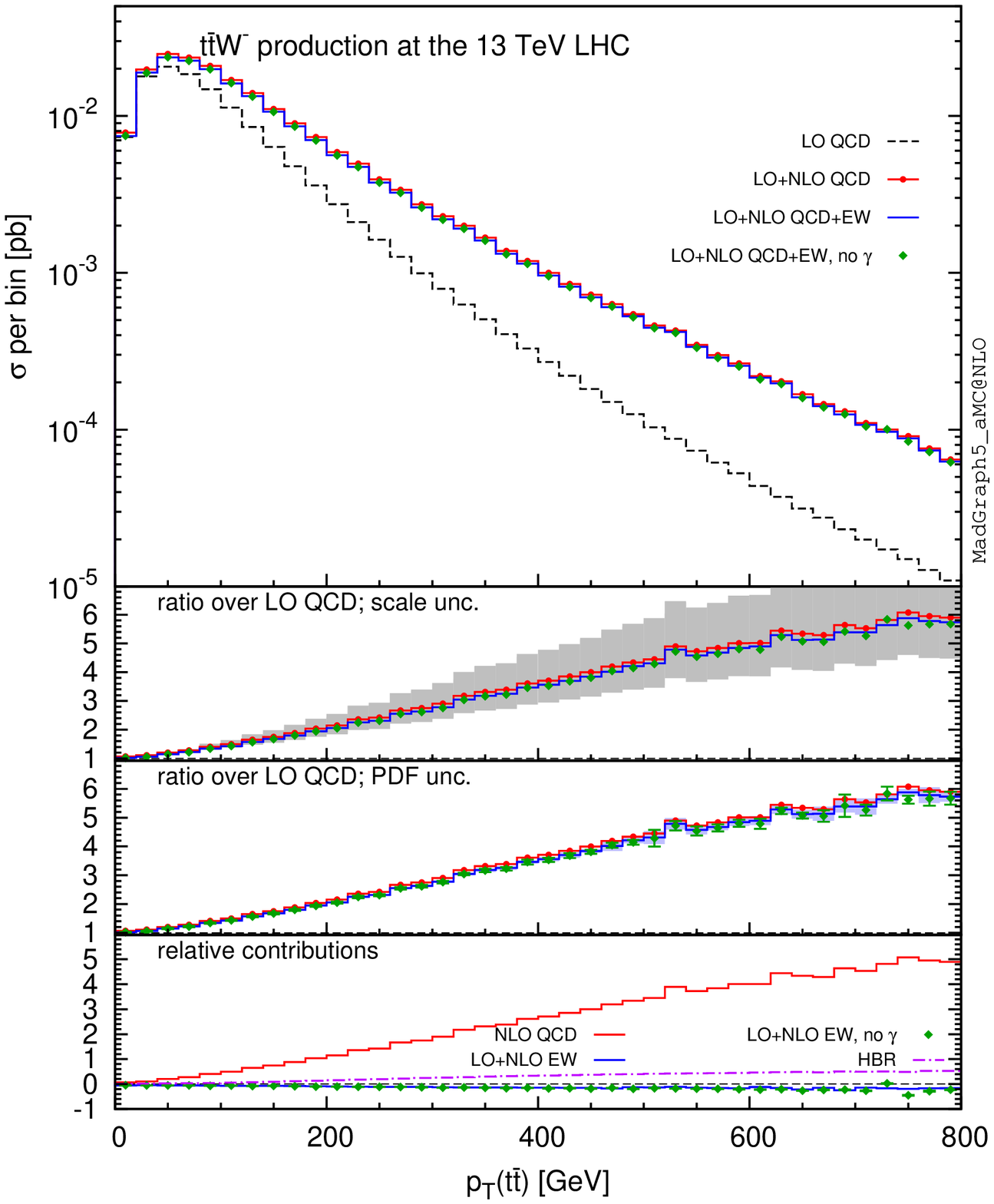}
    \includegraphics[width=0.42\textwidth]{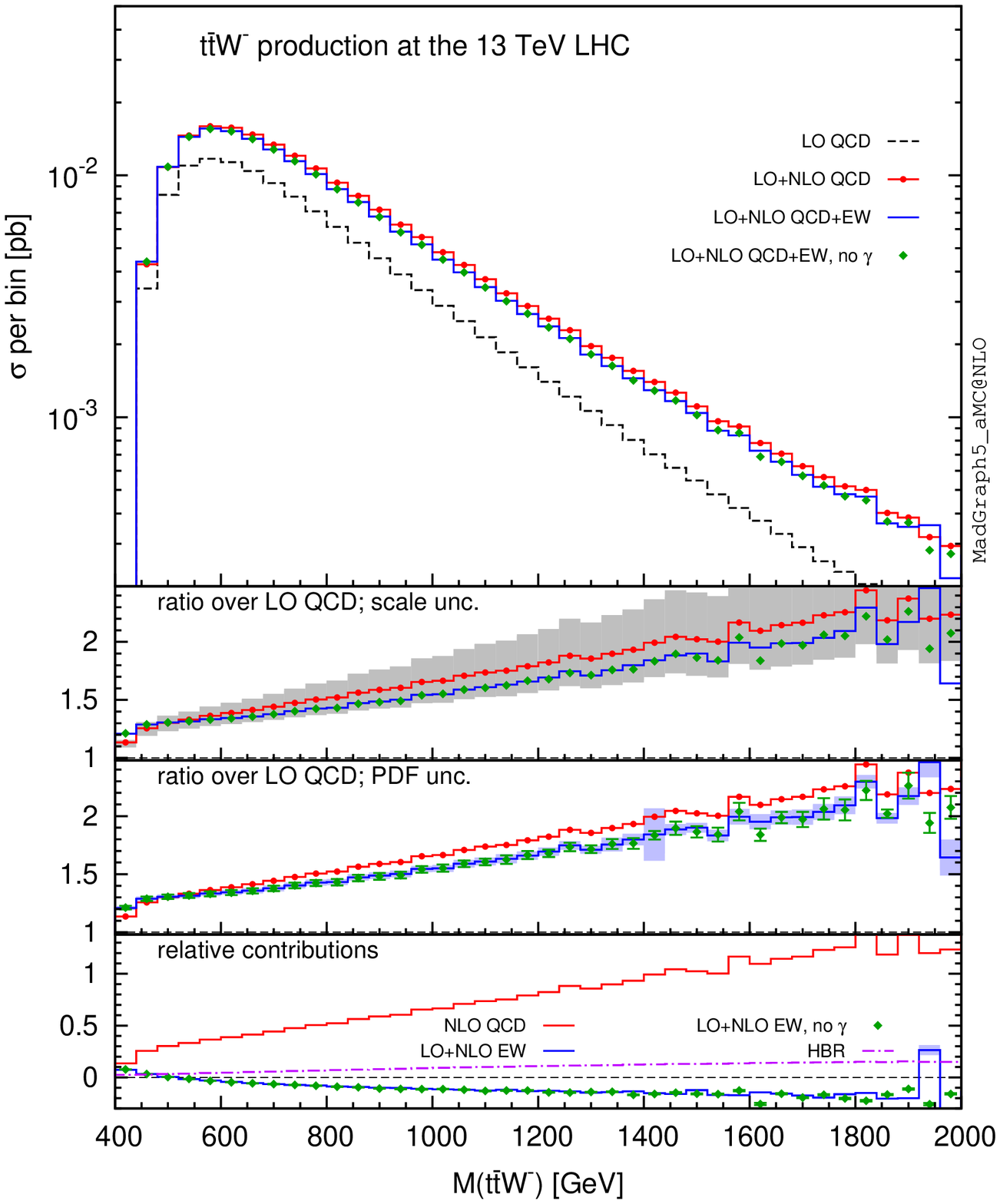}\\
    \includegraphics[width=0.42\textwidth]{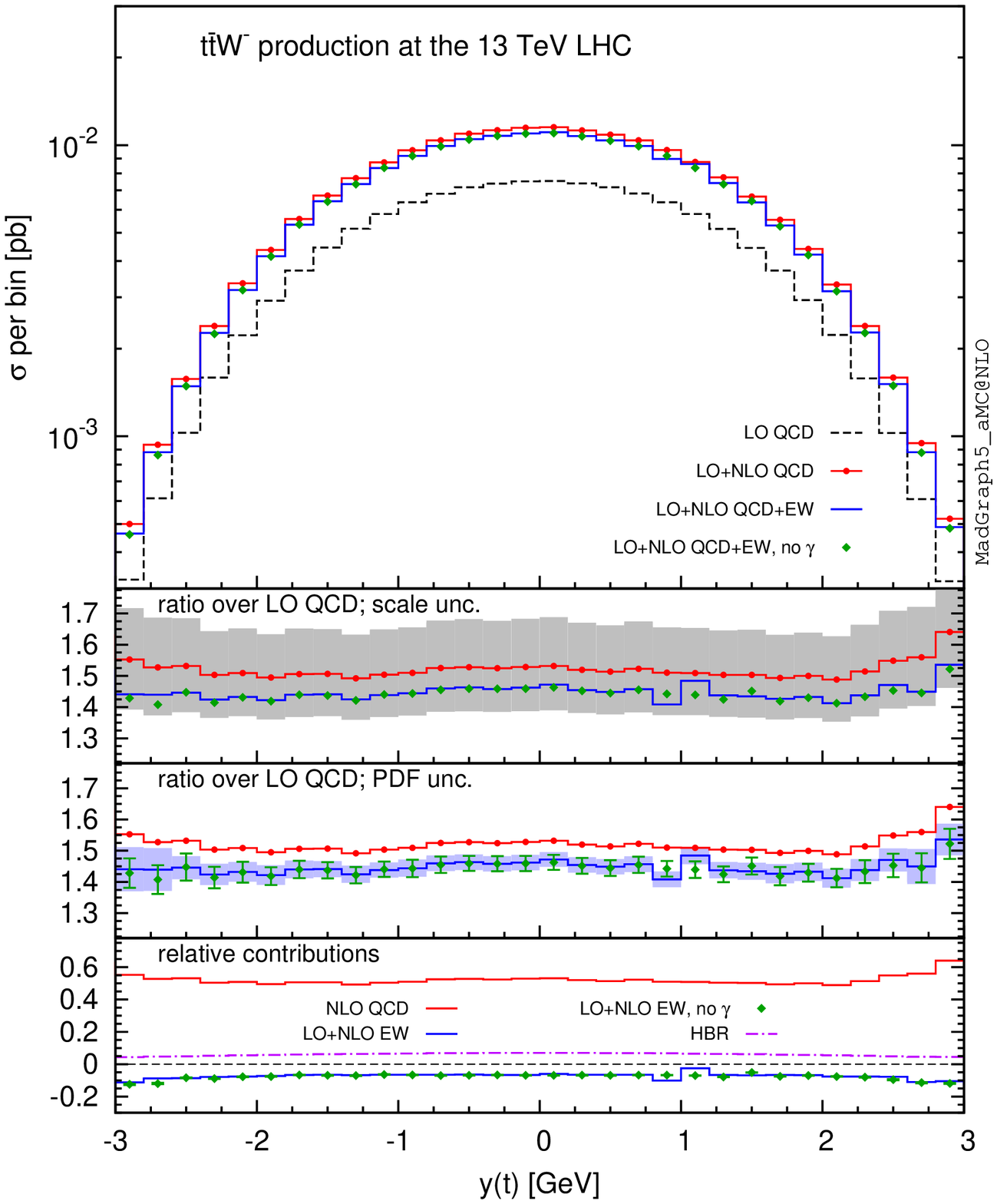}
    \includegraphics[width=0.42\textwidth]{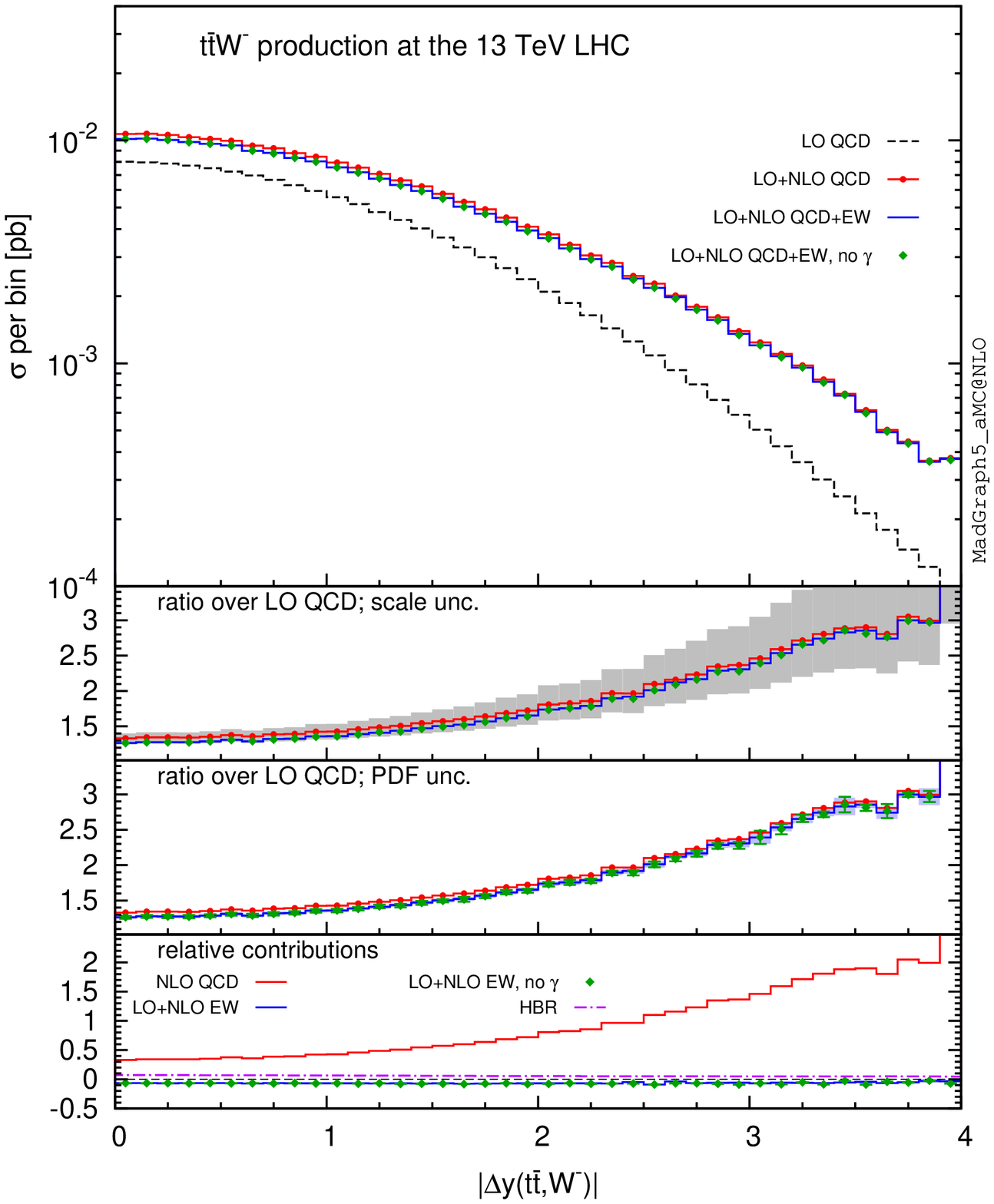}
    \caption{Same as in fig.~\ref{fig:tth13-nocuts}, for $\ttwm$ production.}
    \label{fig:ttwm13-nocuts}
\end{figure}

\begin{figure}[t]
    \centering
    \includegraphics[width=0.42\textwidth]{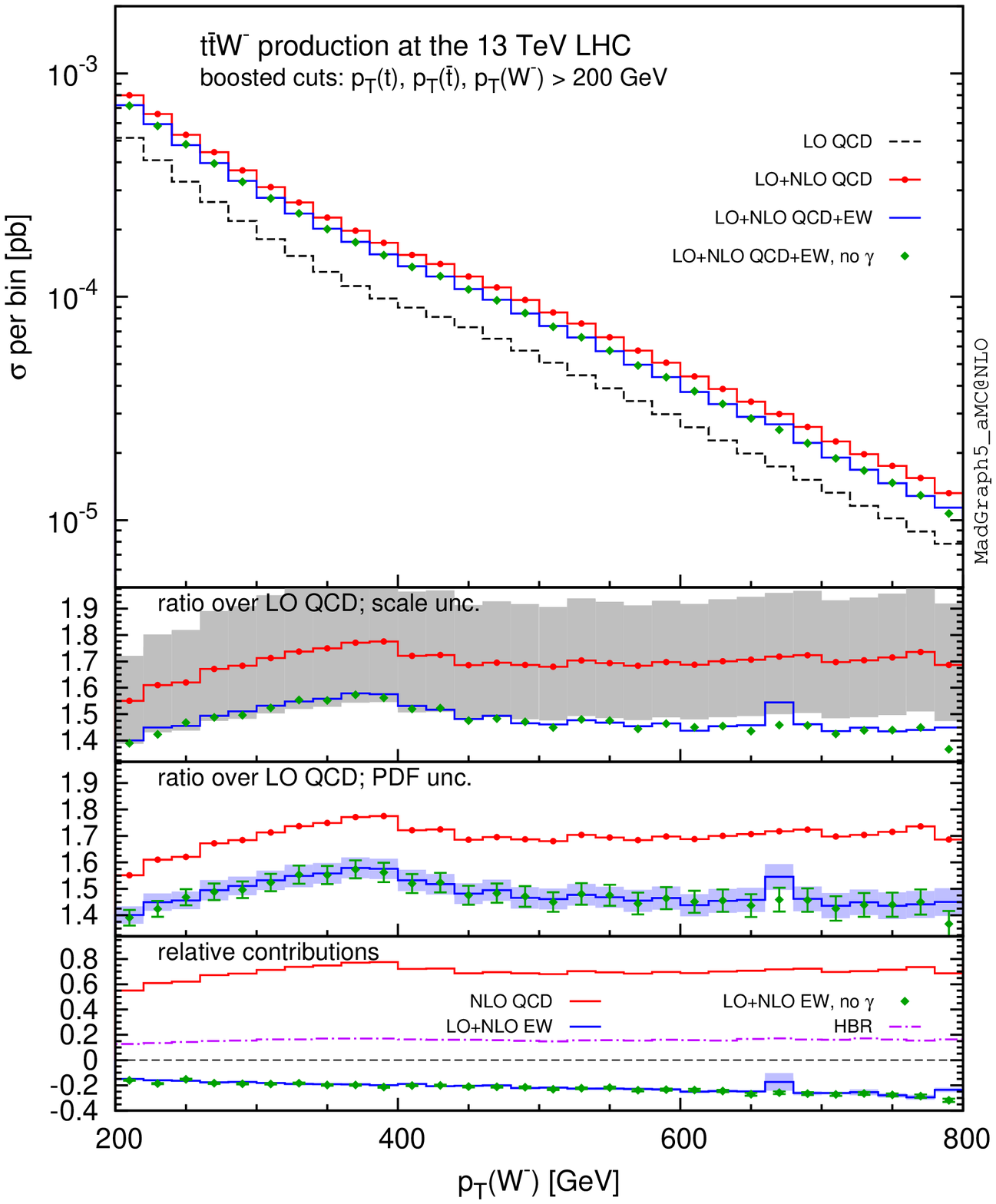}
    \includegraphics[width=0.42\textwidth]{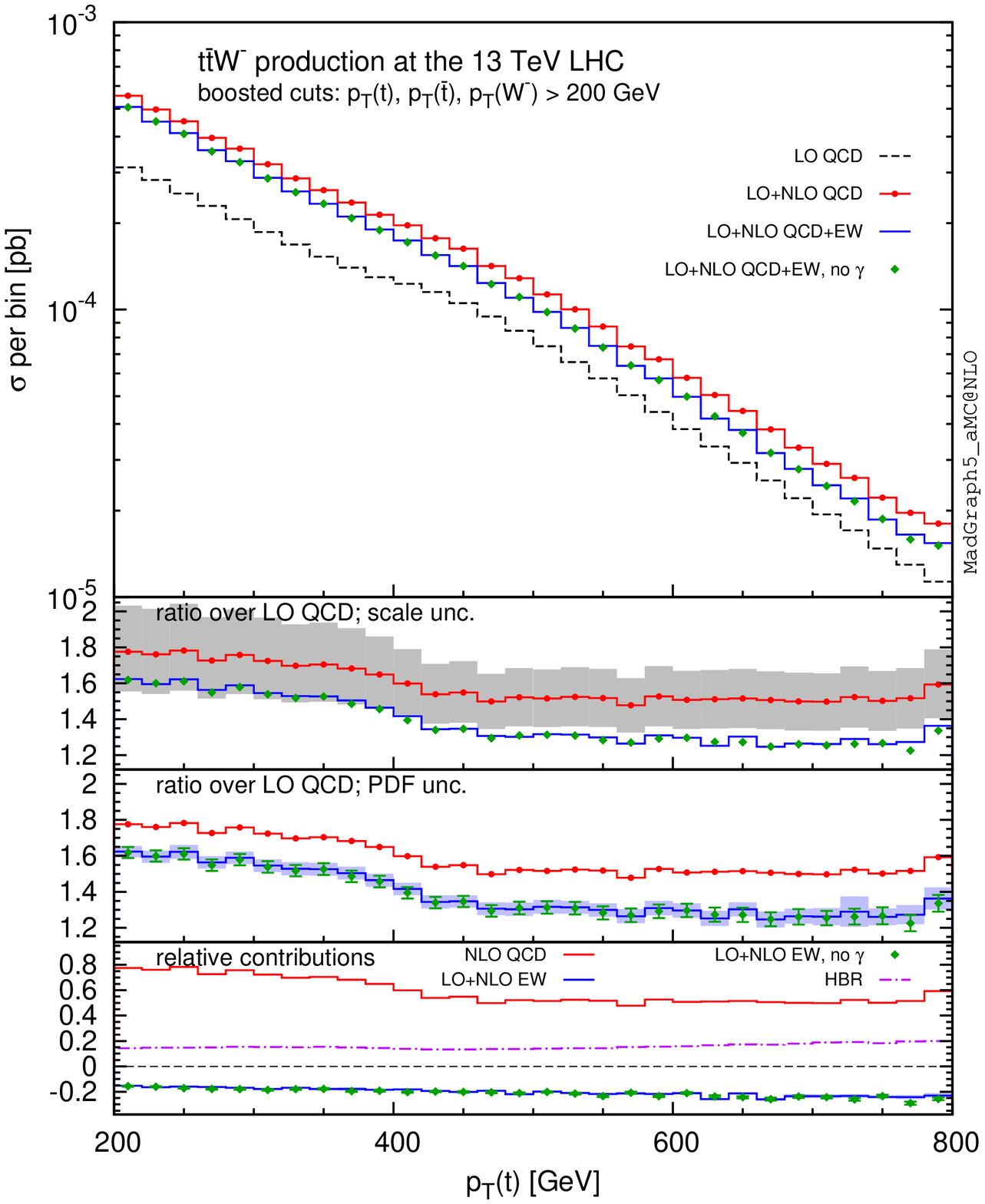}\\
    \includegraphics[width=0.42\textwidth]{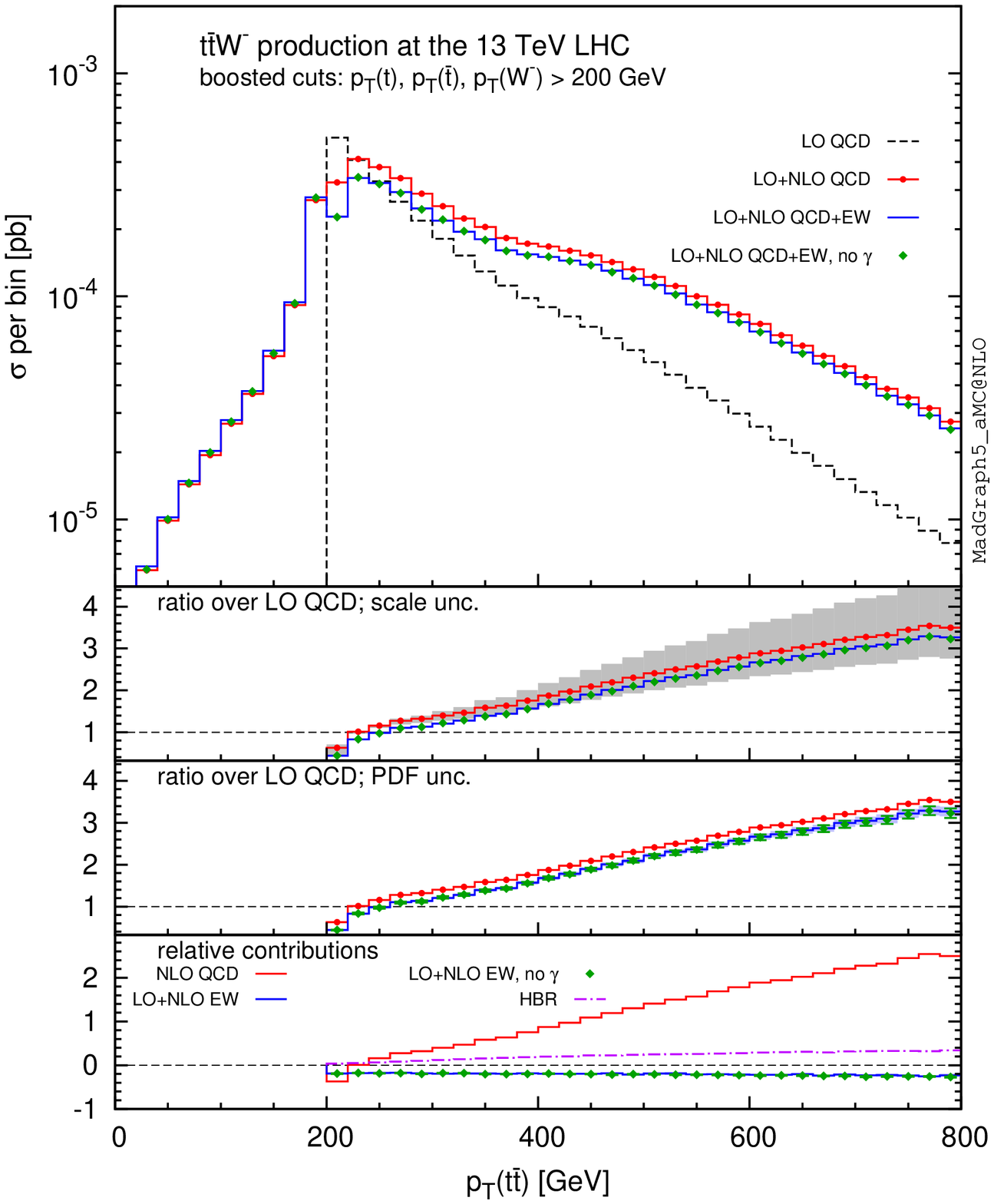}
    \includegraphics[width=0.42\textwidth]{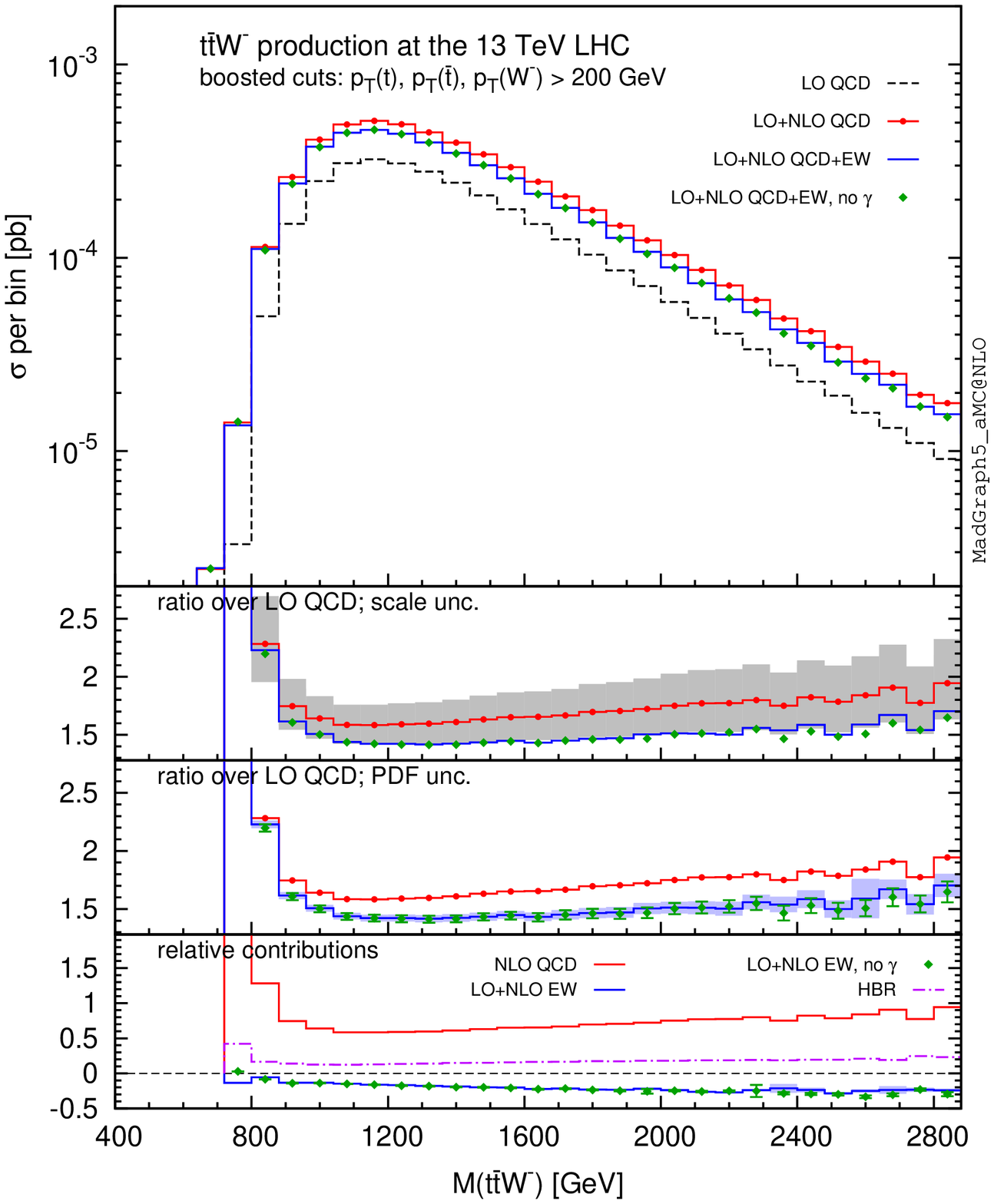}\\
    \includegraphics[width=0.42\textwidth]{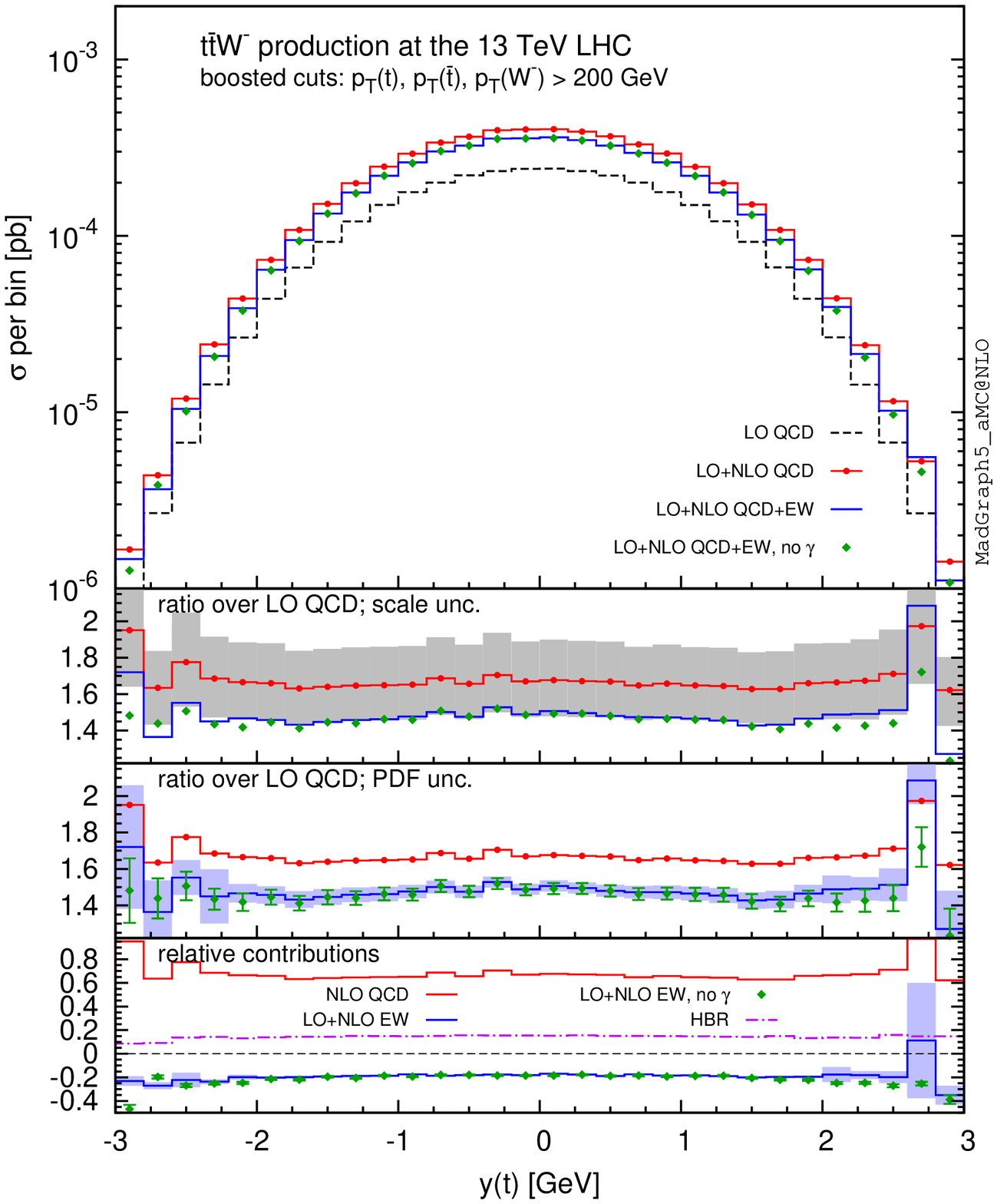}
    \includegraphics[width=0.42\textwidth]{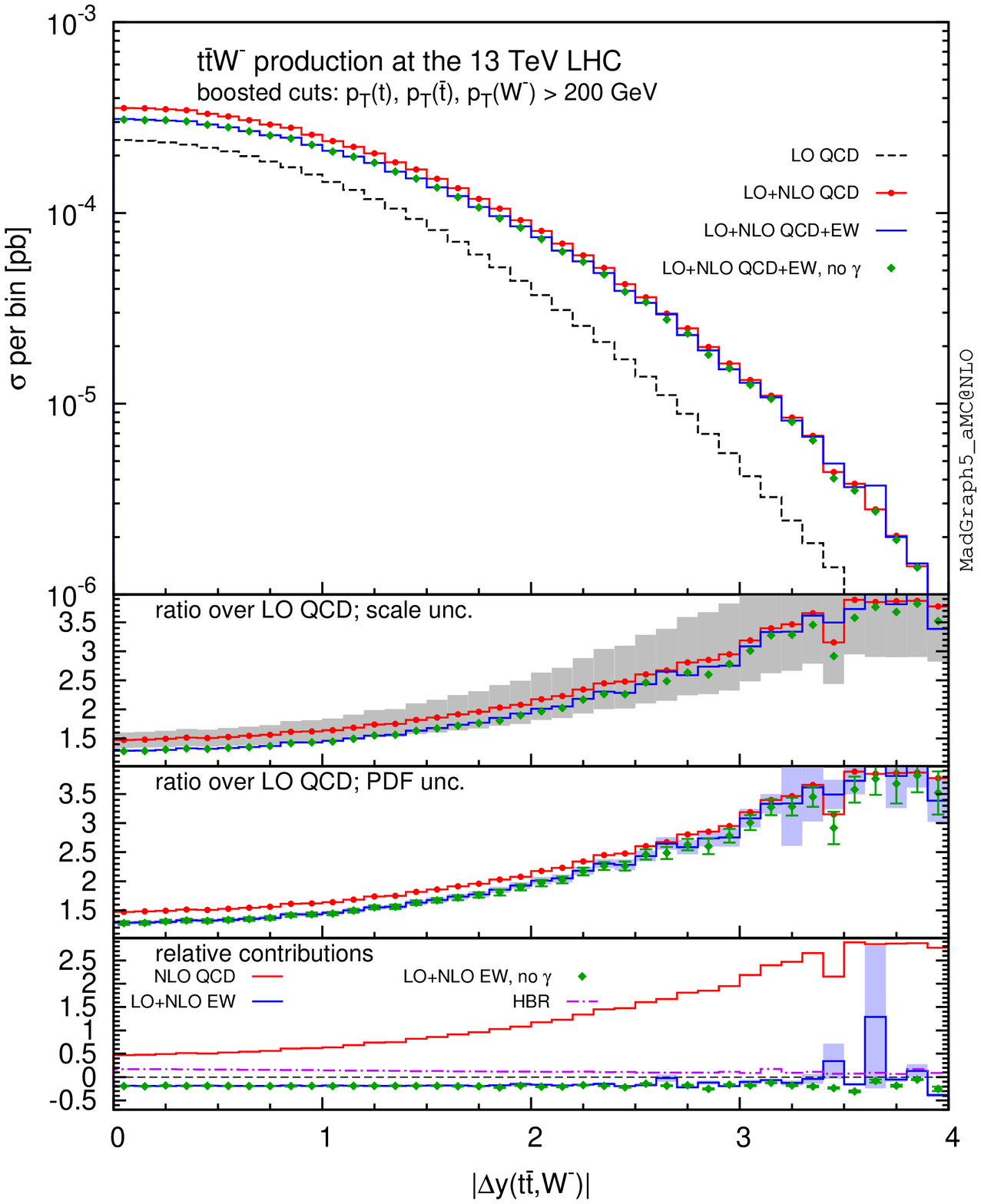}
    \caption{Same as in fig.~\ref{fig:tth13-cuts}, for $\ttwm$ production.}
    \label{fig:ttwm13-cuts}
\end{figure}

In analogy with ref.~\cite{Frixione:2014qaa}, we have considered
the following observables:
\begin{itemize}
\item the transverse momentum of the heavy boson $\pt(V)$;
\item the transverse momentum of the top quark $\pt(t)$; 
\item the transverse momentum of the top-antitop pair $\pt(t\bt)$;
\item the invariant mass of the top-antitop-heavy boson system $M(\ttv)$;
\item the rapidity of the top quark $y(t)$; 
\item the rapidity separation between the top-antitop pair and the 
heavy boson $\Delta y(t\bt,V)$.
\end{itemize}
We present these six observables for each of the four production
processes at the 13-TeV LHC, without and with the cuts of
eq.~(\ref{eq:boosted}): $\ttH$ in figs.~\ref{fig:tth13-nocuts} 
and~\ref{fig:tth13-cuts}, $\ttz$ in figs.~\ref{fig:ttz13-nocuts} 
and~\ref{fig:ttz13-cuts}, $\ttwp$ in figs.~\ref{fig:ttwp13-nocuts}
and~\ref{fig:ttwp13-cuts}, and $\ttwm$ in figs.~\ref{fig:ttwm13-nocuts}
and~\ref{fig:ttwm13-cuts}, respectively.
We use an identical layout for all the plots, with a main frame and three 
insets; we employ the labelling convention introduced in table~\ref{tab:names},
and used in sect.~\ref{sec:rates} for the total rates.
Four histograms appear in the main frame, that represent the differential
cross sections; they are the predictions for $\xLOo$
(dashed black), $\xLOo+\xNLOo$ (solid red overlayed with full circles),
and $\xLOo+\xNLOo+\xLOt+\xNLOt$ (solid blue and green diamonds, 
with and without photon density respectively).
In the upper and middle insets, the bin-by-bin ratios of the latter three 
histograms over the first one (i.e.~$\xLOo$) are presented, by using 
the same patterns as in the main frame. The upper insets also display
a grey band, centered around the $\xLOo+\xNLOo$ prediction, which represents
the fractional scale variation of this cross section. Conversely, the 
middle insets show the fractional PDF uncertainties that affect the full 
NLO cross section $\xLOo+\xNLOo+\xLOt+\xNLOt$: with (blue band) and
without (green error symbols) photon density. Finally,
in the bottom insets we present the ratios of the following three 
quantities over $\xLOo$: $\xNLOo$ (red solid), $\xLOt+\xNLOt$ 
(solid blue with photon density, and green diamonds without photon
density), and HBR (purple dot-dashed). PDF-uncertainty bands
for the relative $\xLOt+\xNLOt$ predictions are also given, with the
same patterns as their analogues in the middle insets.
Note that HBR processes that feature two identical vector bosons 
in the final state might give up to two contributions per event to 
the $\pt(V)$, $M(\ttv)$, and $\Delta y(t\bt,V)$ distributions.

The general features of these differential results are largely
independent of the specific process considered. When no cuts
are applied (figs.~\ref{fig:tth13-nocuts}, \ref{fig:ttz13-nocuts},
\ref{fig:ttwp13-nocuts}, and~\ref{fig:ttwm13-nocuts}), regions
close to threshold (i.e.~associated with small transverse momenta)
are dominated by QCD contributions; by adding EW effects one generally
shifts downwards the cross sections, but still within the theoretical
uncertainties. By moving towards large $\pt$'s, the $K$ factors due
to the leading NLO term tend to increase, for some observables and
processes in a truly dramatic manner. The relative size of EW 
contributions also increases in absolute value in the same regions;
since the corresponding cross sections are negative, this partly compensates
the growth induced by QCD terms. Furthermore, at variance with what
happens at threshold, such a compensation is quite often significant,
being of the same order as, or larger than, the theoretical uncertainty;
the largest effects are seen in $\ttw$ production. This fact, together
with the observation that $K$ factors are not flat, implies that both QCD
and EW higher-order effects need to be taken into account at a fully
differential level for a precision study of $\ttV$ production.
As was already observed in the case of total rates, the impact of
photon-initiated processes is not large on NLO-accurate results; there 
is a slight fractional increase when moving towards large transverse 
momenta, but given the current theoretical uncertainties this is
hardly significant; a similar conclusion applies to the impact of the
photon density uncertainty on the overall PDF errors. The largest
effects are seen in $\ttH$ production. A notable exception to 
these statements is to be found at large top-rapidity values in $\ttH$
and $\ttz$ production, where the cross sections with or without the
photon-initiated processes exhibit large differences. However,
such differences are offset by a very significant increase of the
PDF uncertainty, which is driven by the poorly known photon density.

Given these results, it is not surprising that the various higher-order
effects are enhanced when one imposes the cuts of eq.~(\ref{eq:boosted}) --
see figs.~\ref{fig:tth13-cuts}, \ref{fig:ttz13-cuts},
\ref{fig:ttwp13-cuts}, and~\ref{fig:ttwm13-cuts}. In this case,
the importance of taking into account EW effects is obvious; this
includes the HBR cross sections which, according to our results,
are particularly large for certain observables in $\ttw$ production,
in keeping with what has been already found for total rates.
The sharp thresholds in $\pt(t\bt)$ and $M(\ttV)$, and the knee
at $\pt(t)\sim 400$~GeV, are LO features common to all processes,
which become less dramatic when NLO corrections are included.
Their origins have been discussed in ref.~\cite{Frixione:2014qaa}
(for $\ttH$ production, but those arguments apply to $\ttz$ and $\ttw$
production as well), and therefore will not be repeated here.
%%%%%%%%%%%%%%%%%%%%%%%%%%%%%%%%%%%%%%%%%%%%%%%%%%%%%%%%%%%%%%%%%%%%%%%
%%%%%%%%%%%%%%%%%%%%%%%%%%%%%%%%%%%%%%%%%%%%%%%%%%%%%%%%%%%%%%%%%%%%%%%
%%%%%%%%%%%%%%%%%%%%%%%%%%%%%%%%%%%%%%%%%%%%%%%%%%%%%%%%%%%%%%%%%%%%%%%
%%%%%%%%%%%%%%%%%%%%%%%%%%%%%%%%%%%%%%%%%%%%%%%%%%%%%%%%%%%%%%%%%%%%%%%
%%%%%%%%%%%%%%%%%%%%%%%%%%%%%%%%%%%%%%%%%%%%%%%%%%%%%%%%%%%%%%%%%%%%%%%
%%%%%%%%%%%%%%%%%%%%%%%%%%%%%%%%%%%%%%%%%%%%%%%%%%%%%%%%%%%%%%%%%%%%%%%
%%%%%%%%%%%%%%%%%%%%%%%%%%%%%%%%%%%%%%%%%%%%%%%%%%%%%%%%%%%%%%%%%%%%%%%
%%%%%%%%%%%%%%%%%%%%%%%%%%%%%%%%%%%%%%%%%%%%%%%%%%%%%%%%%%%%%%%%%%%%%%%
%%%%%%%%%%%%%%%%%%%%%%%%%%%%%%%%%%%%%%%%%%%%%%%%%%%%%%%%%%%%%%%%%%%%%%%
%%%%%%%%%%%%%%%%%%%%%%%%%%%%%%%%%%%%%%%%%%%%%%%%%%%%%%%%%%%%%%%%%%%%%%%
%%%%%%%%%%%%%%%%%%%%%%%%%%%%%%%%%%%%%%%%%%%%%%%%%%%%%%%%%%%%%%%%%%%%%%%
%%%%%%%%%%%%%%%%%%%%%%%%%%%%%%%%%%%%%%%%%%%%%%%%%%%%%%%%%%%%%%%%%%%%%%%
%%%%%%%%%%%%%%%%%%%%%%%%%%%%%%%%%%%%%%%%%%%%%%%%%%%%%%%%%%%%%%%%%%%%%%%
%%%%%%%%%%%%%%%%%%%%%%%%%%%%%%%%%%%%%%%%%%%%%%%%%%%%%%%%%%%%%%%%%%%%%%%
%%%%%%%%%%%%%%%%%%%%%%%%%%%%%%%%%%%%%%%%%%%%%%%%%%%%%%%%%%%%%%%%%%%%%%%
%%%%%%%%%%%%%%%%%%%%%%%%%%%%%%%%%%%%%%%%%%%%%%%%%%%%%%%%%%%%%%%%%%%%%%%

\section{Conclusions and outlook\label{sec:outlook}}
We have studied the production of a $t\bt$ pair in association with
a heavy SM boson (Higgs, $Z$, or $W^\pm$) at a $pp$ collider with
three different c.m.~energies (8, 13, and 100~TeV). Our predictions
are obtained by computing the two dominant terms at both the leading
and the next-to-leading order in a mixed perturbative expansion in
the QCD ($\as$) and EW ($\aem$) couplings. Such terms factorise the
coupling combinations $\as^2\aem$ and $\as\aem^2$ at the LO, and
$\as^3\aem$ and $\as^2\aem^2$ at the NLO; the latter two contributions are 
usually denoted as QCD and EW NLO corrections, respectively. 
The ${\cal O}(\as^2\aem^2)$
results for $\ttH$ production had been previously presented in the
literature in refs.~\cite{Frixione:2014qaa,Yu:2014cka} (with the
former paper ignoring QED effects); those for $\ttz$, $\ttwp$, 
and $\ttwm$ are given here for the first time.

These $\ttv$ processes are characterised by tiny cross sections, the total
rates being smaller than 1~pb at LHC energies, and of the order of
10~pb at a 100-TeV collider. However, in view of the luminosity
foreseen at the LHC Run II and at future colliders, it will become possible 
to measure them with a good accuracy; furthermore, $\ttz$ and $\ttw$ are 
significant backgrounds to several BSM searches, that typically feature 
multi-lepton final states. There are thus compelling phenomenological 
motivations to increase the precision of the theoretical predictions for 
$\ttV$ production, which is what we have done in this paper. From 
a technical viewpoint, our calculations have been performed with 
\aNLO~\cite{Alwall:2014hca}, and are fully automated. They are 
the first public results obtained with such a code that include
QED subtractions, and they constitute part of the validation
procedure that will lead to the public release of an upgraded
\aNLO\ capable of handling mixed-coupling expansions at the NLO.
We point out that no part of the code has been specifically constructed
or modified in order to handle $\ttV$ production.

The main findings of our study are the following. The fully differential
computation of higher-order corrections is essential in order to attain
a realistic description of the processes at hand. 
$K$ factors are large and not flat, and tend to grow 
with the collider c.m.~energy; such a growth is particularly spectacular
in the case of $\ttw$ production, owing to the impact of gluon-initiated
partonic processes which are absent at the LO (at variance with what 
happens for $\ttH$ and $\ttz$). At a given collider energy, a similar
feature is observed when at least one of the final-state particles has
a large transverse momentum. Theoretical uncertainties are predominantly
due to scale variations; PDF errors are smaller but not negligible.
For the same reason as explained above, scale uncertainties become
very significant in $\ttw$ production at high energies and/or transverse
momenta. The considerable size of higher-order corrections stems predominantly
from ${\cal O}(\as^3\aem)$ terms. However, effects of EW origin do 
change the pure-QCD results in a way which is significant 
(i.e.~of the same order as, or larger than, the theoretical uncertainties)
at large energies and $\pt$'s. The precise impact of such effects depends 
on the observable (and, of course, on the process), which implies again
the necessity of performing one's calculations in a fully differential manner.
QED corrections lead to the inclusion of processes with initial-state
photons. We find that these give a modest fractional contribution to 
the NLO ${\cal O}(\as^2\aem^2)$ term, but a very large one to the LO
${\cal O}(\as\aem^2)$ term (for $\ttH$ and $\ttz$, being identically 
zero in $\ttw$ production); however, the latter does not induce
a visible change in the physical NLO-accurate cross section, since
the corresponding contribution is quite small in absolute value.
These facts imply that, although the large uncertainties on the photon
PDF are sizable when one only considers second-leading LO and NLO 
terms, they essentially become irrelevant as far as the overall
uncertainties on the NLO-accurate results are concerned (with the 
exception of large top rapidities in $\ttH$ and $\ttz$ production).
Finally, we find that processes with a $t\bt$ and a heavy-boson
{\em pair} in the final state, which we have called HBR, {\em might} be 
responsible for detectable effects, particularly in the case of $\ttw$
production: a definite conclusion on this point can only be reached
with a realistic acceptance study (either by including HBR contributions
in the $\ttV$ cross section inclusive in $V$, or by subtracting them
if the $\ttV$ cross section is exclusive in a single $V$).
\vfill

\section*{Acknowledgements}
We are grateful to J.~Rojo for clarifications concerning NNPDF2.3QED
and for comments on the manuscript, to S.~Uccirati for producing 
pointwise results with {\small\sc RECOLA} for cross-checks,
and to R.~Frederix, F.~Maltoni, and M.~Mangano for discussions and
collaboration at different stages of this work. This work is 
supported in part (DP and HSS) by, and performed in the framework 
of, the ERC grant 291377 ``LHCtheory: Theoretical predictions 
and analyses of LHC physics: advancing the precision frontier". 
The work of VH is supported by the SNF with grant PBELP2 146525.
The work of MZ is supported by the ERC grant ``Higgs@LHC'', in
part by the Research Executive Agency (REA) of the European Union under 
the Grant Agreement number PITN-GA-2010-264564 (LHCPhenoNet), and
in part by the ILP LABEX (ANR-10-LABX-63), in turn supported by French 
state funds managed by the ANR within the ``Investissements d'Avenir'' 
programme under reference ANR-11-IDEX-0004-02.

\bibliographystyle{JHEP}
\bibliography{ttV}
\end{document}